\documentclass[preprint,prd,tightenlines,nofootinbib,eqsecnum,superscriptaddress]{revtex4-2}

\usepackage{amsmath}
\usepackage{amsfonts}
\usepackage{amssymb}
\usepackage{amsthm}
\usepackage{amstext}

\allowdisplaybreaks

\usepackage{mathrsfs}
\usepackage{bm}
\usepackage{graphicx}
\usepackage{float}
\usepackage{xcolor}
\usepackage{array}
\usepackage{booktabs}
\usepackage{verbatim}
\usepackage{pagecolor}
\usepackage{enumerate}

\usepackage{mathtools,xcolor}
\usepackage{multirow}
\usepackage{orcidlink}
\usepackage[normalem]{ulem}
\usepackage{scalerel}
\usepackage{textcomp}
\usepackage{ulem}
\usepackage{url}
\usepackage{comment}

\usepackage{hyperref}
\hypersetup{
    colorlinks=true,
    linktocpage,
    unicode,
    linkcolor=blue,
    filecolor=magenta,      
    urlcolor=cyan
}

\newcolumntype{C}{>{$}c<{$}}
\AtBeginDocument{
\heavyrulewidth=.08em
\lightrulewidth=.05em
\cmidrulewidth=.03em
\belowrulesep=.65ex
\belowbottomsep=0pt
\aboverulesep=.4ex
\abovetopsep=0pt
\cmidrulesep=\doublerulesep
\cmidrulekern=.5em
\defaultaddspace=.5em
}

\definecolor{red  }{rgb}{1,0,0}
\definecolor{blue }{rgb}{0,0,1}
\definecolor{green}{rgb}{0,1,0}
\definecolor{CiteColor}{rgb}{0,0.5,0}
\definecolor{RefColor}{rgb}{0.55,0,0}

\definecolor{CiteColor}{rgb}{0,0,0.35}
\definecolor{URLColor}{rgb}{0,0,0.35}

\newcommand{\mathd}{\mathrm{d}}
\newcommand{\mathpi}{\pi}

\newcommand{\tmop}[1]{\ensuremath{\operatorname{#1}}}

\DeclareSymbolFontAlphabet{\mathrsfs}{rsfs}

\hypersetup{colorlinks,citecolor=CiteColor,urlcolor=URLColor}
\hypersetup{citecolor=CiteColor}
\hypersetup{linkcolor=RefColor}

\newcommand{\beq}{\begin{equation}}
\newcommand{\eeq}{\end{equation}}
\newcommand{\ud}{\mathrm{d}}

\newcommand{\ui}{\mathrm{i}}

\newcommand{\scL}{\mathscr{L}}

\newcommand{\scE}{\mathscr{E}}

\newcommand{\RR}{\mathbb{R}}

\newcommand{\NN}{\mathbb{N}}

\newcommand{\TT}{\mathbb{T}}
\newcommand{\tTD}{\text{\tiny{TD}}}

\newcommand{\fK}{\mathfrak{K}}
\newcommand{\fQ}{\mathfrak{Q}}

\newcommand{\mcT}{\mathcal{T}}
\newcommand{\mcJ}{\mathcal{J}}

\newcommand{\mcL}{\mathcal{L}}
\newcommand{\mcN}{\mathcal{N}}
\newcommand{\mcM}{\mathcal{M}}
\newcommand{\mcP}{\mathcal{P}}
\newcommand{\mcG}{\mathcal{G}}

\newcommand{\sqf}{\sqrt{f}}
\newcommand{\length}{\text{length}}

\newcommand{\quand}{\quad \text{and} \quad}

\begin{document}

\title{Symplectic mechanics of relativistic spinning compact bodies. II.\texorpdfstring{\\}{}Canonical formalism in the Schwarzschild spacetime}

\author{Paul Ramond,\orcidlink{0000-0001-7123-0039}}\,\email{paul.ramond@obspm.fr}
\affiliation{IMCCE, Observatoire de Paris, Université PSL, 77 Avenue Denfert-Rochereau, FR-75014
Paris}

\author{Soichiro Isoyama,\orcidlink{0000-0001-6247-2642}}\,\email{isoyama@yukawa.kyoto-u.ac.jp}
\affiliation{Department of Physics, National University of Singapore, 
21 Lower Kent Ridge Rd, 119077 Singapore}

\date{\today}

\begin{abstract}
This work constitutes the second part of a series of studies that aim to utilise tools from Hamiltonian mechanics to investigate the motion of an extended body in general relativity. 
The first part of this work [Refs.~\cite{Ra.PapI.24,Ra.PRL.24}] constructed 
a ten-dimensional, covariant Hamiltonian framework encompassing all the linear-in-spin corrections from the secondary spin to the geodesic motion in arbitrary spacetime. This framework was proven to be integrable in the Schwarzschild and Kerr spacetimes, specifically.   
The present work translates this abstract integrability result into tangible applications for linear-in-spin Hamiltonian dynamics of a compact object in a Schwarzschild spacetime. In particular, a canonical system of coordinates is constructed explicitly, which exploits the spherical symmetry of the Schwarzschild spacetime. These coordinates are based on a relativistic generalization of the classical Andoyer variables of Newtonian rigid body motion. This canonical setup allows us to derive ready-to-use formulae for action-angle coordinates and gauge-invariant Hamiltonian frequencies, which automatically include all linear-in-spin effects. No external parameters or ad hoc choices are necessary, and the framework can be used to find complete solutions by quadrature of generic (bound or unbound), linear-in-spin orbits, including orbital inclination, precession and eccentricity, as well as spin precession. The efficacy of the formalism is demonstrated here in the context of circular orbits with arbitrary spin and orbital precession, with the results validated against known results in the literature.  
\end{abstract}

\maketitle


\clearpage

%

\section*{Introduction and Summary}

\subsection{Introduction}
This paper is the second part of our series of work [Refs.\cite{Ra.PapI.24,Ra.PRL.24}], 
which aim at formulating the motion of a small body in curved spacetime 
as a covariant Hamiltonian system, while accounting for the proper rotation of the body, 
or \textit{spin}. The general goal of this programme is to develop a precise yet efficient model 
of the relativistic orbital dynamics of such two-body systems, making use of classical tools from Hamiltonian mechanics. 

The main motivation of our efforts comes from extreme mass-ratio inspirals (EMRIs)  
– the gradual inspiral of a stellar-mass compact object 
(typically a neutron star or black hole or even an ``exotic'' compact object) 
into a supermassive black hole~\cite{Amaro-Seoane:2012lgq}. 
These binary inspirals are promising sources of gravitational waves 
for an interferometric detector in space  
(e.g., the Laser Interferometer Space Antenna~\cite{Audley:2017drz,Colpi:2024xhw}, 
as well as DECIGO~\cite{Kawamura:2020pcg}, 
TianQin, and Taiji~\cite{TianQin:2020hid,2021PTEP.2021eA108L,Gong:2021gvw,Li:2024rnk}) 
and they will eventually provide an exquisite probe of the strongly gravitating region 
close to a black hole; 
for example, the source’s astrophysical parameters, 
such as the masses and the spin of the larger central black hole, 
can be measured with unparalleled precision~\cite{Babak:2017tow,Fan:2020zhy}.
However, the gravitational-wave signals from EMRIs will be so intricate 
and sufficiently weak that they can only be extracted 
from the noisy detector data with modelled waveforms, 
that match the phase of the signal over its full duration 
(scaling like the inverse of the mass ratio) 
across all possible orbital and spin configurations 
around a central black hole~\cite{LISAWHITE.23}. 
As a result, the scientific promise described above will be attained 
only with a highly accurate and efficient model of relativistic two-body systems, 
built on the detailed understanding of their orbital dynamics 
in the (multidimensional) EMRI parameter space.  

In what follows, we give some more details about the relativistic two-body problem 
in the extreme mass-ratio regime, 
and motivate our work by emphasising the need for a Hamiltonian treatment 
of the secondary spin.


Over the years, exact numerical relativity simulations~\cite{
Jani:2016wkt,Duez:2018jaf,Boyle:2019kee,Healy:2022wdn,Hamilton:2023qkv,SACRA,Wittek:2024pis}, 
analytical approximation schemes~\cite{FerrGual.08,Poisson:2011nh, Bl.14,BaPo.18,PoWa.21,Porto:2016pyg,Levi:2018nxp,Bern:2019nnu,Kalin:2020fhe,Driesse:2024xad,Long:2024ltn} 
and effective models~\cite{Damour:2012mv,Damour:2016gwp,Pratten:2020fqn,Pratten:2020ceb,Pompili:2023tna,Khalil:2023kep,Albertini:2023aol,Buonanno:2024byg} 
have been devised to tackle 
the relativistic two-body problem~\cite{LISAConsortiumWaveformWorkingGroup:2023arg}, 
each specializing to some type of source, 
but also overlapping with the other in their respective domain of validity~\cite{Le2.14,vandeMeent:2020xgc,Gralla:2021qaf,Albertini:2022rfe,Albertini:2022dmc,Barack:2023oqp}. 
%
A point mass in general relativity travels on a pre-defined geodesic in the curved geometry of spacetime. However, this feature is only an approximation in the test-particle limit, as revealed by multipolar expansion schemes to a small, extended, compact object (surrounded by a vacuum region) ~\cite{Di.74,StPu.12,Ha.12,Ha.15}. 
This modelling approach represents the extended object as a point particle 
but endowed with a finite number of multipoles, 
encompassing mass, linear and angular momentum, deformability coefficients, among others. 
The granularity of this description increases with the number of multipoles included, 
providing a more realistic description of the object.

In this multipolar expansion approach to the two-body dynamics, 
the associated small parameter is the typical
length scale of the object over the typical length scale of the background spacetime, 
which is therefore tailor-made for modelling asymmetric binaries, in particular EMRIs. 
At the ``zeroth'' order, the motion of a smaller compact object is well approximated 
by a test body moving along a geodesics of the fixed background spacetime. 
However, this is only a crude approximation. Additional effects arise and lead to corrections of this geodesic motion~\cite{Ha.12}. These include (i) conservative and dissipative self-field effects of the small object, owing to the non-linearity of general relativity (GR) and the emission of gravitational waves, and (ii) extended-body effects, due to the compact object not being exactly a point, but rather an extended object which can spin and deform. All these effects are important for measuring EMRIs by space-based gravitational-wave detectors, 
as all of them have an impact of quadratic order in the mass ratio in the waveform phase \cite{PoWa.21,MaPoWa.22}, 
which is considered to be the minimal accuracy required for parameter estimation \cite{LISAWHITE.23,Burke:2023lno}. 

In the present paper (and its foundation \cite{Ra.PapI.24,Ra.PRL.24}),
we leave the self-force effects aside 
(i.e. we neglect the back-reaction of the small object’s gravitational field),
and solely focus on the leading order, linear-in-spin corrections 
from the secondary's extended body to the geodesic motion. 
%
At this linear-in-spin order, although it is specific to GR, the equations that describe the evolution of linear and angular momenta along the particle's worldline are purely kinematical 
(no body-specific multipolar forces or torques)
and universal (all bodies move the same way) \cite{Ha.15,Ha.23}. 
Body-specific forces and torques only arise at the next quadrupolar order, 
with quadratic-in-spin effects \cite{Ha.20}.  

Analytically solving the relevant linear-in-spin evolution equations, 
now known as the Mathisson-Papapetrou-Tulczyjew-Dixon (MPTD) equations, 
on an astrophysically relevant Kerr black-hole background has been the subject 
of many past studies using a whole spectrum of methods.
Among them, state-of-the-art methods use the powerful formalism of the Marck tetrad \cite{Marck.83,VdM.20} for the spin to solve the MPTD equations directly in a ``perturbed geodesic'' philosophy. This method is very natural given the quasi-geodesic nature of linear-in-spin dynamics, especially leading to purely analytical results for generic orbits in the Schwarzschild spacetime \cite{WiPi.23} and simple orbital configurations in Kerr~\cite{10.1093/mnras/189.3.621,Skoupy:2021asz,Skoupy:2022adh,MaPoWa.22,Mathew:2024prep}. 
A hybrid numerical/analytical result exists for more generic orbits~\cite{10.1046/j.1365-8711.1999.02754.x,DruHug.I.22,DruHug.II.22,Drummond:2023wqc}. 
As such, there is a sense in which this Marck formalism offers a simpler implementation 
than other means of obtaining the solutions to the linear-in-spin equations of motion. 
Having the solution trajectory, however, is not the only way to understand  
the linear-in-spin dynamics of the MPTD system. 
Many other useful characterizations such as \textit{integrability} of the system, 
in which the equations of motion can be solved by quadrature 
(as in, e.g., Refs.~\cite{WitzHJ.19,WiPi.23}) for any orbital configurations, 
may require another approach to the MPTD problem. 

A Hamiltonian formulation just offers such an approach. In the test-mass approximation (i.e. with no spin), the powerful machinery of Hamiltonian mechanics has been already proven its value 
in the context of motion in curved spacetime. A non-exhausted list of examples includes 
(but not limited to) the Hamilton-Jacobi equation \cite{Carter.68}, action-angle variables \cite{Schm.02,WitzAA.22}, the phase space analysis \cite{Dean.99}, 
Hamiltonian frequencies and resonances \cite{Schm.02,BrGeHiPRL.15}, 
the first law of mechanics~\cite{GrLe.13,ALT.14,Fu.al.17}, 
the perturbation (KAM-type) theory \cite{HiFl.08,Xue.20}, Birkhoff normal forms (through Lie series) \cite{Kera.al.23} symplectic integrators \cite{Wu.al.21,Wang.al.21}. 
From the mathematical point of view, these tools are indeed brought by the unique integrability of Kerr geodesic dynamics as a Hamiltonian system~\cite{HuSo.73}. 

Nonetheless, we are nowhere near having a Hamiltonian understanding of these Kerr, linear-in-spin dynamics as thorough as that of the Kerr geodesics dynamics. The main issue with including the spin is that the treatment becomes necessarily perturbative, since the MPTD equations themselves are always truncated to some multipolar order, for physical consistency (see details in Sec.~\ref{subsec:recap}.) This perturbative treatment makes ordinary Hamiltonian concepts such as ``an integrable system'' or the definition of action-angle variables perturbative as well. We shall follow this approach here as well, and refer to Ref.~\cite{RaGi.25} for details on the mathematical well-posedness of these concepts. On top of this difficulty, the requirement for a spin-supplementary condition (SSC) when the spin is involved in the dynamics implies the existence of algebraic constraints in addition to the differential equations of motion. From a symplectic mechanics point of view, this means that one must reduce the original, non-constrained phase space, to a lower-dimensional one, where the SSC holds. Hamiltonian schemes including linear-in-spin effects exist, for example those constructed in 
Refs.~\cite{Souri.70,Damour:2024mzo,Souriau:2024qis,Ba.al.09,WiStLu.19} 
(see also Refs.~\cite{Ra.PapI.24} for detailed references). In particular, based on Ref.~\cite{WiStLu.19}, the framework presented in Ref.~\cite{WitzHJ.19} allowed a well-posed definition of frequencies of motion and (analytical) solutions for spherically symmetric backgrounds in Ref.~\cite{WiPi.23,Skoupy:2024jsi} 
as well as those for the Kerr background in Ref.~\cite{Piovano:2024yks} 
under the Tulzyjew-Dixon (TD) spin supplementary condition. It is, however, based on the Marck tetrad formalism in Ref.~\cite{Marck.83}, which intrinsically relies on describing spinning trajectories as perturbation of (non-spinning) geodesics. In a similar vein, Refs.~\cite{Ba.al.09,KuLeLuSe.16} (and references therein) established the linear-in-spin integrability of the system under another spin supplementary condition, the Newton-Wigner one. 

It is not trivial to extend these linear-in-spin formalisms to the quadratic-in-spin case: the Marck tetrad formulation relies on the parallel transport equation for the spin, which is not valid beyond linear order, while the Newton-Wigner spin supplementary condition does not lead to the existence of extra constants of motion, which are key to exploit the quasi-integrability of the dynamics. The present paper, and its foundation in Refs.~\cite{Ra.PRL.24,Ra.PapI.24}, aim at providing a rigorous treatment of the linear in spin dynamics, backed up by mathematical theorems on Poisson geometry, for the Hamiltonian analysis at linear order in spin, in a Schwarzschild background. It will provide a basis on which to extend the analysis to quadrupolar, quadratic-in-spin dynamics in subsequent works.

\subsubsection{This work: the canonical formalism in the Schwarzschild spacetime}

With these motivations in mind, as a preliminary analysis, we examine Schwarzschild geometry and consider a \textit{spinning} compact object moving within this fixed background spacetime (i.e., without self-field effects) to the linear order in spin. Future publications will present extensions to the more astrophysically relevant Kerr case.

The classical multipolar expansion schemes (reviewed in Sec.~\ref{subsec:recap}) assert that this object can be modelled as a point particle endowed with a linear momentum 1-form $p_a$ and spin tensor $S^{ab}$. Their evolution along the particle's worldline $\scL$, at linear order in $S^{ab}$ and under the Tulczyjew-Dixon (TD) spin supplementary condition (which fixes a unique representative worldline for the extended body), is given 
by the following set of (linearized) MPTD equations
\beq \label{MPTDintro}
\nabla_{u} p_a = R_{abcd}S^{bc}u^d \,, \quad \nabla_{u} S^{ab}=0 \quand p_a S^{ab}=0 \,,
\eeq
where $u^a=p^a/\mu$ is the particle's four-velocity, tangent to $\scL$. 
The system described in Eq.~ \eqref{MPTDintro} is made of $10$ differential equations and $3$ algebraic equations, for $13$ unknowns in $u^a,p_a$ and $S^{ab}$. It encodes everything that we have to know about the object's dynamics at linear order in spin: its worldline $\scL$ in $x^\alpha$ and 
how its energy-momentum content evolves along $\scL$ in $(p_a, S^{ab})$.

As previously outlined by one of us (PR) in Ref.~\cite{Ra.PapI.24} 
(hereafter referred to as ``Paper I'')
the algebraic-differential system in Eq.~\eqref{MPTDintro} can be cast 
into a covariant Hamiltonian formulation on a $10$ dimensional ($10$D) phase space $\mcP$. 
Most importantly, Paper I demonstrates that, when applied to a background Kerr spacetime, 
the resulting Hamiltonian system is integrable in the usual, Arnold-Liouville sense [cf Ref.~\cite{Liou1855,Arn}].

At the level of foundations, the integrability can be proved without any explicit coordinate system in $\mcP$. At the level of practical implementation, however, 
we yet have to identify the coordinate system (particularly the canonical one) 
to translate the mathematical property of the integrability 
into the powerful and unique tools in Hamiltonian mechanics.  

In the present paper we take on that endeavour, and construct canonical coordinates 
on $\mcP$ explicitly (see also Ref.~\cite{Witz.MG.23} for other attempts in this
direction). 
Based on these coordinates, we also derive a manifestly integrable Hamiltonian system which is effectively 1D system. This allows us to construct action-angle coordinates in a straightforward way. These action-angle coordinates permit the calculation of frequencies associated with oscillatory or rotational motion without the need to solve the equations of motion. This makes them a valuable tool in practical calculations.

It is important to acknowledge that there are several independent research projects aiming to derive action-angle variables for spinning secondaries orbiting a primary black hole; 
see, for example, Refs.~\cite{Tanay:2021bff,WitzHJ.19,WiPi.23,Gonzo:2024zxo,Witzany:2024ttz}.  
These works generalize our own in the sense that they are 
for primary Kerr black holes, not just for Schwarzschild primaries. 
They are also restricted versions of our own, 
in the sense that they either are concerned with unbound orbits, assume the weak-field, post-Newtonian gravity, or are based on the Marck tetrad formalism which is limited to the linear-in-spin approximation, while our framework \cite{Ra.PRL.24,Ra.PapI.24} can be extended to the quadratic-in-spin case.
In the context of EMRI modelling, we expect that the Hamiltonian frequencies 
will effectively characterize the periodicity of the orbit, 
in a manner analogous to that observed in EMRI systems with a non-spinning secondary, 
as described in Refs.~\cite{Schm.02,HiFl.08}. 

\subsection{Outline of the work}
\label{intro:outline}

We shall provide in this section an overview of our paper and its main results. 
We begin in Sec.~\ref{sub2sec:intro-reduction} by outlining the construction 
of the canonical coordinate of the physical phase space $\mcP$, 
followed by an examination of the associated canonical formalism 
of the Hamiltonian system on phase space $\mcP$ in Sec.~\ref{sub2sec:intro-P}. 
The action-angle coordinates and Hamilton frequencies of the system are presented 
in Sec.~\ref{sub2sec:intro-AA}.

As previously stated, the present work is based on our general, covariant 
Hamiltonian framework for the dynamics of an extended (`test') body in curved spacetime. 
See already Sec.~\ref{sec:Ham-Schw} a review of the framework presented 
in Paper I~\cite{Ra.PapI.24}; for a more detailed overview, 
readers may also consult the letter version of Paper I: Ref.~\cite{Ra.PRL.24}.


\subsubsection{Brief overview of the construction}
\label{sub2sec:intro-reduction}

This will be the content of Sec.~\ref{sec:Ando}-~\ref{sec:Reduc} 
of the body of the paper. We outline the main points here.

\begin{figure}[H]
    \begin{center}
    	\includegraphics[width=0.75\linewidth]{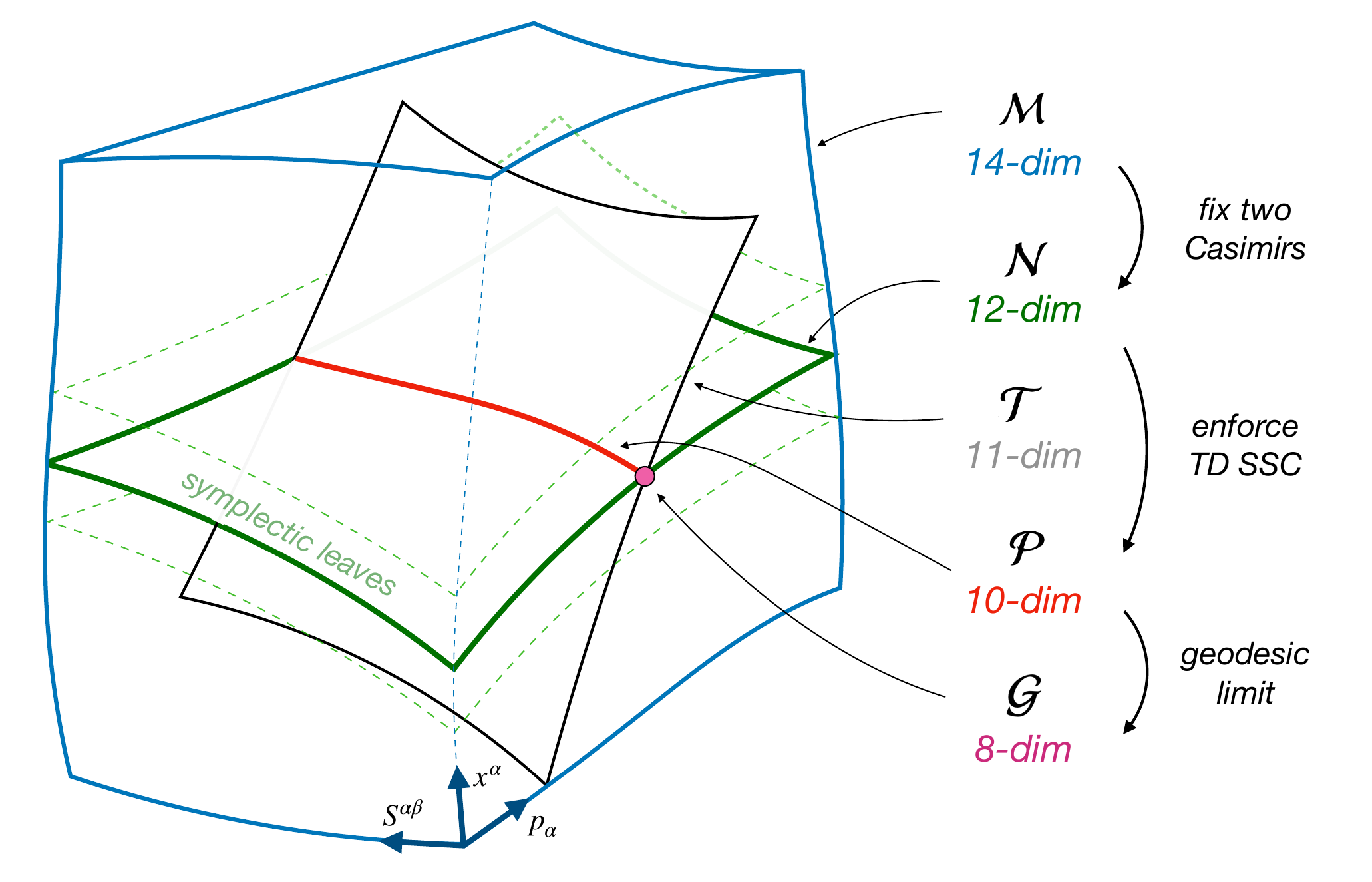}
        \caption{
        The relevant phase space and its sub-manifolds for the covariant Hamiltonian formulation of the linear-in-spin system 
        presented in Refs.~\cite{Ra.PapI.24,Ra.PRL.24}. 
        We use different sub-manifolds to lift the degeneracies associated 
        with the existence of Casimir invariants 
        $(\mcM\rightarrow\mcN)$ and correctly implement 
        the TD spin supplemental conditions $(\mcN\rightarrow\mcT\rightarrow\mcP)$. 
        Geodesics dynamics are recovered in the non-spinning limit, 
        spanning the $8$D phase space $\mcG$.}
        \label{fig:PS}
    \end{center}
\end{figure}

Like previous studies on the covariant Hamiltonian formulation 
for the dynamics of spinning secondary bodies 
(e.g. Refs.~\cite{WiStLu.19,Witzany:2019dii,WitzHJ.19}),  
our framework begins by casting the MPTD system in Eq.~\eqref{MPTDintro} 
into the most general $14$D Poisson manifold $\mcM$, which is degenerate. 
This Poisson system is mapped to a non-degenerate $12$D symplectic manifold $\mcN$, 
which does not respect the spin supplemental condition. 
This symplectic manifold is then projected onto a non-degenerate 
$10$D physical phase space $\mcP$,  
where the spin supplemental conditions are identically satisfied. 
For an illustration of this reduction process, see also FIG.~\ref{fig:PS}, 
along with the accompanying discussion in Sec.~\ref{subsec:Ham-N}.

In parallel, the construction of the canonical coordinate~\footnote{We shall use 
interchangeably ``variable'' and ``coordinates'' when there is no danger of confusion.} 
is based on the generic coordinate $(x^\alpha,p_\beta,S^{\gamma\delta})$ 
associated with the $14$D Poisson manifold $\mcM$ (for details, see Sec.~\ref{subsec:Ham-N}). 
From this fundamental coordinate, the canonical coordinate $y_{\mcN}$ 
of the $12$D symplectic manifold $\mcN$ [cf. Eq.~\eqref{eq:def-vecSD}]
is derived as described in Paper I. 
It should be noted that these coordinates can be utilized in any curved background spacetime.
However, due to this versatility, they may not always be the most optimal coordinates 
for a specific spacetime, especially if it has symmetry. 
This is the focus of our investigation in this paper. 

In order to fully exploit the spherical symmetry inherent to Schwarzschild geometry, 
we put forth the canonical transformation of 
$y_{\mcN} \mapsto y_{\mcN}^A$ in Sec.~\ref{sec:Ando}, 
which generalises the conventional orbital elements for Schwarzschild geodesics dynamics 
to those applicable to the dynamics of a spinning secondary.
We call the resulting $12$ coordinates $y_{\mcN}^A$: 
\textit{the (relativistic) Andoyer variables}, 
and they are presented in Eq.~\eqref{eq:def-rel-Andoyer} of Sec.~\ref{subsec:Andospin}, 
%
%
%

The following Sec.~\ref{sec:Reduc} is dedicated to apply the theory of 
constrained Hamiltonian systems which enables the transition from 
$12$D symplectic manifold $\mcN$ to its $10$D physical submanifold $\mcP$,  
where the spin supplementary condition is held to be true.
Because the Andoyer variables $y_{\mcN}^A$ inherited from $\mcN$ are no longer canonical 
on $\mcP$, not merely due to the differing dimensions of the phase spaces involved, 
the construction of canonical coordinate on $\mcP$ will proceed in two step. 
In Sec.~\ref{subsec:DBalgo}, we introduce the $10$ `complete' (yet non-canonical) 
variables $y_{\mcP}^C$ [cf. Eq.~\eqref{choiceP}] as intermediate coordinates. 
In the second step in Sec.~\ref{subsec:tocano}, we then transform $y_{\mcP}^C$ 
to the final $10$ canonical coordinate: $y_{\mcP}^F$.  
This use of $y_{\mcP}^C$ as intermediate non-canonical variables is for purely pragmatic, 
in order to facilitate our search for the final form of $y_{\mcP}^F$.

In summary, we have the following reduction procedure, also illustrated 
in Eq.~\eqref{eq:intro-reduction-proc}:  
\begin{equation}\label{eq:intro-reduction-proc}
    \left( \mathcal{M}; x^\alpha,p_\beta,S^{\gamma\delta} \right) \underset{\substack{\color{red}\uparrow \\ 
                       \mathclap{\textup{\tiny Sec.~\ref{sec:Ham-Schw}}} \\
                       \mathclap{\textup{\tiny (or Paper I)}}}}
     {\mapsto }
    \left( \mcN; y_{\mcN} \right)
    \underset{\substack{\color{red}\uparrow \\ 
                       \mathclap{\textup{\tiny Sec.~\ref{sec:Ando}}}  }}
    {\mapsto }
    \left( \mcN; y_{\mcN}^{\textrm{A}} \right)
    \underset{\substack{\color{red}\uparrow \\ 
                       \mathclap{\textup{\tiny Sec.~\ref{subsec:DBalgo}}} }} 
   {\mapsto }
    \left( \mcP; y_{\mcP}^{\textrm{C}} \right)
    \underset{\substack{\color{red}\uparrow \\ 
                       \mathclap{\textup{\tiny Sec.~\ref{subsec:tocano}}} }}
    {\mapsto }
   \left( \mcP; y_{\mcP}^{\textrm{F}} \right)\,.
\end{equation}
It should be highlighted that all of the coordinate systems 
in Eq.~\eqref{eq:intro-reduction-proc} that emerged during the reduction process 
are connected to one another through one-to-one mappings. 
No additional assumptions have been made, apart from the linear-in-spin approximation. 
Therefore, it is a simple process to move between each of the variables. 

Consequently, any given Hamiltonian associated with distinct variables 
($y_{\mcN}, y_{\mcN}^{\textrm{A}}, y_{\mcP}^{\textrm{F}}$, etc.) 
and a different phase space ($\mcN, \mcP$, etc.) can be derived 
by simply inserting the mapping 
in Eq.~\eqref{eq:intro-reduction-proc} 
into the original Hamiltonian $H_\mcM: \mcM\rightarrow\RR$ 
of the initial $14$D Poisson system with $\mcM$, 
expressed in terms of the original variables $(x^\alpha,p_\beta,S^{\gamma\delta})$. 
To illustrate, for example, the final Hamiltonian 
$H_c: \mcP \rightarrow \RR$ 
of the $10$D physical phase space $\mcP$ can be obtained schematically 
by $H_c := H_\mcM [x^\alpha(y_{\mcP}^{\textrm{F}}), 
p_\beta(y_{\mcP}^{\textrm{F}}), S^{\gamma\delta}(y_{\mcP}^{\textrm{F}})]$, 
whose explicit expression will be shortly presented in Eq.~\eqref{Hcanoresume} 
in Sec.~\ref{sub2sec:intro-P}.

The reader is directed to Table.~\ref{table:coordinates} 
at the end of this introduction [cf. Sec.~\ref{subsec:conventions}], 
which sets out the precise location of each mapping in explicit forms.


%

\subsubsection{Canonical formalism}
\label{sub2sec:intro-P}
The content of the present subsection, along with the following one, 
will be the content of Sec.~\ref{sec:canofin} 
of the main body of the paper.
Again, we outline the main points here.

We construct $5$ pairs of canonical coordinates $y_{\mcP}^F$ 
on the $10$D physica; phase space $\mcP$, 
here displayed in the form $(\text{$5$ coordinates};\text{$5$ momenta})$ as:
\begin{equation} 
\label{coordcanoresum}
    y_{\mcP}^F
    := 
    \left( \tilde{t},r,\nu,\tilde{\ell},\varpi ; \pi_t,\tilde{\pi}_r,\pi_{\nu},\pi_\ell,\pi_\varpi \right)\,. 
\end{equation}
These coordinates have a simple physical interpretation: $\pi_t$ is (minus) the energy of the particle; $r$ is Schwarzschild orbital radius; $\pi_\nu,\pi_\ell$ are the vertical-component and norm of the total angular momentum; and $\pi_\varpi$ is the projection of the particle's spin onto the vertical direction, traditionally denoted $S_\parallel$ in the literature. These canonical variables are evolved using Hamilton's equation with the following Hamiltonian 
\begin{equation} \label{Hcanoresume}
H_{c} := 
- \frac{\pi_t^2}{2 f} + \frac{f\tilde{\pi}_r^2}{2}  + \frac{\pi_\ell^2}{2 r^2} 
+ \left( 1 - \frac{3 M}{r} \right) \frac{\mu_0 \pi_t \pi_{\ell}\pi_{\varpi}}{f r^2 \nu_0^2} \,,
\end{equation}
where $f = 1-2M/r$ with the mass $M$ of the primary Schwarzschild black hole, 
and we introduced the two functions $(\mu_0, \nu_0)$ 
by $\mu_0^2 = \pi_t^2/f - f\tilde{\pi}_r^2 - \pi_\ell^2 / r^2 = \nu_0^2 - f\tilde{\pi}_r^2$. 
The subscript $c$ in $H_c$ stands for canonical. 
The variables $(\tilde{t}, {\tilde{\ell}})$ are spin-corrected variants of the Schwarzschild time coordinate $t$ and the true anomaly $\ell$ along the orbit, respectively. Similarly, $\tilde{\pi}_r$ is $r$'s conjugated momentum, and $\nu$ is a constant angle defining the line of nodes for the invariant plane of mean orbital motion. Lastly, ${{\varpi}}$ plays the role of a precession angle for the particle's spin.

The four quantities $(\pi_t,\pi_\nu,\pi_\ell,\pi_{\varpi})$ are \textit{integrals} of motion (or \textit{first integrals}), since their conjugated coordinate $(\tilde{t},\nu,{\tilde{\ell}},{{\varpi}})$ do not enter in Eq.~\eqref{Hcanoresume} explicitly; they are \textit{cyclic} coordinates.
This is a notable feature of our covariant formalism. 
The quantities $\pi_{\varpi}$ and $\mu$ are first integrals (scalar fields on $\mcP$), 
not external parameters on $\mcP$ (like a mass $M$ of the primary black hole).

All covariant quantities $(p_\alpha, S^{\alpha\beta})$ can be explicitly given as simple algebraic expressions involving only the canonical coordinates, 
as defined in Eq.~\eqref{coordcanoresum}. 
Consequently, the analytical solutions of the (integrable) Hamiltonian 
in Eq.~\eqref{Hcanoresume} just lead to those of the original system 
in Eq.~\eqref{MPTDintro} through algebraic manipulations. 
In light of the Hamiltonian dynamics, it can be observed that specific configurations 
of the solution trajectory and/or the spin within Schwarzschild spacetime 
are associated with certain \textit{fixed points} of the phase space $\mcP$, 
as defined by the coordinates in Eq.~\eqref{coordcanoresum}. 
These are summarized in Table.~\ref{table:special-config.}.

\begin{table}[!ht]
    \vspace{0.2cm}
	\begin{tabular}{ccc}
		\toprule
        \textbf{Phase space}       \quad&\quad \textbf{Orbital config.}     \quad&\quad \textbf{Spin config.}       \\
          \midrule  
        $\tilde{\pi}_r = 0$      \quad & \quad circular  \quad & \quad   ---       \\
        $\pi_{\ell} = 0$   \quad&\quad radial infall  \quad & \quad    ---    \\ 
		$\pi_{\nu} = L$   \quad&\quad near-equatorial   \quad & \quad   ---     \\
        $\pi_{\varpi} = S_{\circ}$  \quad&\quad planar  \quad & \quad   aligned   \\
        $\pi_{\varpi} = -S_{\circ}$  \quad&\quad planar   \quad & \quad   anti-aligned \\
        $\pi_{\varpi} = 0$       \quad&\quad ---   \quad & \quad  perpendicular \\
        \bottomrule
	\end{tabular}
        \caption{Dictionary of special orbital and spin configurations. The alignment of spin is defined between the 4-spin $S^a_\tTD$ and the covariant angular momentum 4-vector $\mcL^a$ [cf. Sec.~\ref{sub2sec:AM}]. It should be noted that planar motion is possible if and only if $S^a_\tTD$ is aligned with $\mcL^a$.  }
    \label{table:special-config.}
\end{table}

\subsubsection{Action-angle variables and Hamiltonian frequencies}
\label{sub2sec:intro-AA}
%
%

The integrability of the Hamiltonian system in Eq.~\eqref{Hcanoresume} directly implies 
the existence of \textit{generalised action-angle} variables $(\vartheta^i,\mathcal{J}_i)_{i\in\{1\ldots 5\}}\in\TT^5\times\RR^5$~\footnote{We note that action-angle variables are denoted here with the angles $\vartheta^i$ first and the action $\mathcal{J}_i$ second. This is conventional in physics.}, in the sense of Ref.\cite{Schm.02,HiFl.08} to handle the usual, non-compact time direction.
Using the classical method based on a type-$2$ generating function 
[cf. Refs.~\cite{HiFl.08,Schm.02,Arn}], 
the action coordinates (for bound orbits) can be constructed as follows
\footnote{Action-angle coordinates are typically presented in the form of distinct patches, each covering a specific region of the phase space. In this study, we will focus on bound orbits.} 
\begin{equation}
    \mcJ_1  =  - E\,, \quad 
    \mcJ_2  = \mathcal{J}_r\,, \quad
    \mcJ_3  = L_z \,, \quad
    \mcJ_4  = L \,, \quad
    \mcJ_5  = S_{\parallel}\,,
\end{equation}
where the radial action $\mcJ_r$ is given explicitly by  
\begin{equation}
    {\mathcal J}_{2} :=  \frac{1}{2 \mathpi} \oint V(r)\, \ud r
    \quad \text{with} \quad
     V(r)
    := \pm \sqrt{\frac{E^2}{f^2} - \frac{L^2}{f r^2} - \frac{\mu^2}{f} 
    + \frac{r - 3 M}{r - 2 M} \frac{2 \mu E L S_{\parallel} }{E^2 r^2 - L^2 f}  }\,.
\end{equation}
Here, the integral is over the closed contour in the 2D phase space $(r,\tilde{\pi}_r)$ 
corresponding to the bound orbit of parameters $\mu, E, L$ and $S_\parallel$. 

From action-angle variables, we can straightforwardly define the Hamiltonian frequencies $\omega^i(\mathcal{J}) := \partial {H_c} / \partial \mathcal{J}_i$, 
which leads to the following explicit expressions 
\begin{equation}
    \omega^{1} =  \omega^2  \frac{\partial {\mathcal J}_{2}}{\partial E}\,, \quad
    \omega^{2}  = \left( \frac{\partial {\mathcal J}_{2}}{\partial H_c} \right)^{- 1}\,, \quad
    \omega^{3}  = 0\,, \quad
    \omega^{4}  = - \omega^2  \frac{\partial {\mathcal J}_{2}}{\partial J}\,, \quad
    \omega^{5}  = - \omega^2  \frac{\partial {\mathcal J}_{2}}{\partial S_{\parallel}}\,.
\end{equation}
All the right-hand side will be given by ready-to-use expressions in Eqs.~\eqref{eq:dJdP}.



A more detailed Hamiltonian analysis of generic orbital and spin dynamics, as well as thorough comparisons between existing schemes and our results, is desirable, and will appear in a forthcoming paper. 

\subsection{Organisation of this paper and guide to the reader}


We conclude this introduction with an overview of the subsequent sections of this paper a
nd a guide for the reader, who may be interested in a specific subset 
of the results presented in our work.

In Sec.~\ref{sec:Ham-Schw}, we shall give various preliminaries concerning 
the covariant Hamiltonian framework of the spinning secondary 
with a background Schwarzschild metric, up to linear order in the secondary spin.
This section is aligned with the contents presented in Paper I~\cite{Ra.PapI.24}. 
Subsequently, in Sec.~\ref{sec:Ando} we shall introduce the canonical variables on the $12$D phase space $\mcN$, namely the (relativistic) Andoyer variables, which make use of the spherical symmetry of Schwarzschild spacetime. 
The main technical details of the reduction from the $12$D phase space $\mcN$ 
to its $10$D physical sub-manifold $\mcP$, 
where the spin supplementary condition is satisfied identically, 
are provided in Sec.~\ref{sec:Reduc}. 
We finally present in Sec.~\ref{sec:canofin} the Hamiltonian system 
on the $10$D physical phase space $\mcP$ with the canonical coordinates, 
where the action-angle variables and Hamiltonian frequencies are also obtained. 
Some technical aspects of our treatment are relegated to appendices, 
including a thorough Hamiltonian investigation of Schwarzschild geodesics 
in App.~\ref{app:geo}, and a summary of the Marck tetrad formalism in Ref.~\cite{Marck.83}
which has traditionally been used in the literature, in App.~\ref{app:TD-spin-vec}. 
Appendix \ref{app:detailsNtoP} contains some detailed calculations relative 
to the construction of canonical coordinates in Sec.~\ref{sec:Reduc}.

The covariant Hamiltonian framework, as outlined in Sec~\ref{sec:Ham-Schw}, 
is a foundational element of our series of works. 
In particular, Sec.~\ref{subsec:inv} contains all the explicit formulae required 
to apply this general framework to a specific Schwarzschild background, 
with an emphasis on the role of symmetries and related constants of motion. 
Therefore, we advise all our readers to familiarise themselves with this section, 
which sets out the conventions, notations and invariants of motion defined on Schwarzschild. 
However, the reader who have already studied our previous works may choose to skip this section.

The reader only interested in results concerning the newly defined canonical coordinate, 
i.e., the relativistic Andoyer variables on the $12$D symplectic manifold $\mcN$, 
need only read on Sec.~\ref{sec:Ando}, particularly 
Sec.~\ref{sec:dynando} for a comprehensive account 
of the resulting $12$D Andoyer Hamiltonian system 
and Sec.~\ref{subsec:Hill-EOMs} for its practical `Hill' formulation.

Conversely, those with only a passing interest in results pertaining 
to the canonical formulation on the $10$D physical manifold $\mcP$, 
need only refer to Sec.~\ref{sec:canofin} for details. 
Those seeking solely the practical results, 
such as the action-angle variables and Hamilton frequencies, 
are advised to turn directly to Sec.~\ref{subsec:AA}.

\subsection{Conventions and notations}
\label{subsec:conventions}

Throughout the paper, we use geometric units in which $G=c=1$, and our conventions for the Riemann curvature tensor and other geometric tensors are as in Wald's text book \cite{Wald}, 
in particular $\nabla_{a}\nabla_b \omega_c = R_{abc}^{\phantom{abc}d}\omega_d$ for any $1$-form $\omega_a$. We use $a,b,c,d,\ldots$ to denote abstract indices. 
For the components, we use two kinds of indices varying in $\{0,1,2,3\}$ : $\alpha,\beta,\gamma,\delta,\ldots$ are used for components in the natural basis $(\partial_{\alpha})^a$ associated with a spacetime coordinate system $x^\alpha$, and $A,B,C,D,\ldots$ are used for components in an orthonormal tetrad basis $(e_A)^a$. 
An arrow is sometimes used to denote Euclidean $3$-vectors, using 
interchangeably $\vec{v}=(v^1,v^2,v^3)=v^I$, 
as well as usual notations $\cdot$ (dot) for the scalar product 
and $\times$ (cross) for the cross product.

In the phase space, pairs of canonical coordinates are always ``degrees of freedom first, conjugated momenta second'', i.e., of the form $(q_1,q_2,\ldots,\pi_{q_1},\pi_{q_2},\ldots)$, with $\pi_q$ always being the conjugated momentum of some degree of freedom $q$. Poisson brackets are denoted with the brackets $\{\cdot,\cdot\}$. For any integer $N\in\NN$, ``$N$D'' means $N$-dimensional. A table for the frequently used symbols, description and references is given in Table.I of Paper I.

\begin{table}[!ht]
    \caption{List of phase space coordinates used in this paper. 
    The term ``canonical'' is used to indicate that the coordinate 
    verifies the canonical Poisson matrix. 
    The equation numbers indicate the locations within this paper 
    where each coordinate is defined. }
    \vspace{0.2cm}
	\begin{tabular}{cccc}
		  \toprule
		  \textbf{Manifold}          \quad  &\textbf{Coordinate set}  &$\,\,\,$ \textbf{Canonical} $\,\,\,$& \textbf{Definition}  \\
		  \midrule
		  \textbf{14D-Poisson}                    & $ (x^\alpha,p_\beta,S^{\gamma\delta})$            &   No      & Paper I \\ 
            \textbf{manifold} $\mathcal{M}$         & $(x^\alpha,\pi_\beta,S^{I},D^I)$                                                          &   No      & Paper I [Eq.~\eqref{eq:def-vecSD}]  \\ 
                                                                                            &           &           \\
            \textbf{12D-Symplectic}                 & $y_{\mcN} :=
            (t,r,\theta,\phi,\sigma,\zeta, \pi_t,\pi_r,\pi_\theta,\pi_\phi,\pi_\sigma,\pi_\zeta)$     &   Yes     & Paper I  [Eq.~\eqref{spinsympSSC}] \\ 
            \textbf{manifold} $\mathcal{N}$         &$ y_{\mcN}^{\textrm{A}} :=
            (t,r,\nu,\ell,s,\zeta, \pi_t, \pi_r,\pi_\nu,\pi_\ell,\pi_s,\pi_\zeta)$                    &   Yes     & Andoyer  [Eq.~\eqref{eq:def-rel-Andoyer}] \\ 
                                                                                            &           &           \\
            \textbf{10D-Physical}                   &$y_{\mcP}^{\textrm{C}} :=
            (t, r,\nu,\ell,\Gamma,\pi_t,\pi_r,\pi_\nu,\pi_\ell,\pi_{\varpi})$                                    &    No     & Complete  [Eq.~\eqref{choiceP}]\\ 
            \textbf{phase space} $\mathcal{P}$      &$y_{\mcP}^{\textrm{F}} :=
            (\tilde{t}, r, \nu, \tilde{\ell}, {\varpi}, \pi_t, \tilde{\pi}_r, \pi_\nu, \pi_\ell, \pi_{\varpi})$                             &   Yes     & Final  [Eq.~\eqref{eq:def-yc}]  \\ 
            \midrule
	\end{tabular} 
    \label{table:coordinates}
\end{table}

The symbol $O(\epsilon^N)$ indicates an expression of order $N$ in
the small parameter $\epsilon=S_\circ/\mu M$ [cf. Eq.~\eqref{smallp}]. 
Unless otherwise specified, we work consistently at order $O(\epsilon)$, 
with terms of order $O(\epsilon^n)$ with $n \geq 2$ being neglected. 
Furthermore, such terms are most often omitted from equations in order to enhance readability. 

The various coordinate systems employed in the different phase spaces [cf. FIG.~\ref{fig:PS}] described in the main body of the paper are summarized in Table~\ref{table:coordinates}. \\

\section{Covariant Hamiltonian formalism in Schwarzschild} 
\label{sec:Ham-Schw}

This section summarizes the model that we use to describe the motion of a compact object in a given background spacetime, and how the resulting equations can be put into an explicit Hamiltonian form. In Sec.~\ref{subsec:recap}, the compact object is described as a point particle endowed with linear and angular momentum. The evolution equations for these momenta follow from a multipolar expansion, which is truncated at dipolar order. A general Hamiltonian formulation of the resulting equations is presented in Sec.~\ref{subsec:Ham-N}, based on the covariant Hamiltonian formalism exposed in Paper I, where more details can be found. We then apply this general setup to the Schwarzschild spacetime in Sec.~\ref{subsec:inv}, which serves as the basis for all subsequent calculations.

\subsection{Recap of Paper I: the linear-in-spin system} 
\label{subsec:recap}

In this section, we summarize the main points of motion of the relativistic spinning body in an arbitrary spacetime, but at linear order in the body's spin. 
The material in this subsection is well-covered in the literature about general relativistic two-body problems. For an historical account, we refer to Chap.~(1) of Ref.~\cite{2015emrg.book.....P} and Chap.~(2) of Ref.~\cite{PhDHal.21}. The multipolar formalism used here relies on the original works by Dixon \cite{Di.64}, as well as its rigorous and modern extension constructed by Harte \cite{Ha.12,Ha.15}.

\subsubsection{Dipole approximation and Killing vectors}

We consider the motion of a compact object within a given spacetime described by a metric tensor $g_{ab}$. The object is modelled as a point particle with four-momentum 1-form $p_a$ and antisymmetric spin tensor $S^{ab}$. We let $\scL$ be the particle's worldline with proper time $\tau$ and four-velocity $u^a$, normalized such that $u^a u_a = -1$. The body is assumed to evolve within (and not back react on) the background spacetime. The equations governing the evolution of $p_a$ and $S^{ab}$ at dipolar order (neglecting quadrupoles and higher order multipoles) are the so-called Mathisson-Papapetrou-Tulczyjew-Dixon (MPTD) equations \cite{Ma.37,Pa.51,Tu.59,Di.74,Ha.12} 
\begin{subequations}\label{MPTD}
	\begin{align}
        \nabla_u p_a & = R_{abcd} S^{bc} u^d \, , \label{EoM} \\
		\nabla_u S^{ab} & = 2 p^{[a}u^{b]}  \, , \label{EoP} 
 	\end{align}
\end{subequations}
where $R_{abcd}$ denotes the Riemann curvature tensor\footnote{The algebraic symmetries of the Riemann tensor and spin tensor imply $R_{b c d a} S^{b c} = 2 R_{a b c d} S^{b c}$, which may be inserted in Eq.~\eqref{EoM} to give it a more familiar form.} and $\nabla_u:= u^a\nabla_a$ the covariant derivative along $\scL$, with $\nabla_a$ the metric-compatible connection. 
Three positive scalars $\mu,S_\circ$ and $S_\star$, built solely from the geometry and $(p_a,S^{ab})$, are defined by
\begin{equation}\label{eq:def-norms}
    \mu^2 := -p_a p^a\,,
    \qquad
    S_\circ^2 := \frac{1}{2}S_{ab}S^{ab}\,,
    \quand 
    S_\star^2 := \frac{1}{8}\varepsilon_{abcd}S^{ab}S^{cd}\,,
\end{equation}
with $\mu,S_\circ$ specifically referred to as the \textit{dynamical mass} and \textit{spin norm}, respectively.  
It should be noted that these quantities are not conserved under the general system 
described in Eq.~\eqref{MPTD}. 
However, conserved quantities can be built from the symmetries of the background spacetime, and given in terms of Killing vector fields. More precisely, let $k^a$ be a Killing vector field on $g_{ab}$, such that it satisfies Killing's equation $\nabla_{(a}{k}_{b)}=0$. Then, along the particle's worldline $\scL$, the quantity
\beq \label{Killinginv}
    \Xi 
    := 
    p_{a} {k}^{a} + \frac{1}{2} S^{ab} \nabla_{a} {k}_{b} 
\eeq
is exactly conserved [cf.~Refs.\cite{Rasband:1973zz,Tod:1976ud,Ha.15}]. 
Such Killing invariants will be an essential ingredient in this paper.

In a given basis, the number of independent components in Eqs.~\eqref{MPTD} are $4$ for $p^a$, 
$6$ for $S^{ab}$ (due to the anti-symmetry) and $3$ for $u^{a}$ (due to the normalization $u_au^a=-1$). If we view these components as functions of the proper time $\tau$ along $\scL$, then the system in Eq.~\eqref{MPTD} is equivalent to $10$ ordinary differential equations (ODEs), for a total of $13$ unknowns (given the geometry). It is, therefore, not well-posed. To fix this problem, one adopts a so-called \textit{spin supplementary condition} (SSC). Physically, this condition is a choice of the frame in which some components of the spin tensor (the mass dipole, usually) will vanish. Consider an arbitrary, timelike, unit vector $v^a$ along $\scL$. Then it is always possible to decompose the spin tensor $S^{ab}$ 
into two vectors $S^b_{(v)}$ and $D^b_{(v)}$ orthogonal to $v^a$, as follows:  
\beq\label{eq:SD-decomposition}
	S^{ab} = \varepsilon^{ab}_{\phantom{ab}cd} v^c S^d_{(v)}+ 2 D_{(v)}^{[a} v^{b]}
    \quad \Leftrightarrow \quad 
	\begin{cases}
		\,\,\, S^b_{(v)} := \frac{1}{2} \varepsilon^{abcd} v_a S_{cd} \, , \\
		\,\,\, D^b_{(v)} := v_a S^{ab}  \,.
	\end{cases}
\eeq
In this construction, the vectors $S^a_{(v)}$ and $D^a_{(v)}$ depend on $v^a$, and 
they are physically interpreted as the spin $4$-vector and mass dipole $4$-vector measured in the rest frame of an observer with four-velocity $v^a$, respectively. 
They are both spacelike vectors because they are orthogonal to $v^a$, by definition. 
Different choices of $v^a$ will lead to different decomposition 
of $S^{ab}$ in Eq.~\eqref{eq:SD-decomposition}, or, in physical terms, 
different observers will measure different spins and mass-dipoles. 

Now we can state what ``choosing an SSC'' means: it amounts to a choice of observer (i.e., a choice of $v^a$) for whom the mass dipole $D_{(v)}^b$ vanishes. For this observer, all the content of $S^{ab}$ can be obtained from $S_{(v)}^a$ because $D^b_{(v)} = 0$ implies 
$S^{ab} = \varepsilon^{ab}_{\phantom{a b} c d} v^{c} S^d_{(v)}$
according to Eq.~\eqref{eq:SD-decomposition}.
However, this does not mean that the spin and mass dipole vectors defined in another frame are irrelevant for the analysis. In fact, it is essential in our framework to make a distinction between a frame defining the SSC and that to carry out the mathematical analysis.

\subsubsection{Linearization, Tulczyjew-Dixon SSC and Killing-Yano tensors}

In this series of work, we will always adopt the Tulczyjew-Dixon SSC (TD SSC), 
obtained by setting the mass dipole $4$-vector of an observer 
with four-velocity $\bar{p}^a := p^a/\mu$ to zero.
More precisely, we shall set 
\begin{equation} \label{TDSSC}
    C^b := p_a S^{ab}=0 \,,
\end{equation}
where the first equality defines the quantity $C^a$ and the second defines the TD SSC as the condition $p_a S^{ab}=0$. 
By virtue of the general decomposition in Eq.~\eqref{eq:SD-decomposition}, 
the TD SSC readily implies $D_{(\bar{p})}^a = 0$, 
and defines what we refer to as the \textit{TD spin} $4$-vector, as
\begin{equation}\label{eq:def-TD-spins} 
    S^{b}_{\tTD}
    := \frac{1}{2}\varepsilon^{abcd} {\bar {p}}_{a} S_{cd}\,,
\end{equation}
to emphasize its dependence on the TD SSC. Combining Eqs.~\eqref{eq:def-norms} and \eqref{eq:SD-decomposition} as well as the identity $\varepsilon^{abcd}\varepsilon_{afgh} = - 6 \,\delta^{[b}_f \delta^{\phantom{|}\!c}_g \delta^{d]}_h$ 
[cf. a text book~\cite{Wald}], 
the TD spin vector is found to satisfy
\begin{equation}\label{eq:norm-TD-spins}
    {\bar {p}}_{a} S^{a}_{\tTD} = 0
    \quand
    S^{a}_{\tTD}\,S_{a}^{\tTD} = S_{\circ}^2 \,,
\end{equation}
where $S_{a}^{\tTD}:=g_{ab}S^{b}_{\tTD}$.

In the context of our Hamiltonian formulation, the $4$ equations in Eq.~\eqref{TDSSC} will be treated as \textit{algebraic relations} satisfied the components $(p_\alpha,S^{\alpha\beta})$. Moreover, only three of them are linearly independent, 
because the contraction of Eq.~\eqref{TDSSC} with $p_b$ results in a mere $0=0$. 

Mathematically, the algebraic-differential system made 
of Eqs.~\eqref{MPTD} and~\eqref{TDSSC} is now well-posed: it has $13$ equations 
(i.e. $10$ ODEs and $3$ algebraic equations) for $13$ unknowns.
Physically, however, it is not self-consistent. 
Indeed, the MPTD system defined by Eq.~\eqref{MPTD} is the truncation, at dipolar order, of a more general system that results from the application of a multipolar expansion 
in Refs.~\cite{Di.74,Ha.12,Ha.15}. 
Yet, it is well-known that, at the quadrupolar level, quadratic-in-spin corrections arise 
to the right-hand side of these evolution equations. Therefore, working at the dipolar level means that we have implicitly omitted quadratic-in-spin terms in Eq.~\eqref{MPTD}, by assumption. To be consistent, we must apply a truncation of Eq.~\eqref{MPTD} at linear-in-spin order; 
otherwise, the MPTD system becomes ambiguous due to the partial description of quadratic-in-spin effects. The TD SSC allows us to perform this necessary linearization. This procedure can be rigorously performed as an expansion in powers of the small parameter $d/\mathcal{R}\ll 1$ involved in the multipolar expansion of arbitrary extended objects in Refs.~\cite{Di.74,Ha.15}, with $d$ the typical length scale of the body and $\mathcal{R}$ that of the background spacetime. 
Since the spin angular momentum of an \textit{extended} object scales as $S_\circ\sim\mu d$, we have the following small parameter at hand \cite{Moller.49}
\beq \label{smallp}
    \epsilon := \frac{S_\circ}{\mu L} \ll 1 \,.
\eeq
For the particular case of a compact object, $d \sim\mu$, from which $S_\circ\sim\mu^2$ follows, we have $\epsilon=\mu/\mathcal{R}\ll 1$. In particular, if the compact object orbits a black hole of mass $M$, the length scale is $\mathcal{R}\sim M$ and the parameter $\epsilon$ becomes nothing but the (standard) mass ratio $\mu/M \ll 1$ used in black hole perturbation theory 
(see, e.g., Ref.~ \cite{PoWa.21}).

In any case, neglecting non-linear terms in the $\epsilon$ expansion, Eqs.~\eqref{MPTD} and \eqref{TDSSC} lead us to the equations of interest in this paper, 
namely the \textit{linear-in-spin, MPTD system}:
\begin{subequations}\label{EElin}
	\begin{align}
        \nabla_u p_a & = R_{abcd} S^{bc} u^d \, , \label{EoMlin} \\
		\nabla_u S^{ab} & = 0  \,, \label{EoPlin} \\
        p^a &= \mu u^a \,. \label{momvellin} 
 	\end{align}
\end{subequations}
Referring to Eq.~\eqref{eq:def-norms}, the linearized system implies 
that all scalars $\mu,S_\circ$ and $S_\star$ are conserved along $\scL$. 
The system in Eq.~\eqref{EElin}, with the TD SSC imposed at all times, 
now produces the physically consistent dipole approximation of the MPTD system.  

{From this point onward, all results will be derived at linear order in spin 
and we will discard all $O(\epsilon^2)$ or higher terms, 
in the sense of Eq.~\eqref{smallp}, unless otherwise specified.}

The linearized system defined in Eq.~\eqref{EElin}, 
along with the TD SSC in Eq.~\eqref{TDSSC}, implies the remarkable fact that 
there exists other invariants of motion along $\scL$ 
in addition to the Killing-vector invariants $\Xi$ defined in Eq.~\eqref{Killinginv}.
For example, 
the existence of a Killing-Yano tensor $Y^{ab}$~\cite{1981RSPSA.375..361D,1982RSPSA.381..315D}, 
defined as a rank-$2$, antisymmetric tensor $Y^{ab}$ that satisfies $\nabla_{(a}Y_{b)c}=0$, 
gives rise to the conservation of two quantities for the system of Eq.~\eqref{EElin}. 
They are known as the \textit{R\"{u}diger invariants}, 
originally found by R. R\"{u}diger in Refs.~\cite{Rudiger.I.81,Rudiger.II.83}. 
We shall adopt for them an expression 
(inspired by those in Refs.~\cite{ComDru.22,ComDruVin.23})  
\begin{subequations} \label{Ruds}
    \begin{align}
        \fK &:= \frac{1}{4} \varepsilon_{abcd} Y^{ab} S^{cd} \,,  \label{Rudcov} \\
        \fQ &:= 
         K_{ab} p^a p^b + 4  \varepsilon_{ade[b} Y^{e}_{~c]} \xi^a S^{db} p^c\,,
        %
        \label{quadRudcov}
    \end{align}
\end{subequations} 
where $\xi^a$ and $K_{ab}$ are two tensors uniquely associated with $Y_{ab}$ via
\begin{equation}\label{eq:def-KS-tensor}
    \xi^a := -\frac{1}{6} \varepsilon^{abcd}\nabla_bY_{cd}
    \quand
    K_{ab}:= Y_{ac} Y_b^{\phantom{b}c}\,.
\end{equation}
From the properties of $Y_{ab}$, it follows that $K_{ab}$ is symmetric and verifies $\nabla_{(a}K_{bc)} = 0$. It is, therefore, a Killing-St\"ackel\footnote{There are two conventions to define a Killing-St\"{a}ckel tensor 
from contracting two copies of $Y^{ab}$, which only differ by an overall minus sign. We follow the convention used in Eq.~(2.16) of Ref.~\cite{DruHug.I.22}.} (or simply Killing) tensor; 
see, e.g., Refs.~\cite{Car.Hou.09,Frolov.17}. 
Similarly, since $\nabla_{a}Y_{bc}$ is a totally antisymmetric rank-3 tensor, $\xi^a$ is the $4$-vector such that $\nabla_aY_{bc} = \varepsilon_{abcd}\xi^d$, 
which is equivalent to the left-hand side of Eq.~\eqref{eq:def-KS-tensor}.
In a Kerr background, the invariant $\fQ$ can be regarded as a generalization to the linear-in-spin case of the Carter constant $K := K_{a b} p^{a} p^{b}$ for geodesics, initially identified in Ref.~\cite{Carter.68}. In contrast, the other invariant $\fK$ simply vanishes in the non-spinning (geodesic) limit.

The R\"{u}diger invariants in Eqs.~\eqref{Ruds} are referred to 
as \textit{quasi-invariants} in Ref.~\cite{ComDru.22} 
in the sense that (i) they are only conserved at linear order in spin [i.e., $\nabla_u\fK$ and $\nabla_u\fQ$ are $O(\epsilon^2$)] and (ii) their conservation holds only under the TD SSC in Eq.~\eqref{TDSSC}. 

\subsection{Formulation as a covariant Hamiltonian system}
\label{subsec:Ham-N}

In this section, we cast the differential system defined in Eq.~\eqref{EElin} 
into a covariant Hamiltonian system. 
Our treatment is based on the framework detailed in Paper I \cite{Ra.PapI.24}, which uses tools from symplectic geometry to reduce a 14D, degenerate Hamiltonian formulation to a 10D, non-degenerate formulation. This reduction is done in several steps, summarized in Fig.~\ref{fig:PS}. We refer readers to Ref.~\cite{Ra.PRL.24} for an executive summary and Sec.~II. and IV. of Paper I for an extended discussion and details. In this subsection, the background spacetime is arbitrary. Restriction to the Schwarzschild spacetime will be done in Sec.~\ref{subsec:inv}.

%
%
\subsubsection{Geometric preliminaries}

Consider a spinning test body that evolves within a background spacetime $(\scE,g_{ab})$, where the 4D manifold $\scE$ is covered with some coordinates $x^\alpha$. We will consider two vector bases at each event of $\scE$: one, denoted, $\{(\partial_\alpha)^a\}_{\alpha\in\{0,\ldots,3\}}$, is the natural basis associated with $x^\alpha$. The other, denoted $\{(e_A)^a\}_{A\in\{0,\ldots,3\}}$, is an orthonormal tetrad field, whose components in the natural basis are $(e_A)^\alpha$ ($A$ labels the four different vector fields, and $\alpha$ the components of that vector). 
Following Wald's textbook \cite{Wald}, we denote by $\omega_{aBC}:= g_{bc}(e_B)^b \nabla_a (e_C)^c$ the six independent connection $1$-forms associated with the tetrad, 
which satisfy the antisymmetric relation $\omega_{aBC} = -\omega_{aCB}$. 
Their components in the natural basis, $\omega_{\alpha BC}$, are the connection coefficients, while the tetrad components $\omega_{ABC}$, are the Ricci rotation coefficients. 
All these objects are defined globally in $\scE$.

We now focus on the particle's worldline $\scL$, where we can expand the momentum $p_a$ and spin tensor $S^{ab}$ along either of the natural or tetrad bases. 
Of particular interest to us are the quantities $\pi_\alpha$ and $S^{AB}$ defined by
\begin{subequations} \label{pStopiS}
    \begin{align}
    p_\alpha &= \pi_\alpha +\frac{1}{2} \omega_{\alpha BC} S^{BC} \,, \label{ptopi}\\
    S^{\alpha\beta} &= S^{AB} (e_A)^\alpha (e_B)^{\,\beta} \,. \label{StoS}
    \end{align}
\end{subequations}
The physical interpretation of Eq.~\eqref{pStopiS} is simple: $\pi_\alpha$ is a correction to the linear momentum $p_\alpha$ that accounts for the coupling between the spin and the background curvature (through the connection coefficients); and $S^{AB}$ are the tetrad components of the spin tensor. 
Following the general decomposition of the spin tensor 
in Eq.~\eqref{eq:SD-decomposition}, the six $S^{AB}$ can be decomposed into the spin and mass dipole four vectors in the rest frame of an observer with four-velocity $(e_0)^a$, i.e., we set 
\begin{subequations}\label{eq:def-vecSD}
    \begin{align}
        \vec{S} &:= \left( S^1, S^2, S^3 \right) = \left( S^{23}, S^{31}, S^{12} \right)  
        \quad \Leftrightarrow \quad 
        S^I = \tfrac{1}{2} \varepsilon^{I}_{\phantom{I}JK} S^{JK}\,, \\
        \vec{D} &:= \left( D^1, D^2, D^3 \right) = \left( S^{01}, S^{02}, S^{03} \right)  
        \quad \Leftrightarrow \quad 
        D^I = S^{0I} \,.
    \end{align}
\end{subequations}

While all these quantities have perfect covariant meaning in general relativity, 
from now on, we will view them as the fundamental building blocks 
for the construction of the Hamiltonian formulation in the phase space geometry. 
Even though we leave the realm of Lorentzian geometry, we shall still use $x^\alpha$ without risk of confusion. In addition, we keep using the indices $\alpha,\beta,\gamma,\ldots$ and $A,B,C,\ldots$ inherited 
from GR, even though they now label the phase space coordinates. 
We also keep using Einstein's summation convention, and, should any ambiguity arise, 
it will be pointed out in the text explicitly.  

\subsubsection{The $14$D Poisson manifold $\mcM$}
\label{sub2sec:mcM}

The variables $(x^\alpha,p_\alpha,S^{\alpha\beta})$ 
or the variables $y_{\mcM} := (x^\alpha,\pi_\alpha,S^I,D^I)$ 
are two different charts for the same $14$D phase space, 
denoted $\mcM$ 
[cf. TABLE~\ref{table:coordinates} in Sec.~\ref{subsec:conventions}].
%
Using the latter variables $y_{\mcM}$, which are more useful for practical calculations, 
the Poisson brackets for two arbitrary functions $F(y_{\mcM})$ and $G(y_{\mcM})$ on $\mcM$ 
are defined by [cf. Sec. II of Paper I] 
\begin{equation} \label{PBnonsymp}
    \{F,G\}_{\mcM} := \sum_{i,j} \,\Lambda^{ij}(y_{\mcM}) \,  
    \frac{\partial F}{\partial y_{\mcM}^i} \, \frac{\partial G}{\partial y_{\mcM}^j} \,,
\end{equation}
where the coefficients $\Lambda^{ij}(y_{\mcM})$ are the coordinate expressions 
for the \textit{Poisson structure} $\Lambda$ on $\mcM$.
Because Eq.~\eqref{PBnonsymp} implies 
$\{y_{\mcM}^i,y_{\mcM}^j\}_{\mcM}=\Lambda^{ij}(y_{\mcM})$, 
the following Poisson brackets for $y_{\mcM}^i$ define $\Lambda^{ij}(y_{\mcM})$: 
\begin{subequations}\label{PBsSD}
    \begin{align}
    \{x^\alpha,\pi_\beta\}_{\mcM} &= \delta^\alpha_\beta
    \,, \\
    \{S^I,S^J\}_{\mcM} &= \varepsilon^{IJ}_{\phantom{IJ}K} S^K 
    \,, \\
    \{D^I,D^J\}_{\mcM} &= -\varepsilon^{IJ}_{\phantom{IJ}K} S^K 
    \,, \\
    \{D^I,S^J\}_{\mcM} &= \varepsilon^{IJ}_{\phantom{IJ}K} D^K \,.
    \end{align}
\end{subequations}
The $(S^I,D^I)$ part of these brackets is identical to the commutators 
of the $SO(1,3)$ Lorentz algebra. 
This is a consequence of the local Lorentz invariance of general relativity, 
and is built-in the orthonormal tetrad field that we use. 

The Poisson brackets defined in Eqs.~\eqref{PBsSD} are \textit{degenerate,} i.e., the antisymmetric matrix $\Lambda^{ij}(y_{\mcM})$ has (non-maximal) rank $12$. 
The difference between dim$(\mcM)$ and rank$(\Lambda)$ is precisely 
the number of \textit{Casimir invariants} that exists for the brackets. 
These $14 - 12 = 2$ Casimir invariants, which we denote 
by $\mathcal{C}_\circ$ and $\mathcal{C}_\star$, 
have the unique (and defining) property that their Poisson bracket 
with any other function on $\mcM$ identically vanishes. 
In our case, these Casimirs are given by [cf. Sec.~II.D of Paper I]
\begin{equation}\label{eq:def-Casimir}
    \mathcal{C}_\circ := \vec{S}\cdot\vec{S} - \vec{D}\cdot\vec{D} 
    \quad \text{and} \quad
    \mathcal{C}_\star := \vec{S}\cdot\vec{D}\,,
\end{equation}
with classical Euclidean notations for three-vectors. 
It should be noted that the Casimir invariants are inherent in the phase space $\mcM$ 
and are not dependent on any choice of the Hamiltonian, which will be introduced below.  

A Hamiltonian on $\mcM$ is a scalar field $H_\mcM: \mcM\rightarrow\RR$, 
or, equivalently, a function of the $14$ coordinates 
$y_{\mcM} = (x^\alpha,\pi_\alpha,S^I,D^I)$. 
It will generate a particular set of curves (or trajectories) in $\mcM$. 
For these trajectories to match the solutions of the linear MPTD system in Eq.~\eqref{EElin}, 
we choose the Hamiltonian such that 
\begin{equation} \label{HM}      
    H_{\mcM} : ( x^{\alpha}, \pi_\alpha,S^I,D^I) \mapsto 
    \frac{1}{2} g^{\alpha\beta} \pi_{\alpha} \pi_{\beta} 
    + 
    \frac{1}{2} g^{\alpha\beta} \pi_{\alpha} \omega_{\beta C D}S^{CD} \,.
\end{equation}
The three ingredients: 
the phase space $\mcM$, the Poisson structure $\{\,,\,\}_{\mcM}$  
and the Hamiltonian $H_\mcM$ define a $14$D Poisson system. 
The phase space trajectories generated by $H_\mcM$ are described 
by the ordinary differential equation (ODE) 
\beq \label{HameqM}
    \forall F:\mcM\mapsto\RR \,, \quad \frac{\ud F}{\ud \bar{\tau}} = \{F,H_\mcM\}_{\mcM}\,,
\eeq
where the ``time'' parameter $\bar{\tau}$ in this equation corresponds physically 
to the dimensionless proper time $\tau/\mu$. 
Since $\mu$ is conserved along $\scL$, 
the parameter $\bar{\tau}$ is an affine parameter. 
When we insert $F = y_{\mcM}^i$ into Eq.~\eqref{HameqM} and apply the Leibniz rule, 
it becomes evident that the resulting $14$ ODEs are in fact equivalent 
to the complete linear MPTD system presented in Eq.~\eqref{EElin}. 
This is discussed in greater detail in Section II.B of Paper I.

\subsubsection{The $12$D symplectic leaves $\mcN$ }
\label{sub2sec:mcN}

The 14D phase space $\mcM$ is naturally foliated by 12D sub-manifolds, denoted $\mcN$ 
[cf. Fig.~\ref{fig:PS}]. 
They are defined as the level sets of the two Casimir invariants defined in Eq.~\eqref{eq:def-Casimir}, and called \textit{symplectic leaves}. 
Comparing the (phase-space) definitions in Eq.~\eqref{eq:def-Casimir} with the (spacetime) spin norms $S_\circ$ and $S_\star$ in Eq.~\eqref{eq:norm-TD-spins} readily reveals that, numerically, $\mathcal{C}_\circ = S_\circ^2$ and $\mathcal{C}_\star=S_\star^2$, hence the notation. 

From now on, we shall focus on those leaves $\mcN$ where 
\beq \label{Casimir}
\mathcal{C}_1 = S_\circ^2 
\quad \text{and} \quad
\mathcal{C}_2 = 0  \,,
\eeq
for some fixed numerical value of $S_\circ^2$. The reason for setting $\mathcal{C}_2 = 0 = S_\star^2$ is that it will be the only case compatible with a choice of SSC (thus with the TD SSC), as was already pointed out in Ref.~\cite{WiStLu.19}. The $12$D symplectic leaves $\mcN$, as their name suggests, are symplectic, i.e., non-degenerate. In particular, it is possible to endow $\mcN$ with 6 pairs of \textit{canonical} coordinates. The construction of such coordinates is done in Sec.~III of Paper I, and we only summarize the results here.

We endow $\mcN$ with $12$ canonical coordinates 
$y_{\mcN} := ( x^{\alpha}, \sigma, \zeta, \pi_{\alpha}, \pi_{\sigma}, \pi_{\zeta} )$.  
The ordering is important: 
\textit{first} the $6$ degrees of freedom $(x^\alpha,\sigma,\zeta)$, 
and \textit{then} their $6$ respective conjugated momenta $(\pi_\alpha,\pi_\sigma,\pi_\zeta)$. \footnote{Here there is an implicit sum over $\alpha\in\{0,1,2,3\}$ in these expressions. For example, in Cartesian coordinates $x^\alpha=(t,x,y,z)$, one would read them as $(x^\alpha,\sigma,\zeta)=(t,x,y,z,\sigma,\zeta)$.}  
The very phrasing ``canonical coordinates'' leaves no choice 
for the Poisson structure on $\mcN$: it is the canonical one, 
associated with the Poisson brackets 
between two arbitrary functions $F(y_{\mcN})$ and $G(y_{\mcN})$ given by 
\begin{equation} \label{PBCano}
\{F,G\} = \sum_{i,j} \,\Lambda^{ij} \,  
\frac{\partial F}{\partial y_{\mcN}^i} \, \frac{\partial G}{\partial y_{\mcN}^j} \,,
\quad \text{with} \quad 
\Lambda^{ij} := 
\mathbb{J}_{12} = \left( \begin{array}{cc}
  0 & \mathbb{I}_6 \\
-\mathbb{I}_6 & 0 
\end{array} \right)\,,
\end{equation}
where $\mathbb{I}_6$ is the $6 \times 6$ identity matrix, such that $\mathbb{J}_{12}$ is the canonical $12 \times 12 $ Poisson matrix. 
The Hamiltonian $H_{\mcM}$, when restricted to $\mcN$, simply becomes the following scalar field $\mcN \rightarrow \RR$
\begin{equation} \label{Hstart}      
    H_{\mcN} : ( x^{\alpha}, \sigma, \zeta, \pi_{\alpha}, \pi_{\sigma}, \pi_{\zeta} ) \mapsto
    \frac{1}{2} g^{\alpha\beta} \pi_{\alpha} \pi_{\beta} 
    + 
    \frac{1}{2} g^{\alpha\beta} \pi_{\alpha} \omega_{\beta C D}S^{CD} \,.
\end{equation}
Even though the right-hand sides of Eqs.~\eqref{HM} and \eqref{Hstart} are seemingly identical, their left-hand sides indicate that one is a function of $14$ coordinates $\mcM$, while the other of canonical coordinates on $\mcN$, leaving no ambiguity. More precisely, in the right-hand side of Eq.~\eqref{Hstart}, the 4 coordinates $x^\alpha$ are hidden in $g^{\alpha\beta}$ ad $\omega_{\alpha BC}$, their momenta $\pi_\alpha$ appear explicitly, and the two canonical pairs $(\sigma,\pi_\sigma),(\zeta,\pi_\zeta)$ are hidden in the spin and mass dipole $3$-vectors $(\vec{S},\vec{D})$ via Eq.~\eqref{eq:def-vecSD} and the following formulae
\begin{subequations} \label{spinsympSSC}
    \begin{align}
        S^1 &= {\pi_\sigma}            \,, \\ 
        S^2 &= \sqrt{\pi_\zeta^2-\pi_\sigma^2} \cos\sigma        \,, \\
        S^3 &= \sqrt{\pi_\zeta^2-\pi_\sigma^2} \sin\sigma        \,, \\
        D^1 &= -\sqrt{1-S_\circ^2/\pi_\zeta^2} \sqrt{\pi_\zeta^2-\pi_\sigma^2} \sin \zeta     \,, \label{D1111}\\
        D^2 &= \sqrt{1-S_\circ^2/\pi_\zeta^2} \, \bigl( {\pi_\sigma} \sin \zeta \cos\sigma + {\pi_\zeta} \cos\zeta \sin\sigma \bigr)       \,, \\
        D^3 &= \sqrt{1-S_\circ^2/\pi_\zeta^2} \, \bigl( {\pi_\sigma} \sin \zeta \sin\sigma -  {\pi_\zeta} \cos\zeta \cos\sigma \bigr)         \,.
    \end{align}
\end{subequations}
where the Casimir $S_\circ$ appears as a parameter here; it labels the leaf $\mcN$. 
The upper labels $I \in \{1,2,3\}$ for $(\vec{S},\vec{D})$ in Eq.~\eqref{spinsympSSC} 
are shifted from those adopted in Eqs.~(3.2) in Paper I, by the cyclic permutation $(S^1,S^2,S^3,D^1,D^2,D^3) \mapsto (S^3,S^1,S^2,D^3,D^1,D^2)$. This corresponds to a symplectic transformation, interpreted as a $2\pi/3$-rotation with respect to the direction $\text{Span}(\vec{e}_1 + \vec{e}_2 + \vec{e}_3)$ in the Euclidean triad $(e_I)^a$.  We only account for this permutation to have simpler expressions in the subsequent developments. The overall formalism developed in Paper I stays completely unchanged. 

As can be seen in Eqs.~\eqref{spinsympSSC}, 
the two degrees of freedom $(\sigma,\zeta)$ are dimensionless, 
and their conjugate momenta $(\pi_\sigma,\pi_\zeta)$ have 
the dimensions of the angular momentum, $[\length]^{+2}$.
Their physical meaning is closely related with the Euler angles 
of $(\vec{S},\vec{D})$ within the Euclidean triad $(e_I)^a$. More on their physical interpretation can be found in Sec.~III.B in Paper I. 

Because the $12$ individual coordinates $y_{\mcN}$ on $\mcN$ are canonical, 
their evolution is simply given by the classical \textit{Hamilton equations}. 
In terms of the Poisson brackets in Eq.~\eqref{PBCano}, 
setting $F = y_{\mcN}^i$ and $G = H_{\mcN}$, we have 
\begin{equation} \label{Hameq}
\frac{\ud y_{\mcN}^i}{\ud \bar{\tau}} 
=
\{ y_{\mcN}^i, H_{\mcN}\}
= 
\sum_{j}\, \Lambda^{ij} \, \frac{\partial H_{\mcN}}{\partial y_{\mcN}^j} \,.
\end{equation}

Before leaving this subsection, it is very important to appreciate the distinction 
between the role played by the body's mass $\mu$ 
and that by the spin norm $S_{\circ}$ in our Hamiltonian formalism, 
even though from the point of view of the MPTD equations, they have the same status. 
The constancy of $\mu$ along solutions is a consequence of our Hamiltonian in Eq.~\eqref{Hstart} which is autonomous (independent of $\bar{\tau}$). 
Inserting Eqs.~\eqref{ptopi} into Eq.~\eqref{Hstart}, 
and making the substitution in the definition of the dynamical mass $\mu$ 
in Eq.~\eqref{eq:def-norms}, 
we find that this ``on-shell'' value of $H_{\mcN}$ along trajectories is numerically equal to $-\mu^2/2$. In other words, the mass $\mu$ is conserved because it is a scalar field on $\mcN$ corresponding to a \textit{first integral} of the system (we refer to Sec.~IV of Paper I for a reminder about first integrals and other classical notions of Hamiltonian mechanics). On the other hand, the spin norm $S_{\circ}$ is a constant because as a result of the symplectic structure of $\mcN$ itself; 
it is the property of the phase space geometry. 
It would be constant for any other Hamiltonian, 
and may as well be considered as an external parameter on $\mcN$. 
In particular, it is \textit{not a first integral}, 
as it is not a scalar field on $\mcN$ in the first place, and therefore should be counted in the list of integrals of motion (to verify integrability, for example). We note that $S_\circ$ being a Casimir of the Poisson bracket also happens for the Newton-Wigner SSC, for which the algebra is SO$(3)$ [cf. \cite{KuLeLuSe.16}].

\subsubsection{The $10$D-Physical phase space $\mcP$}
\label{sub2sec:mcP}

In the previous subsection, we defined a Hamiltonian system 
on a $12$D phase space $\mcN$ endowed with canonical coordinates 
$( x^{\alpha}, \sigma, \zeta, \pi_{\alpha}, \pi_{\sigma}, \pi_{\zeta} )$ 
and the Hamiltonian $H_{\mcN}$ in Eq.~\eqref{Hstart}.
However, this system is physically incomplete for two reasons. 
First, dim$(\mcN)=12$ while the number of independent unknowns in the linear-in-spin MPTD + TD SSC equations is $10$. Second, not all the solution trajectories generated by $H_{\mcN}$ in Eq.~\eqref{Hstart} are physically meaningful because they do not satisfy the TD SSC throughout the motion. 

The issue here is that the TD SSC in Eq.~\eqref{TDSSC} \textit{cannot} be recovered from the linear-in-spin system in Eq.~\eqref{EElin}, even though the latter was originally derived with the help of the TD SSC in Sec.~\ref{subsec:recap}. It is a one way procedure, and indeed Eqs.~\eqref{EElin} only imply 
$\nabla_u (p_a S^{ab})=0$, not the TD SSC itself. 
This is not a problem when working directly at the level of the algebraic-differential system defined by Eqs.~\eqref{EElin}-\eqref{TDSSC}, i.e., outside Hamiltonian mechanics. 
Suppose that the TD SSC $p_a S^{ab} = 0$ is imposed at a point on the worldline $\scL$. Then, the parallel transport $\nabla_u (p_a S^{ab}) = 0$ of the SSC (as implied by Eqs.~\eqref{EElin} up to quadratic corrections), guarantees that the TD SSC is maintained throughout the worldline.

However, this is not the case in the Hamiltonian picture, because there is \textit{a priori
} no way to ``choose an initial condition'' of the trajectory in a phase space.  
By definition, the phase space is the space of all possible trajectories. 
At the level of the phase space $\mcN$, some of them pass through points where $p_\alpha S^{\alpha\beta}=0$, and some do not, making them just physically irrelevant for our purpose. 
Instead, one must limit the analysis to the sub-manifold $\mcT\subset\mcM$ 
where $p_\alpha S^{\alpha\beta}=0$ holds \textit{identically}.  
To give a precise formulation, let $\mcP := \mcT \cap \mcN$ be the subpart of $\mcN$ where the TD SSC holds identically,  
and consider the \textit{constrained Hamiltonian system} on the phase space $\mcP$, 
in which the TD SSC is treated as an algebraic constraint \footnote{We are thus in the domain of ``singular non-degenerate theory'', in the terminology of Ref.~\cite{Derigl.22}.}.
The properties of $\mcP$ are worked out in Sec.~IV.A of Paper I, 
and the conclusion there is twofold:
\begin{itemize}
    \item $\mcP$ is \textit{invariant} under the flow of $H_{\mcN}$, thus ensuring that any trajectory starting on $\mcT$ will never leave it (see Sec.~III.A.1 in Paper I).
    \item $\mcP$ is \textit{$10$-dimensional}, which matches the numbers of unknowns in the linear-in-spin system in Eq.~\eqref{EElin} (see Sec.~III.A.2 in Paper I).
\end{itemize}
This implies, in particular, that only two out of the four\footnote{This was already pointed out in Ref.~\cite{WiStLu.19}, based on the fact the $p_aS^{ab}=0$ only leads to 3 independent equations.
Indeed, one of them is equivalent to $S_\star = 0$, which we enforced by choosing a symplectic leaf in Eq.~\eqref{Casimir}.} TD SSCs in Eq.~\eqref{TDSSC} suffice to define $\mcP$ as a sub-manifold of $\mcN$. 

For the purpose of later computations, 
it is useful to introduce the required constraints in terms of the tetrad components 
of the TD SSC $C^A := \pi_A S^{AB}$, where $\pi^I =  \eta^{IJ}(e_J)^\alpha\pi_\alpha$ 
is the tetrad component of $\pi_{\alpha}$. 
Making use of Eq.~\eqref{eq:def-vecSD},  
we then adopt the first two components of $C^A$ as the constraints in the form 
\begin{subequations}\label{D0D1}
    \begin{align}
        C^0 &= -\pi_1 D^1 - \pi_2 D^2 - \pi_3 D^3 = 0\,, \\
        C^1 &= \pi_0 D^1 - \pi_2 S^3 - \pi_3 S^2  = 0\,,
    \end{align}
\end{subequations}
where $\pi_I$ are functions of $(x^\alpha,\pi_\alpha)$ and $(S^I,D^I)$ are given by Eq.~\eqref{spinsympSSC}. We also used the approximation for the momentum: 
$p_a = \pi_a + O(\epsilon)$.
%
%
We note that any other pair $(C^A, C^B)$ of constraints from the TD SSC (or any combinations thereof), for fixed $A$ and $B$ in $0,\ldots,3$, can also be imposed as constraints. The reason for our specific choice in Eq.~\eqref{D0D1} will become clear when we concretely apply it to the Schwarzschild case in subsequent developments. The restriction from $\mcN$ to $\mcP$ via the TD SSC sub-manifold $\mcT\subset\mcM$ is best summarized in Fig.~\ref{fig:PS}.

\subsubsection{The $8$D geodesic phase space $\mathcal{G}$}

Thus far, our Hamiltonian formalism has been concerned exclusively with a spinning body. 
Which part of the physical phase space $\mcP$ then corresponds to geodesic motion of a non-spinning body? In the view of the original linearized system in Eq.~\eqref{EElin} 
under the TD SSC in Eq.~\eqref{TDSSC}, the spin is encoded in the spacelike four-vector $S^a_\tTD$, whose norm is $S_\circ$ (recall the discussion around Eq.~\eqref{eq:def-TD-spins}).
Thus, the non-spinning limit is defined by $S_{\circ} = 0$.

However, in the Hamiltonian formulation, the parameter 
$S_{\circ}$ is a Casimir invariant built into the definition of both the $12$D symplectic leaves $\mcN$ and the $10$D physical phase space $\mcP$ bound to $\mcN$. Geodesics will thus be confined to the symplectic leaves with $S_{\circ} = 0$. On such leaves, the canonical coordinates in Eq.~\eqref{spinsympSSC} admit the well-defined subcase $(\pi_\sigma,\pi_\zeta)=(0,0)$, which also make all spin components $S^I,D^I$ vanish. The associated angles $\sigma,\zeta$ can be chosen freely, setting them to zero being the simplest choice. These four conditions
\beq
    \left( \sigma, \pi_\sigma, \zeta, \pi_\zeta \right) = \left( 0, 0, 0, 0 \right)
    \quad
    \text{: Non-spinning limit,}
\eeq
define a well-posed, $8$D sub-manifold of the particular leaf $\mcN(S_\circ=0)$, which we denote $\mathcal{G}$. 

By construction, $\mathcal{G}$ is invariant under the flow of the Hamiltonian (which becomes the usual geodesic Hamiltonian $H_\mcG(x^\alpha, p_\alpha) = \tfrac{1}{2}g^{\alpha\beta} p_\alpha p_\beta$ in this limit, which can be found in, e.g., Refs.~\cite{Schm.02,HiFl.08}). 
In particular, $\mcG$ contains all (and only) the phase space trajectories that correspond to geodesics of the background metric.

\subsection{The Schwarzschild spacetime and its symmetries} 
\label{subsec:inv}

We now specialize the covariant Hamiltonian formalism, summarized in Sec.~\ref{subsec:Ham-N} to the Schwarzschild spacetime, and review its geometrical features with an emphasis on symmetries and related invariants of motion.

\subsubsection{Geometric generalities}

To apply the general formalism reviewed in Sec.~\ref{subsec:recap} to the case of the Schwarzschild spacetime, we make a choice of coordinates $x^\alpha$ covering that spacetime. We choose the Schwarzschild-Droste coordinates $x^\alpha=(t,r,\theta,\phi)$, in which the contravariant components of the metric read [cf. Ref.~\cite{GourgoulhonBH}]
\beq \label{metrics}
    g^{tt} = -\frac{1}{f(r)} \,, \quad 
    g^{rr} = f(r) \,, \quad g^{\theta\theta} = \frac{1}{r^2} \,, \quad 
    g^{\phi\phi} = \frac{1}{r^2\sin^2\theta} \,,
\eeq
where $f(r) := 1 - 2M/r$, and $M\geq0$ is the Schwarzschild mass parameter. Because the metric is diagonal, an orthonormal tetrad can be obtained by normalizing the natural basis $(\partial_\alpha)^a$ associated with the coordinates $x^\alpha$. 
With regard to this basis, we identify the tetrad as follows: 
\beq \label{tetrads}
(e_0)^a = \frac{1}{\sqrt{f}} (\partial_t)^a \,, \quad (e_1)^a = \sqrt{f} (\partial_r)^a \,, \quad (e_2)^a = \frac{1}{r} (\partial_\theta)^a \,, \quad (e_3)^a = \frac{1}{r\sin\theta} (\partial_\phi)^a \,.
\eeq
We shall use this tetrad to define the spin components in Eq.~\eqref{eq:def-vecSD} 
and the connection coefficients $\omega_{\alpha BC}$.  
A direct calculation shows that, out of the $48$ coefficients of $\omega_{\alpha BC}$, 
only eight are not vanishing for the tetrad in Eq.~\eqref{tetrads}. 
The explicit listing is (see also Ref.~\cite{STCatalog} but with different conventions):
\beq \label{Riccis}
\omega_{t10}=\frac{M}{r^2} \,, \quad \omega_{\theta21} = \sqrt{f} \,, \quad \omega_{\phi31}=\sqrt{f} \sin\theta \,, \quad \omega_{\phi32} = \cos\theta \,,
\eeq
and the remaining four coefficients are obtained 
by anti-symmetry $\omega_{\alpha BC}=-\omega_{\alpha CB}$. 

We can now express the general convariant Hamiltonian on $\mcN$ 
for a Schwarzschild background. 
Inserting all the ingredients Eqs.~\eqref{metrics} and \eqref{Riccis} 
into Eq.~\eqref{Hstart}, we obtain 
\begin{equation} \label{Htot}
H_{\mcN} = - \frac{\pi_t^2}{2 f} + \frac{f \pi_r^2}{2} + \frac{1}{2 r^2}  \left(
\pi_{\theta}^2 + \frac{\pi_{\phi}^2}{\sin^2 \theta} - \frac{2 \pi_{\phi} \cos
\theta}{\sin^2 \theta} S^1 \right) + \frac{\pi_{\phi}  \sqrt{f}}{r^2 \sin
\theta} S^2 - \frac{\sqrt{f} \pi_{\theta}}{r^2} S^3 + \frac{M \pi_t}{r^2 f} D^1\,,
\end{equation}
where, again, it is understood that the components $(S^I,D^I)$ are functions of  $(\sigma,\zeta,\pi_\sigma,\pi_\zeta)$ using Eqs.~\eqref{spinsympSSC}. We note that $\pi_t,\pi_r$ have physical dimension $[\length]^{+1}$, while $\pi_\theta,\pi_\phi,S^I,D^I$ have dimension $[\length]^{+2}$, just like $H_\mcN$.

The two constraints in Eq.~\eqref{D0D1} defining the $10$D physical phase space $\mcP$ can 
also be expressed in terms of the canonical coordinates on $\mcN$ as
\begin{subequations} \label{twoconstraints}
    \begin{align}
        C^0 &= 
        -\sqrt{f} \pi_r D^1 - \frac{\pi_{\theta}}{r} D^2 - \frac{\pi_{\phi}}{r \sin \theta} D^3 = 0 \,, \label{constraint1} \\ 
        C^1 &= 
        \frac{\pi_t}{\sqrt{f}} D^1 + \frac{\pi_{\phi}}{r\sin \theta} S^2  - \frac{\pi_\theta}{r} S^3 = 0 \,. \label{constraint2}
    \end{align}
\end{subequations}

It is now straightforward to derive Hamilton's equations by inserting Eq.~\eqref{Htot} into Eq.~\eqref{Hameq}. However, we shall not provide them here because their expressions are not required in the context of this paper.

\subsubsection{Invariants from Killing vectors}

Next we look at the symmetries of the Schwarzschild spacetime, which are in correspondence with four Killing vector fields. Given in the natural vector basis $(\partial_\alpha)^a$ associated with $x^\alpha=(t,r,\theta,\phi)$, they read [cf. Ref.~\cite{Poi}]
\begin{subequations} \label{KVs}
    \begin{align}
        k_{(t)}^a &:= (\partial_t)^a \,, \\
        k_{(x)}^a &:= -\sin\phi \,(\partial_\theta)^a - \cos\phi \cot\theta \,(\partial_\phi)^a \,, \\
        k_{(y)}^a &:= \cos\phi \,(\partial_\theta)^a - \sin\phi \cot\theta \, (\partial_\phi)^a \,, \\
        k_{(z)}^a &:= (\partial_\phi)^a \,.
    \end{align}
\end{subequations}
The first one is timelike outside the event horizon (the only place we are interested in for now), and is associated with the invariance under time translations. The last three are spacelike and associated with invariance under spherical (spatial) rotations. Making the substitution within Eq.~\eqref{Killinginv}, we obtain four Killing-vector invariants 
in terms of the canonical coordinates on $\mcN$:  
\begin{subequations} \label{eq:KillingV-invariants}
    \begin{align}
        \Xi_{(t)} &:= \pi_t \,,\\
        \Xi_{(x)} & := - \sin \phi \,\pi_{\theta} - \cos \phi \cot \theta \,\pi_{\phi} 
        + \cos \phi \csc \theta \, S^{1}\,,\\
        \Xi_{(y)} & := \cos \phi \,\pi_{\theta} - \sin \phi \cot \theta \,\pi_{\phi} 
        + \sin \phi \csc \theta \,S^1\,,\\
        \Xi_{(z)} &:= \pi_\phi \,,
    \end{align}
\end{subequations}
They encapsulate both orbital and spinning degrees of freedom, and there is no unique way of splitting them into a geodesic (orbital) and a spinning parts. 

The invariants in Eq.~\eqref{eq:KillingV-invariants} have a natural physical interpretation. 
The first, $\Xi_{(t)}$, is (minus) the total energy of the particle $E$, 
and the remaining components allow us to introduce 
an ``Euclidean'' angular momentum $3$-vector 
(as described in Refs.~\cite{ZeLuWi.20,Ha.20}: 
\begin{equation}\label{eq:def-vecJ}
    \vec{L} = ( L_{x},\,L_{y},\,L_{z} ) := (\Xi_{(x)},\,\Xi_{(y)},\,\Xi_{(z)} )\,.
\end{equation}
In addition, we notice that only three of Eqs.~\eqref{eq:KillingV-invariants} are in involution. We can readily verify it with the Poisson bracket in Eq.~\eqref{PBCano}; 
\begin{equation}
\{ L_x, L_y \} = L_z \,, 
\quad
\{ L_y, L_z \} = L_x\,,
\quand
\{L_z, L_x\} = L_y \,,
\end{equation}
which features the SO$(3)$ nature of these brackets associated 
with the spherical symmetry of the Schwarzschild metric. 
As is customary, we pick the following three, and denote them
\begin{subequations} \label{EIJ}
    \begin{align}
        E &
        := -\pi_t  \,,\label{eq:E} \\
        L_z &
        := \pi_\phi\,,\label{eq:Jz}\\
        L^2 &
        := \pi_{\theta}^2 + \frac{\pi_\phi^2}{\sin^2 \theta} - \frac{2 \pi_\phi \cos \theta}{\sin^2 \theta} S^1 \,, \label{eq:Jsq}
    \end{align}
\end{subequations}
where $L^2 = L_x^2 + L_y^2 + L_z^2$ is the  (squared) norm of $\vec{L}$. 
We stress that $E,L_z,L^2$ are scalar fields on $\mcN$. For example, Eq.~\eqref{eq:Jsq} defines a function $L^2(\theta,\pi_\theta,\pi_\phi,\pi_\sigma)$ of canonical coordinates on $\mcN$. 

It follows from the definitions in Eq.~\eqref{EIJ} and the Hamiltonian in Eq.~\eqref{Htot} that $E$, $L_z$ and $L^2$ are all \textit{first integrals} of the system on the $12$D phase space $\mcN$ (i.e., functions $\mathcal{I}:\mcN\rightarrow\RR$ such that $\{\mathcal{I},H_\mcN\}=0$). 
While this is rather trivial for $E$ and $L_z$ because $H_{\mcN}$ in Eq.~\eqref{Htot} is independent of $t$ and $\phi$ 
(i.e. the latter are ``cyclic'' coordinates), 
the fact that $L^2$ is a first integral follows from an explicit (but easy) calculation 
of the Poisson bracket $\{ L^2, H_{\mcN} \} = O(\epsilon^2)$ with Eq.~\eqref{PBCano}.

\subsubsection{Invariants from Killing-Yano tensors}

In the Schwarzschild spacetime, there exists a rank-2 Killing-Yano tensor $Y^{ab}$. In the natural basis, it is given by [cf. Refs.~\cite{Floyd.73,BiGeJa.11}]
\beq\label{eq:def-KY-tensor}
    Y^{ab} = \frac{2}{r \sin\theta} (\partial_\theta)^{[a} (\partial_\phi)^{b]}\,,
\eeq
and it is manifestly anti-symmetric. 
Similarly, the Schwarzschild expression for the Killing-St\"{a}ckel tensor $K_{ab}$ 
follows from the definition in Eq.~\eqref{eq:def-KS-tensor}. 
Using Eq.~\eqref{eq:def-KY-tensor}, we find
\begin{equation}\label{eq:K-tensor}
        K_{ab} =  r^4 (\ud\theta)_a (\ud\theta)_b + r^4 \sin^2 \theta \, (\ud\phi)_a (\ud\phi)_b \,.
\end{equation} 
When we substitute Eqs.~\eqref{pStopiS},~\eqref{eq:def-KY-tensor} and~\eqref{eq:K-tensor}
within Eqs.~\eqref{Ruds}, 
we obtain the explicit expressions for R\"{u}diger invariants $\fK$ and $\fQ$ 
in terms of the canonical coordinate on $\mcN$:
\begin{subequations} \label{RudsPS}
    \begin{align}
        \fK &= r D^1 \,,  \label{Rud} \\
        \fQ &= {\pi_{\theta}^2 + \frac{\pi_\phi^2}{\sin^2 \theta} - \frac{2 \pi_\phi \cos \theta}{\sin^2 \theta} S^1}\,,  
        \label{quadRud}
    \end{align}
\end{subequations} 
without any constraints in Eq.~\eqref{twoconstraints}. 
Whether $\fK$ and $\fQ$ are first integrals for $H_{\mcN}$ in Eq.~\eqref{Htot}
can, again, be checked by calculating their Poisson brackets with $H_{\mcN}$
using Eq.~\eqref{PBCano}. We find
\begin{equation} 
\label{kHqH}
\{ \mathfrak{K},H_\mcN \} = -\sqf \, {C}^0  +O(\epsilon^2) \neq 0\,, 
\quand 
\{ \mathfrak{Q},H_\mcN \} = O(\epsilon^2) \,.
\end{equation}
where $C^{0}$ is a constraint introduced in Eq.~\eqref{twoconstraints}.

Our first observation in Eq.~\eqref{kHqH} is that $\fK$ is not an invariant of 
the $12$D phase space $\mcN$ 
because $C^{0}$ on the right-hand side does not vanish there. 
In fact, \textit{$\fK$ will only be invariant on the $10$D physical phase space $\mcP$},  
where the TD SSC is satisfied globally. 
This should be contrasted with the four Killing-vector invariants in Eq.~\eqref{eq:KillingV-invariants} that are first integrals on $\mcN$ \textit{and} on $\mcP$.
Our second observation in Eq.~\eqref{kHqH} is that $\fQ$ is a first integral on $\mcN$.\footnote{} Although unexpected (since $\fQ$ was built to be invariant only under the TD SSC \cite{Rudiger.I.81,Rudiger.II.83,ComDru.22}, we observe that this feature is but a consequence of the equality 
\begin{equation}\label{eq:QvsJ}
\fQ = L^2\,,
\end{equation}
which follows from comparing Eqs.~\eqref{eq:Jsq} and \eqref{quadRud}. This means that the Killing-\textit{tensor} invariant $\fQ$ corresponds to the Killing-\textit{vector} invariant $L^2$. The latter being an invariant irrespective of any SSC, the equality \eqref{eq:QvsJ} explains the result in \eqref{kHqH}.

Although one might be inclined to posit that the Schwarzschild spherical symmetry can be used to explain Eq. ~\eqref{eq:QvsJ} and thus the result in Eq. ~\eqref{kHqH}, 
it is important to note that the latter is also true for the Kerr spacetime 
[cf. paper I.~\cite{Ra.PapI.24}], 
%
%
where one loses spherical symmetry and one does not have Eq.~\eqref{eq:QvsJ}. 
We shall leave any justification of this unexpected fact for the time being. 

The only worthwhile comment we can make is that 
the correspondence in Eq.~\eqref{eq:QvsJ} is probably reminiscent 
of another well-known fact in the non-spinning limit; 
in the case of Schwarzschild spacetime, the Carter constant 
$K := K_{a b} p^{a} p^{b}$ [cf. Ref.~\cite{Carter.68}], 
which is identical to the ``non-spinning part'' of $\fQ$, 
also agrees with the total orbital angular momentum 
$L^2 = p^2_{\theta} + \csc^2\theta \, p_{\phi}^2 $ 
[cf. App.~\ref{app:geo}]. 

Regardless, we conclude that the Killing-Yano tensor in Schwarzschild spacetime gives rise to only one non-trivial R\"{u}diger invariant $\fK$ exclusively at the level of $\mcP$, not $\mcN$. The other invariant $\fQ$ is conserved on $\mcN$ but it is equivalent to  the Killing-vector invariant $L^2$, already introduced in the preceding subsection.

\subsubsection{Angular momentum and spin configurations}
\label{sub2sec:AM}

It is perhaps helpful to note that $\fK$ and $\fQ$ in the physical phase space $\mcP$   
can be given an alternative representation 
in terms of a covariant notion of angular-momentum four vector 
[cf. Sec.~IV.C.\textit{3} in Paper I], namely
\begin{equation}\label{def:KY-AM}
    {\mathcal {L}}^{b} := Y^{ab} p_{a}\,,
\end{equation}
constructed from the Killing-Yano tensor $Y^{ab}$ [cf. Refs.~\cite{DiRu.I.81,DiRu.II.82}].
The vector ${\mathcal {L}}^{a}$ has the dimension of angular momentum $[\length]^{+2}$, 
and its explicit expression in the Schwarzschild spacetime is 
\begin{equation}\label{eq:def-KYL}
{\mathcal {L}}^{a} 
=
- \frac{1}{r \sin\theta} 
\left(p_{\phi}\,(\partial_\theta)^{a} - p_{\theta}\, (\partial_\phi)^{a} \right)\,.
\end{equation}
The sign convention in Eq.~\eqref{def:KY-AM} is the one of Drummond and Hughes~\cite{DruHug.I.22,DruHug.II.22}, 
for which the spatial tetrad components 
${\mathcal {L}}^{I}$
become $(\mcL^1,{\mathcal L}^2, {\mathcal L}^3) = (0,-p_\phi / \sin\theta, p_\theta)$ 
and coincide the standard angular momentum $3$-vector $\vec{L}$ 
in the spherical basis in the non-spinning limit [cf. Eq.~\eqref{Crtf} or the Newtonian limit].
This convention is opposite to the one adopted in, e.g., Ref.~\cite{VdM.20,WitzHJ.19}.

When we insert the inverse relation in Eq.~\eqref{eq:def-TD-spins} 
between the TD spin and the spin tensor in Eq.~\eqref{Ruds}, 
and make use of the relation 
$K_{ab} p^a p^b = {\mathcal {L}}^{a} {\mathcal {L}}_{a} = {\mathcal {L}}^2$ 
(recall $K_{ab}:= Y_{ac} Y_b^{\phantom{b}c}$), 
we find the correspondence
\begin{equation}\label{eq:KQ-L}
    {\mathcal {L}}^{a} S_{a}^{\tTD} \longleftrightarrow -\mu\,\fK  
    \quand
    {\mathcal {L}}^2 \longleftrightarrow \fQ + 2 E \fK 
\end{equation}
where we used $\pi_{a} = p_{a} + O(\epsilon)$,  the ``on-shell'' value $-2H_{\mcN} = \mu^2$, and, crucially, we used the TD SSC in Eq.~\eqref{twoconstraints} in the calculation, 
making Eq.~\eqref{eq:KQ-L} an equality on $\mcP$ but not on $\mcN$. 
We observe that the R\"{u}diger invariants are closely related 
to a covariant angular momentum vector ${\mathcal {L}}^{a}$ 
[cf. Ref.~\cite{baleanu2004angular}] 
and the coupling between the TD spin $S_{a}^{\tTD}$ and itself.

The expressions for $\fK$ and $\fQ$ in Eq.~\eqref{eq:KQ-L} sometimes offer 
a practical advantage over those in Eq.~\eqref{RudsPS} 
because their interpretation is intuitively clear. 
The form of ${\cal L}^2$ bears a formal resemblance to the Carter constant 
of the non-spinning body in the linearized Kerr spacetime:
${\mathcal {L}}^2 = (C + L_z^2) - 2 E a L_z + O(a^2)$, 
where $a$ is the Kerr parameter and $C := K - (L_z - a E)^2$ is an alternative form 
of the Carter constant [cf. Refs.~\cite{Drasco:2005is,Sago:2005gd,IsoAl.19}].
As the form suggests, $\fK$ represents the spin-orbit coupling 
between $S_{a}^{\tTD}$ and ${\mathcal {L}}^{a}$.\footnote{This may be viewed as a relativistic extension of the conserved spin-orbit coupling 
identified in the post-Newtonian formalism [cf. Refs.~\cite{Racine:2008qv,Tanay:2020gfb}].}
In particular, the correspondence \eqref{eq:KQ-L} motivates us to introduce special configurations 
for the spin $S^a_\tTD$, depending on whether it is perpendicular to or aligned with the angular momentum ${\mathcal {L}}^{a}$. 
Taking Eqs.~\eqref{eq:norm-TD-spins} and~\eqref{eq:KQ-L} into account 
and that $L^2 = \fQ = {\cal L}^2 + O(\epsilon)$ with Eq.~\eqref{eq:QvsJ}, 
we then \textit{define} the following spin configurations: 
\begin{equation}\label{eq:spin-config}
    {\mathcal {L}}^{a} S_{a}^{\tTD}
    \longleftrightarrow 
    \begin{cases}
		 L\,S_{\circ} \quad & \text{[aligned]} \,, \\
        0 \quad & \text{[perpendicular]} \,, \\
		- L\,S_{\circ} \quad & \text{[anti-aligned]}  \,.
    \end{cases}
\end{equation}
Notice that this link is no longer limited to the TD spin. 
The $\fK$ is an invariant of $\mcP$, and the link now applies to any other parameterization 
of the spin that can be related to it. 

We also note, however, that the interpretation of the vector ${\mathcal {L}}^a$ is subtle. 
According to Eqs.~\eqref{eq:QvsJ} and Eq.~\eqref{eq:KQ-L}, 
the norm of ${\mathcal {L}}^{a}$ is found to be 
\begin{equation}\label{eq:KYL-sq}
    {\mathcal {L}} = L + \frac{E}{L} \fK + O(\epsilon^2)\,.
\end{equation}
Both quantities ${\mathcal {L}}$ and $L$ have the dimension of the angular momentum $[\length]^{+2}$, 
but they differ from each other by $E \fK / L$. 
Indeed, the ``orbital angular momentum'' in terms of constants of motion  
can be defined in a multitude of ways, and it remains ambiguous 
without establishing a link between ${\mathcal {L}}$, $L^2$   
and other definitions of the angular momentum 
--- the ADM or the Bondi angular momentum, for example  ---
in general relativity.
We shall therefore not speculate on the interpretation of ${\mathcal L}^{a}$ any further;
what is important for our analysis is (i) that $\fK$ and $\fQ$ are invariant on $\mcP$, 
and (ii) the covariant definition of particular spin configurations in \eqref{eq:spin-config}.

\section{Hamiltonian system on $\mcN$ and Andoyer variables} 
\label{sec:Ando}

Andoyer variables, named after the French astronomer Henri Andoyer (1862-1929), are a set of canonical coordinates used in the Hamiltonian mechanics of Newtonian rigid body motion: see, e.g., Refs.~\cite{Ando.1915,Heard.08}. 
They are a generalization of the more well-known Hill and Delaunay variables used in the reduction of the Newtonian two-body problem [cf. Refs.~\cite{Lask.And.17,Arn}]. 
The introduction of Andoyer-type variables in general relativity for spinning dynamics presented here is new, and similar to the definition of orbital elements defined in the geodesic case in Sec.~\ref{subsec:Andogeo}, although more geometry is involved.
Quite remarkably, this ``classical'' change of coordinates can also be applied to the present relativistic context.  It is, in essence, the only means by which we are practically able to transition to the physical phase space $\mcP$.

\textit{Note}: Throughout Sec.~\ref{sec:Ando}, we are not dealing with Lorentzian geometry and/or motion within the Schwarzschild spacetime $\textit{per se}$. As always with interpreting coordinate-dependent expressions, we rely on an abstract Euclidean 3D space where, for example, the angular momentum vector $\vec{L}\,$ lives. These Euclidean geometric considerations serve only to obtain a set of formulae that define a symplectic change of coordinates. This will serve as a useful guide in the relativistic generalization of the next section.

\subsection{Schwarzschild geodesics} 
\label{subsec:Andogeo}

The motion of a non-spinning test particle in a given background spacetime is described by geodesics. These geodesics have a well-known Hamiltonian formulation 
[cf. Refs.~\cite{Schm.02,HiFl.08}] on a $8$D phase space $\mcG$ [cf. FIG.~\ref{fig:PS}], 
spanned by the $4$ spacetime coordinates $x^\alpha$ and their $4$ conjugated momenta $p_\alpha$, corresponding to the particle's momentum $p_a$ components. The Hamiltonian generating the geodesic equations is simply $H_\mcG:=\tfrac{1}{2}g^{\alpha\beta}p_\alpha p_\beta$. For the case of the Schwarzschild spacetime, this reads
\begin{equation} \label{Hgeo}
H_{\mcG}\,(t,p_t,r,p_r,\theta,p_\theta,\phi,p_\phi) 
=  -\frac{p_t^2}{2f}+\frac{f p_r^2}{2}+\frac{1}{2r^2} \left(p_\theta^2+\frac{p_\phi^2}{\sin^2\theta}\right) \,.
\end{equation}
where $(t,r,\theta,\phi)$ are the Schwarzschild-Droste coordinates and $(p_t,p_r,p_\theta,p_\phi)$ are the components of the $1$-form $p_a$ in the natural basis $(\partial_\alpha)^a$. Just like the Kepler problem in Newtonian gravity (of which the Hamiltonian is given by the leading order in a $M/r\ll 1$ expansion of Eq.\eqref{Hgeo}), Schwarzschild geodesics are always spatially confined to a $2$-dimensional plane, reflecting the spherical symmetry. In terms of Hamiltonian mechanics, this geometric feature is nothing but a $1:1$ resonance between the $\theta$ and $\phi$ motion; this fact is not so obvious by direct inspection of Eq.~\eqref{Hgeo}. Because the confinement is rarely presented as a resonance in relativistic mechanics, we present a detailed discussion of it in App.~\ref{app:reso}.

\begin{figure}[H]
    \begin{center}
    	\includegraphics[width=0.5\linewidth]{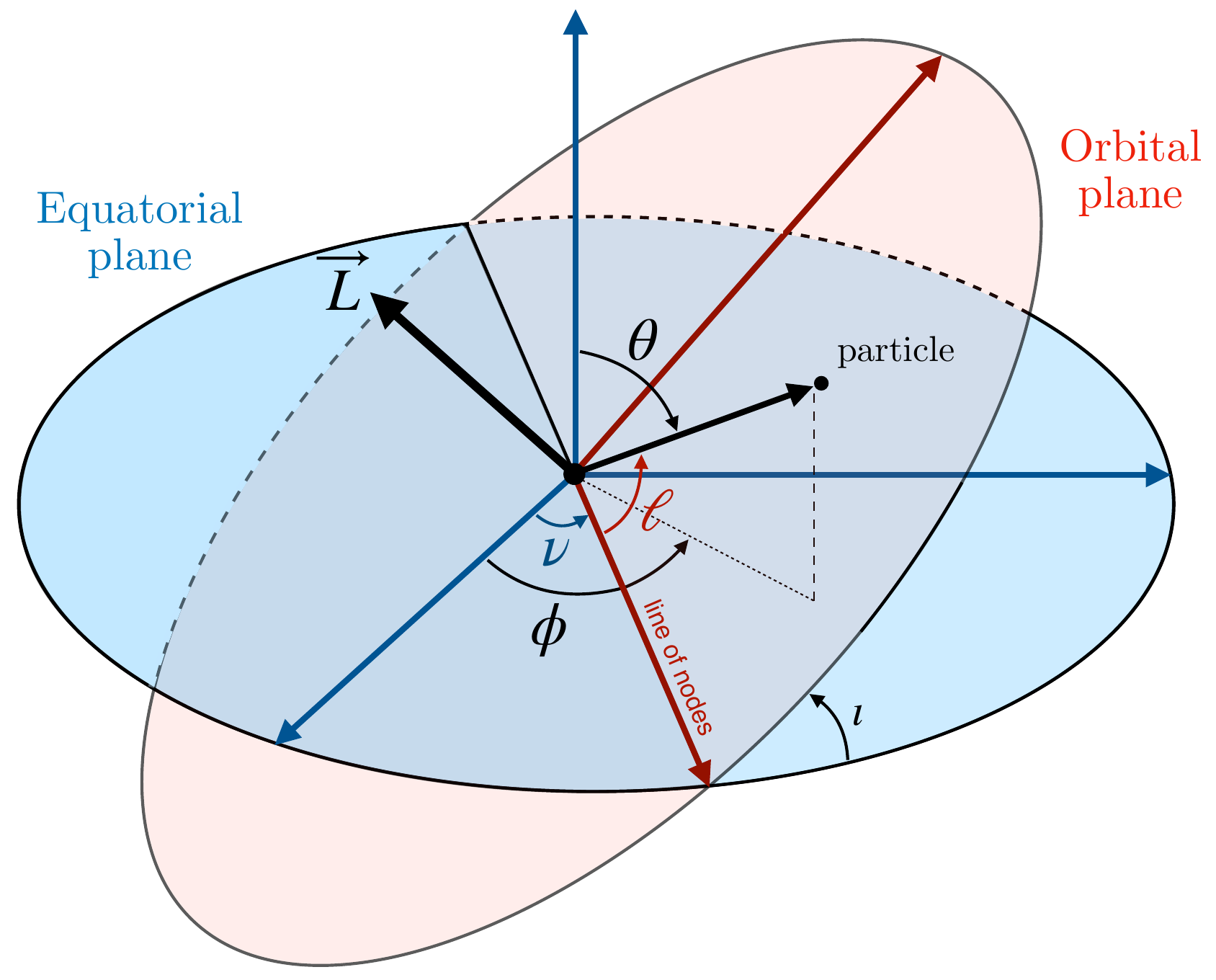}
        \caption{The orbital elements of Schwarzschild geodesics, including the angle $\nu$ defining the line of nodes, the true anomaly $\ell$ and the orbital inclination $\iota$.}
        \label{fig:Andgeo}
    \end{center}
\end{figure} 

A common approach in relativistic mechanics is to limit the geodesic analysis to the equatorial plane ($\theta = \pi/2$). 
However, in the case of an orbit of a spinning body, the spin-curvature coupling will break this resonance and cause orbital precession. 
Furthermore, such a method is not aligned with the philosophy of Hamiltonian mechanics, as it relies on adapting the coordinates to a specific orbit rather than incorporating the symmetry for the entire phase space, encompassing all orbits. 
It is therefore preferable to use tools that are specifically designed to incorporate spherical symmetry into phase space, with the initial step being a canonical transformation. 
In particular, let us consider the following change of coordinates
\begin{equation} \label{symch}
(\theta,p_\theta,\phi,p_\phi) 
\mapsto 
(\ell,p_\ell,\nu,p_\nu) \,,
\end{equation}
where $\nu$ is the angle defining \textit{the line of nodes}, where the orbital plane intersects the equatorial plane, and $\ell$ is \textit{the true anomaly}, which specifies the location of the particle within the orbital plane; see Fig.~\ref{fig:Andgeo}. With the help of some Euclidean trigonometry, as detailed in App.~\ref{app:geo}, we find the following formulae
\begin{subequations}\label{newangles}
    \begin{align}
    \cos \theta & = \sin \iota \sin \ell \,, \\
     \sin \theta \cos (\nu - \phi) & = \cos \ell \,, \\
     \sin \theta \sin (\nu - \phi) & = -\sin \ell \cos \iota \,.
    \end{align}    
\end{subequations} 
These equations define the new coordinate $\ell$ 
(true anomaly) in terms of the old ones $\theta, \phi$, (polar and azimuthal
angles) and two fixed angles $\nu$ (line of nodes) and $\iota$, a constant angle encoding inclination of the orbital angle with respect to the equatorial plane. 
The change of coordinates in Eq.~\eqref{symch} can then be shown 
to induce a canonical transformation on $\mcG$ by setting
\beq \label{newmom}
p_\ell^2 := p_\theta^2+  \csc^2\theta \,p_\phi^2 \,, \quad p_\nu := p_\phi \quand \cos\iota := p_\nu/p_\ell \,,
\eeq
noting that $p_\ell$ coincides with the norm of the angular-momentum $3$-vector $\vec{L}$ defined in Eq.~\eqref{eq:def-vecJ}.
Again, the geometry behind this canonical transformation is summarized in Fig.~\ref{fig:Andgeo}.

In terms of the new coordinates, 
the Schwarzschild geodesic Hamiltonian in Eq.~\eqref{Hgeo} now reads 
\begin{equation} \label{Hgeo-red}
H_{\mcG}(t,p_t,r,p_r,\ell,p_\ell,\nu,p_\nu) 
= 
-\frac{p_t^2}{2f}+\frac{f p_r^2}{2}+\frac{p_\ell^2}{2r^2} \,.
\end{equation}
By direct inspection, this Hamiltonian admits five integrals of motion, i.e., 
functions $\mathcal{I}:\mcG\mapsto\RR$ such that $\{\mathcal{I},H_\mcG\}=0$. 
These are
\begin{itemize}\itemsep0em
    \item $\mu^2=-2H_\mcG$, the (squared) mass of the particle,
    \item $E:=-p_t$, the total energy of the particle,
    \item $L:=p_\ell$, the norm of the angular momentum vector $\vec{L}$,
    \item $L_z:=p_\nu$, the $z$-component of the angular momentum vector $\vec{L}$. 
    \item $\nu$, the angle defining the line of nodes of the fixed plane orthogonal to $\vec{L}$,
\end{itemize}

These five integrals of motion are clearly linearly independent, but only four are \textit{in involution}, i.e., are pairwise Poisson-commuting. This is because $\{\nu,p_\nu\}=1$ by construction. The four first integrals $(H_\mcG,E,L,L_z)$ ensure that the system $(\mcG,\{,\},H_\mcG)$ is integrable, in the sense of Liouville.

\subsection{Construction of relativistic Andoyer variables} 
\label{subsec:Andospin}

%

In the preceding section, we examined the non-spinning (geodesic) situation. 
In the current section, we will generalise the transformation given in Eq.~\eqref{symch} 
to the case of a spinning body, maintaining consistency at the linear order in spin. 
In addition to the orbital angles $\theta,\phi$, 
such a transformation must now involve the spin angle $\sigma$, 
because (the square norm of) the angular momentum $3$-vector $L^2$ in Eq.~\eqref{eq:Jsq}
depends not only on $\pi_\theta$ and $\pi_\phi$,  
but also on the spin conjugated momentum $\pi_\sigma$.

This section's aim is to achieve two main objectives: firstly, to 
to define three new angles $(\nu,\ell,s)$, with their conjugated momenta 
$(\pi_\nu,\pi_\ell,\pi_s)$ related to the angular momentum of the system; 
and secondly, to establish a canonical transformation 
\begin{equation}\label{symch-spin}
    (\theta,\pi_\theta,\phi,\pi_\phi,\sigma,\pi_\sigma) 
    \mapsto 
    (\ell,\pi_\ell,\nu,\pi_\nu,s,\pi_s) \,
\end{equation}
between the old variables and the new coordinates. 
This will facilitate a natural extension of the transformation in the godesic
case in Eq.~\eqref{symch}.

In order to achieve this, it is necessary to ensure that the new conjugated momenta 
$(\pi_\nu,\pi_\ell,\pi_s)$ are related to the angular momenta of the system 
by $(\pi_\nu,\pi_\ell,\pi_s) := (\pi_{\phi}, L, \pi_{\sigma})$. 
There are alternative options that could be considered, but this particular approach provides the most straightforward form of the transformation, which bears resemblance to a well-known transformation in classical mechanics, namely the so-called Andoyer transformation 
[cf. Refs.~\cite{Ando.1915,Heard.08}]. 

Our starting point is the Euclidean, total angular momentum 3-vector $\vec{L}$ 
in Eq.~\eqref{eq:def-vecJ} through its coordinates $(L_x, L_y, L_z)$ 
in some Cartesian basis $(\vec{e}_x,\vec{e}_y,\vec{e}_z)$. 
We then define a secondary basis, associated with the usual transformation from $(x,y,z)$ 
to spherical coordinates $(r,\theta,\phi)$ by 
\begin{subequations}\label{sphtriad}
    \begin{align}
        \vec{e}_r & = \sin \theta \cos \phi \, \vec{e}_x + \sin \theta \sin \phi \, \vec{e}_y + \cos \theta \, \vec{e}_z \,, \\
        \vec{e}_{\theta} & = \cos \theta \cos \phi \, \vec{e}_x + \cos \theta \sin\phi \, \vec{e}_y - \sin \theta \, \vec{e}_z \,, \\
        \vec{e}_{\phi} & = - \sin \phi \, \vec{e}_x + \cos \phi \, \vec{e}_y \,.
    \end{align}
\end{subequations}
Inverting Eqs.~\eqref{sphtriad} and combining the result with Eq.~\eqref{eq:def-vecJ} 
gives the expression of $\vec{L}$ in the spherical basis 
$(\vec{e}_r, \vec{e}_{\theta}, \vec{e}_{\phi})$:
\beq \label{Crtf}
    \vec{L} = \pi_{\sigma} \, \vec{e}_r + \frac{\pi_{\sigma} \cos \theta  - \pi_{\phi}}{\sin\theta}  \vec{e}_{\theta} + \pi_{\theta}  \, \vec{e}_{\phi} \,.
\eeq
Consider now the three planes orthogonal to the vectors $(\vec{e}_z,\vec{L},\vec{e}_r)$, respectively. We refer to them as the Equatorial plane, the Andoyer plane, and the Body plane, respectively. The first two are invariants of the dynamics, since they are orthogonal to a fixed vector~\footnote{This does not mean that the motion is confined into any one of them!}. The body plane is not invariant, and is precessing in general. The Andoyer variables (there are six of them, arranged into three canonical pairs) are entirely constructed from the relations between these three planes, and are defined through geometry in Fig.~\ref{fig:And}. The main results from this construction are as follows.

First, we define the set of Andoyer angles $(\nu,\ell,s)$, 
where $\nu$ and $\ell$ are the same angles as in the geodesic construction 
[cf. Sec.~\ref{subsec:Andogeo}], while $s$ belongs to the Body plane. 
The Andoyer angles can be related to the old ones $(\theta,\phi,\sigma)$ using spherical trigonometry identities (obtained in the spherical triangle formed by the three pairwise intersections of each plane, as in Fig.~\ref{fig:And}). 
These read~\footnote{Compared to Ref.~\cite{Heard.08}, we have shifted the angle $\ell$ by $\pi/2$ so that it coincides with its usual construction. 
This does not break the symplecticity because 
we have $\{\ell, F\} = \{\ell + \pi/2, F\}$ for any function $F$ of the phase space.}

\begin{subequations} \label{Andoyerangles}
    \begin{align}
        \cos \theta                         & = \cos \iota \cos \delta + \sin \iota \sin \delta \sin \ell \,, \\
        \cos (\nu - \phi) \sin \theta    & = \sin \delta \cos \ell \,, \\
        \sin (\nu - \phi) \sin \theta    & = \sin \iota \cos \delta - \cos \iota \sin \delta \sin \ell \,, \\
        \sin (\sigma - s) \sin \theta       & = \sin \iota \cos \ell \,, \\
        \cos (\sigma - s) \sin \theta       & = \cos \iota \sin \delta - \sin \iota \cos \delta \sin \ell \,,   
    \end{align}
\end{subequations}
where $\delta$ and $\iota$ are two intermediate angles defined in Fig.~\ref{fig:And}. Their meaning becomes clear after we introduce the Andoyer momenta, denoted
$(\pi_\nu,\pi_\ell,\pi_s)$, and defined as the projection of the angular momentum vector $\vec{L}$ onto the three axes $(\vec{e}_z,\vec{C},\vec{e}_r)$, respectively. 
From Eqs.~\eqref{eq:Jsq} and \eqref{Crtf}, this leads to the relations 
\beq\label{Andoyermom}
    \pi_\ell^2 := \pi_{\theta}^2+ \frac{(\cos \theta \pi_{\sigma} - \pi_{\phi})^2}{\sin^2 \theta} + \pi_{\sigma}^2\,, \quad 
    \pi_\nu := \pi_\phi\,,  \quad
    \pi_s := \pi_{\sigma} \,,
\eeq
from which, with the help of the geometry in Fig.~\ref{fig:And}, we obtain the expressions for the intermediate $\iota$ and $\delta$ in Eqs.~\eqref{Andoyerangles}. They are given by
\beq \label{iodel}
    \cos\iota = \frac{\pi_\nu}{\pi_\ell} 
    \quand 
    \cos\delta = \frac{\pi_s}{\pi_\ell} \,.
\eeq

The whole set of relations in Eqs.~\eqref{Andoyerangles}, \eqref{Andoyermom} and \eqref{iodel} defines a canonical change of coordinates, expressed in the form of Eq.~\eqref{symch-spin}.
This generalizes the geodesic transformation in Eq.~\eqref{symch}. 
%
%
To prove this, it suffices to either compute the symplectic form just like Andoyer did in classical mechanics (see, e.g., Ref.~\cite{Heard.08}), or to use brute force and check that all Poisson brackets remain canonical. It is worth noting that some formulae, like Eqs.~\eqref{Andoyerangles} or Eq.~\eqref{Andoyermom}, only define the new Andoyer variables up to signs. These signs are all fixed by asking that the transformation be canonical.\footnote{For example, only the square of $\pi_\theta$ can be extracted from Eq.~\eqref{Andoyermom} After imposing $\{\theta,\pi_\theta\}=1$, one readily finds $\pi_\theta=-\pi_\ell \sin\iota \cos(\nu-\phi)$. Similar features happen for other variables.} 

Lastly, we point out two key identities that will be used repeatedly in subsequent developments. They are obtained by combining Eqs.~\eqref{Andoyerangles} and \eqref{Andoyermom} and read 
\begin{equation}\label{magic}
    \pi_\ell \cos s = \frac{\pi_{\phi}}{\sin \theta} \cos \sigma - \pi_{\theta} \sin \sigma  \quand
    \pi_\ell \sin s = \pi_{\theta} \cos \sigma + \frac{\pi_{\phi}}{\sin \theta} \sin \sigma \,.
\end{equation}

\subsection{Interpretation of Andoyer variables} 
\label{sub2sec:Andophys}

The Andoyer variables of (non-relativistic) classical mechanics have to do with the evolution of a free rigid body [cf. Refs.~\cite{gurfil2007serret,Ando.1915,Heard.08}]. 
In our setup, they do not have quite the same Euclidean interpretation, and it is rather the mathematical formulae that they satisfy that we exploit. In this section we get a grasp at what our Andoyer-type variables represent for the relativistic system at hand.

\subsubsection{Orbital elements}

The coordinates $(\ell,\pi_\ell,\nu,\pi_\nu)$ have \textit{almost} the same interpretation as their geodesic counterpart, in terms of orbital elements, as defined in Sec.~\ref{subsec:Andogeo}. More precisely, the set of variables $(\pi_\ell,\nu,\pi_\nu)$ define an invariant plane in the Euclidean $3$-space, orthogonal to the invariant angular momentum vector $\vec{L}$. However, in contrast to geodesics, $\ell$ is not the true anomaly, and the plane is not the orbital plane. Rather, this plane is that of the \textit{mean motion} of the particle, which stays close to it at all times, and $\ell$ is the true anomaly for the \textit{projection} of the position $3$-vector onto this plane. These features are illustrated in Fig.~\ref{fig:And}.

\subsubsection{Planar (and equatorial) orbits}

The physical significance of the angle coordinate $(s,\pi_s)$ is most clearly elucidated by examining the limit as $\pi_s = 0$, 
with the understanding that $\pi_s = \pi_\sigma$ [cf. Eq.~\eqref{Andoyermom}] 
and that this does not render all spin terms as zero [cf. Eq.~\eqref{D1111}].
In this limit Eqs.~\eqref{Andoyerangles} become
~\footnote{In the final two equations of Eqs.~\eqref{Andoyerangles_limit}, 
we also set $s = 0$. This is because the dynamics will ultimately demonstrate that 
if $\pi_s = 0$, then $s = 0$. 
This result will be presented in detail in Sec.~\ref{subsec:Hill-EOMs}} 
%
\begin{subequations} \label{Andoyerangles_limit}
    \begin{align}
        \cos \theta                         & = \sin \iota \sin \ell \,, \\
        \cos (\nu - \phi) \sin \theta    & = \cos \ell \,, \\
        \sin (\nu - \phi) \sin \theta    & = -\cos \iota \sin \ell \,, \\
        \sin \sigma \sin \theta       & = \sin \iota \cos \ell \,, \\
        \cos \sigma \sin \theta       & = \cos \iota \,,  
    \end{align}
\end{subequations}
where, in a similar manner to the inclination angle $\iota$, we assume (without loss of generality) that $\delta\in[0,\pi]$, so that $\pi_s=0$ is equivalent to $\cos\delta=0$ 
[cf. Eq.~\eqref{iodel}], which implies both $\delta = \pi/2$ and $\sin\delta = 1$.

%
%

In the first three equations, we identify the geodesic formulae in Eqs.~\eqref{newangles}, 
for which the motion was planar. Indeed, if $\pi_s = 0$, then it follows from Eq.~\eqref{Crtf} that $\vec{L}$ is orthogonal to $\vec{e}_r$ at all times. 
The motion is confined into the invariant Andoyer plane that is orthogonal to $\vec{L}$. 
The variable $\pi_s$ thus controls the non-planarity of the orbit, and the Andoyer plane can be characterized as a ``mean orbital plane''. In particular, the angle between the position vector $\vec{e}_r$ and the Andoyer plane is $\delta-\pi/2$. 

The angle $s$ is measured within the Body plane, and locates the position of the spin vector $\vec{S}$ [cf Eq.~\eqref{eq:def-vecSD}] with regard to the intersection of the Body and Andoyer plane. The angle $s$ coincides with the spin angle $\sigma$ only for equatorial orbits, where one has $(\theta,\phi,\iota,\sigma) = (\pi/2,\ell+\nu,0,0)$.
Therefore, $s$ is merely the generalization of the angle $\sigma$, which is used to parameterise the spin vector $\vec{S}$ in the spherical triad. For further details on its physical interpretation, see Sec.~III.B in Paper I for its physical interpretation.

%
%
%

Although the orbit of the spinning body is generally non-planar, 
the dynamics still exhibit an invariant plane, which is orthogonal 
to the constant angular momentum vector $\vec{L}$. 
This plane is characterized by a line of nodes $\nu$ and 
an inclination $\iota$ [cf. Eq.~\eqref{iodel}]. 
We stress, again, that the motion of the spinning body is not confined to this plane, 
but stays $O(\epsilon)$-close to it.

\subsubsection{Summary}

To summarize, we have introduced a new set of $12$ canonical coordinate 
for the $12$D phase space $\mcN$. This set, which we call the (relativistic) Andoyer variables, automatically incorporates the spherical symmetry of the Schwarzschild spacetime, and reads
\begin{equation}\label{eq:def-rel-Andoyer}
y_{\mcN}^{\textrm{A}} :=
            (t,r,\nu,\ell,s,\zeta, \pi_t, \pi_r,\pi_\nu,\pi_\ell,\pi_s,\pi_\zeta)\,,
\end{equation}
where the notation $y_{\mcN}^{\textrm{A}}$ includes an $\textrm{A}$ for ``Andoyer''. They are in a 1-to-1 correspondence with the old canonical set $y_\mcN$ defined in Sec.~\ref{sub2sec:mcN} via Eqs.~\eqref{Andoyerangles} and \eqref{Andoyermom}, and they generalize the classical Schwarzschild geodesic elements [cf. Sec.~\ref{subsec:Andogeo}]. 

\begin{figure}[t!]
    \begin{center}
    	\includegraphics[width=0.5\linewidth]{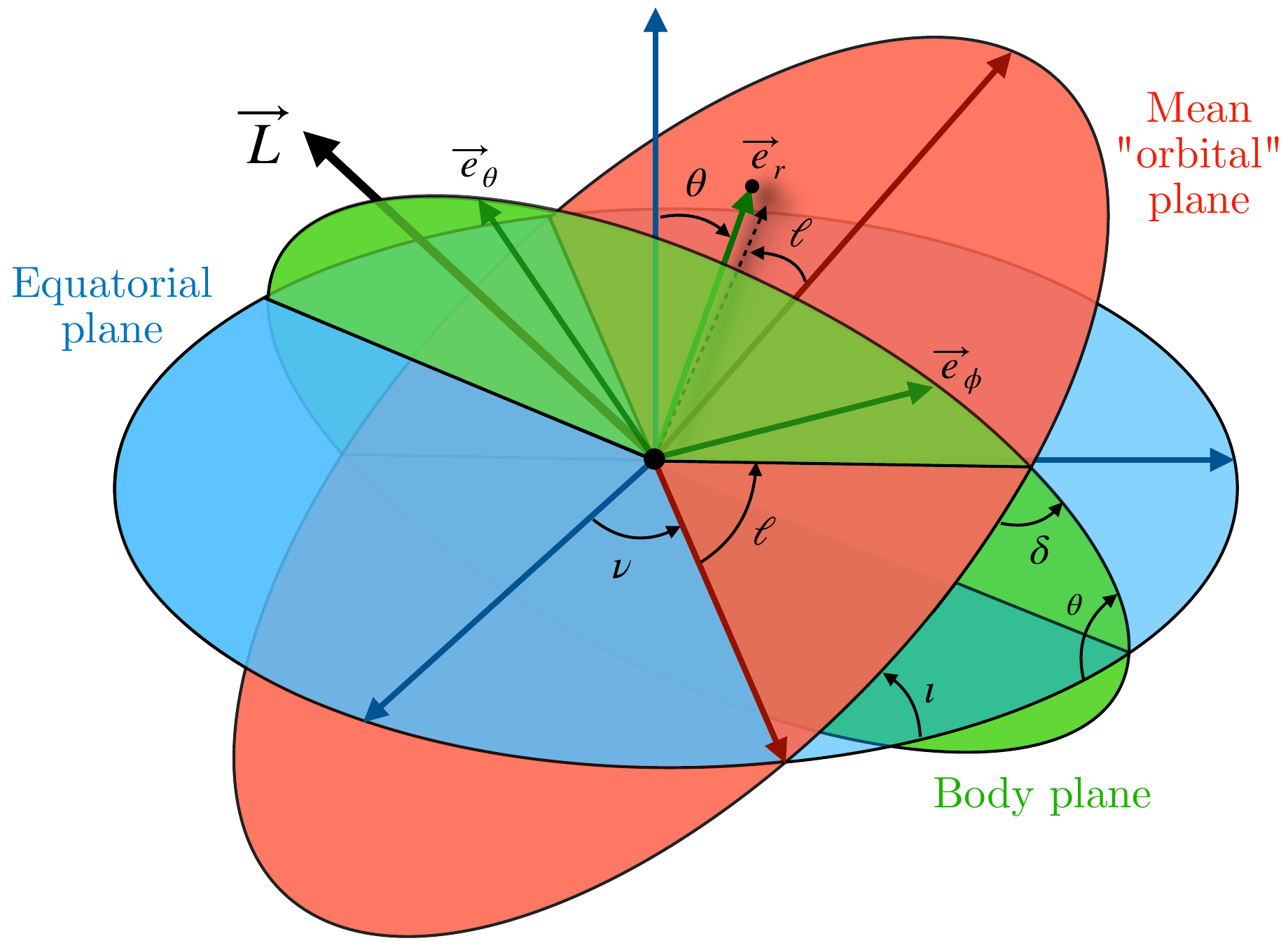}
        \caption{The construction of Andoyer variables for the spinning-case, generalizing the geodesic case depicted in Fig.~\ref{fig:Andgeo}. This time, the position vector $\vec{e}_r$ is not orthogonal to the angular momentum vector $\vec{L}$, the angle between the two being $\pi/2-\delta$. The constant angles $\nu$ and $\iota$ define a constant plane that is orthogonal to $\vec{L}$. This plane is $O(\epsilon)$-close to the particle, but it is not an orbital plane. In general, the orbit is not planar.}
        \label{fig:And}
    \end{center}
\end{figure}

\subsection{Dynamics with Andoyer variables} 
\label{sec:dynando}

The Andoyer variables are much more practical than the original coordinates 
when working on the $12$D phase space $\mcN$. 
This section will discuss the reasons for this. 

\subsubsection{Andoyer Hamiltonian} 
\label{subsec:Andoyer-H}

The Hamiltonian $H_{\mcN}$ in terms of the Andoyer variables is simply 
obtained by inserting Eqs.~\eqref{Andoyerangles} and \eqref{Andoyermom} into the original Hamiltonian in Eq.~\eqref{Htot}. 
Further simplification is achieved by applying the identities in Eq.\eqref{magic}, 
which results in
\begin{equation} \label{Hred}
    H_{\mcN} = 
    - \frac{\pi_t^2}{2 f} + \frac{f \pi_{r}^2}{2} + \frac{\pi_{\ell}^2}{2 r^2} 
    + \left( \frac{\pi_\ell \sqf}{r^2} \cos s - \frac{M \pi_{t} }{r^2 f} 
    \sqrt{1 - \frac{S_{\circ}^2}{\pi_{\zeta}^2}} \sin \zeta \right) 
    \sqrt{\pi_{\zeta}^2 - \pi_{s}^2} \,.
\end{equation}
Once again, it is evident that this Hamiltonian admits five integrals of motion, namely
\begin{itemize}\itemsep0em
    \item $\mu^2 = -2H_{\mcN}$, the mass of the particle,
    \item $E:=-\pi_t$, the total energy of the particle,
    \item $L:=\pi_\ell$, the norm of the total angular momentum vector $\vec{L}$,
    \item $L_z:=\pi_\nu$, the $z$-component of the angular momentum vector $\vec{L}$, 
    \item $\nu$, a constant angle defining the line of nodes of the fixed plane orthogonal to $\vec{L}$.
\end{itemize}
Despite the striking similarity with the first integrals of the geodesic case [cf. the discussion below Eq.~\eqref{Hgeo}], we would like to emphasize two points. First, all these first integrals include a spin contribution, and are not equal to their geodesic counterpart (despite their name). Second, the constant plane, orthogonal to $\vec{L}$, is \textit{not} the orbital plane, in contrast to the geodesic case. Generic spinning orbits are not planar, but they do stay $O(\epsilon)$ to the plane orthogonal to $\vec{L}$, which can be considered to be the ``mean'' orbital plane [cf. Fig.~\ref{fig:And}].

\subsubsection{Constraints and equations of motion on $\mcN$}
\label{sub2sec:EOMs-Ad}

The constraints $C^{0}$ and $C^{1}$ in Eq.~\eqref{twoconstraints} that define 
the $10$D phase space $\mcP$ can be re-written in terms of Andoyer variables. 
Using Eqs.~\eqref{spinsympSSC}, \eqref{Andoyermom} and \eqref{magic}, they read
\begin{subequations} \label{C0C1simple}
    \begin{align}
        C^0 &= -\frac{\sqrt{f} \pi_r}{r} \fK(r,\zeta,\pi_s,\pi_\zeta) 
        - \frac{\pi_\ell}{r} \sqrt{1 - \frac{S_{\circ}^2}{\pi_{\zeta}^2}}
       \left( \pi_s \sin s \sin \zeta -\pi_\zeta \cos s \cos \zeta \right) \,, \\ 
        C^1 &= \frac{\pi_t}{r \sqrt{f}} \fK(r,\zeta,\pi_s,\pi_\zeta)
        + \frac{\pi_\ell}{r} \sqrt{\pi_\zeta^2 - \pi_s^2} \cos s\,,
    \end{align}
\end{subequations}
where we made the dependence on the R\"{u}diger invariant $\fK$ [cf. Eq.~\eqref{Rud}] explicit, even though, on $\mcN$, this quantity is merely function of the Andoyer variables; it is neither a coordinate nor a first integral. 
However, it reveals the practical advantage of choosing $C^0$ and $C^1$ over other choices for the TD SSC constraints: $C^{0}$ and $C^{1}$ can be directly related with the R\"udiger invariant $\fK$, and will be very useful on $\mcP$.

Because the Andoyer variables in Eq.~\eqref{eq:def-rel-Andoyer} are canonical coordinates 
on $\mcN$, their Hamilton's equations are directly obtained 
by the substitution of the Andoyer Hamiltonian in Eq.~\eqref{Hred} into Eq.~\eqref{Hameq}. 
These equations are straightforward to compute, and we find\footnote{We omit the trivial Hamilton equations for the integrals of motion  
$(\pi_{t}, \pi_{\ell}, \nu, \pi_{\nu})$.} 
\begin{subequations}\label{eq:Heq-Ad}
    \begin{align}   
        \frac{\ud t}{\ud \bar{\tau}}  
        &= \frac{1}{f} \left({E} + \frac{M \fK}{r^3} \right)\,, 
        \label{eq:dot-t-Ad} \\
        \frac{\ud r}{\ud \bar{\tau}}  &= f \pi_{r}\,, \\
        \frac{\ud \pi_{r}}{\ud \bar{\tau}}  
        &= -\frac{M E^2}{r^2 f^2} -\frac{M \pi_r^2}{r^2} + \frac{L^2}{r^3}
        + \left(1 - \frac{4M}{r}\right) \left(2-\frac{3M}{r}\right) 
        \frac{E \fK}{r^3 f^2}\,, \\
         \frac{\ud \ell}{\ud \bar{\tau}}  
        &= \frac{1}{r^2} \left( L + \frac{E \fK}{L} \right)\,, 
        \label{eq:dot-psi-Ad}\\
        \frac{\ud \tan s}{\ud \bar{\tau}}  &= 
        - \left(1 - \frac{3M}{r}\right)  \frac{L^2 \pi_s}{E {\fK} \, r^2}\,, 
        \label{eq:dot-tans-Ad}\\
        \frac{\ud \pi_{s}}{\ud \bar{\tau}}  &= \frac{E {\fK}}{r^2}\, \tan s\,, 
        \label{eq:dot-pis-Ad}\\
        \frac{\ud \zeta}{\ud \bar{\tau}}  &= 
        \frac{E M \fK}{r^3 f} 
        \frac{S_{\circ}^2}{ \left( \pi_\zeta^2 - S_{\circ}^2 \right) \pi_{\zeta}}
        +
        \left(1 - \frac{M}{r}\right)  
        \left(1 - \frac{S_{\circ}^2}{\pi_{\zeta}^2}\right)
        \frac{E \pi_{\zeta}}{\fK f}  \sin^2 \zeta  \,,\\
        \frac{\ud \pi_{\zeta}}{\ud \bar{\tau}}  &= \frac{M E \fK}{r^3 f} \cot \zeta\,,
    \end{align}
\end{subequations}
where we have used the constraints in Eq.~\eqref{C0C1simple} to simplify these expressions. We recall that the constraints can be applied only \textit{after} the equations of motion have been computed, as explained in Paper I. Applying them \textit{before} means restricting from $\mcN$ to $\mcP$, and this will be the object of Sec.~\ref{sec:Reduc}.

\subsection{Andoyer-Hill ``effective'' formulation}
\label{subsec:Hill-EOMs}

The equations of motion derived in Eq.~\eqref{eq:Heq-Ad} are just an intermediary step when it comes to solving the MPTD-spacetime dynamics, since the phase space $\mcN$ does not enforce the TD SSC along the trajectories, as we explained in Sec.~\ref{sub2sec:mcP}. However, Eqs.~\eqref{eq:Heq-Ad} have a number of features worth mentioning, which could make them useful in practice. 

\subsubsection{Effective orbital motion}

First, the equations of motion for the ``orbital sector'', i.e., $(t,\pi_t,r,\pi_r,\nu,\pi_\nu,\ell,\pi_\ell)$, depend on the spin only through the R\"{u}diger invariant $\fK$: they completely decouple from the ``rotational sector'', i.e., from the coordinates $(s,\pi_s,\zeta,\pi_\zeta)$. 
We can turn this observation into an effective Hamiltonian~\footnote{The linear-in-$a$ Hamiltonian for Kerr geodesics has the strikingly similar form $H=-\frac{E^2}{2 f}+\frac{f \pi_r^2}{2}+\frac{\fQ}{2 r^2}+\frac{a E L_z}{f r^2}$.}
\begin{equation}\label{eq:def-Heff-Ad}
    H_{\text{eff}} := 
    - \frac{\pi_t^2}{2 f} + \frac{f \pi_r^2}{2} + \frac{{\mathcal {L}}^2}{2 r^2} 
    + \frac{M \pi_t\fK}{r^3 f} \,, 
\end{equation}
where ${\mathcal {L}}^2$ is the square norm of the Killing-Yano angular momentum ${\cal L}^a$ 
in Eq.~\eqref{eq:def-KYL}. If, just in the present effective picture, we treat the eight variables $(t,\pi_t,r,\pi_r,\nu,\pi_\nu,\ell,{\mathcal {L}})$ as four canonical pairs and $\fK$ as a fixed, external parameter, then the corresponding ``effective'' Hamilton equations obtained from Eq.~\eqref{eq:def-Heff-Ad} are identical to the ``true'' equations 
derived in Eq.~\eqref{eq:dot-t-Ad}--\eqref{eq:dot-psi-Ad}.

Moreover, this Hamiltonian can be readily solved by quadrature since it is effectively a $1$D Hamiltonian for the pair $(r,\pi_r)$, all other pairs being of the form (cyclic coordinate, first integral). In a forthcoming article, we shall use this feature to analytically solve the equations of motion by quadrature. This will be compared to available analytical solutions \cite{WiPi.23} based on the (non-Hamiltonian) the Marck tetrad formalism.  

The variables $(t,\pi_t,r,\pi_r,\nu,\pi_\nu,\ell,\pi_\ell)$, we emphasize, are nevertheless \textit{not} canonical on $\mcN$ because (i) $8$ variables are just not sufficient to cover the $12$D phase space $\mcN$, and (ii) they are not even canonical pairs on $\mcN$, since, for example, $\{t,\mathcal{L}\}=-\fK/\mathcal{L}\neq 0$. In particular, the dynamics of old angles $(\theta, \pi_{\theta}, \phi, \pi_{\phi})$  
cannot be recovered from this effective picture, 
in which the spin sector $(s,\pi_s,\zeta,\pi_\zeta)$ is completely separated. 
The Hamiltonian $H_\text{eff}$ in Eq.~\eqref{eq:def-Heff-Ad} can therefore capture only the dynamics of the orbital sector effectively. 

\subsubsection{Hill equation for the spin}

The effective Hamiltonian in Eq.~\eqref{eq:def-Heff-Ad} only recovers the orbital part of Eqs.~\eqref{eq:Heq-Ad}, i.e., the first four ODEs. The spin part, encoded into the two pairs $(s,\pi_s)$ or $(\zeta,\pi_\zeta)$ (last four ODEs in Eq.~\eqref{eq:Heq-Ad}), is required for a complete description of the dynamics. However, only one of these two pairs is sufficient: once we have one, the other can be obtained algebraically from the constraints in Eq.~\eqref{C0C1simple}. We find that the pair $(s,\pi_s)$ is easier to work with than $(\zeta,\pi_\zeta)$. Combining Eqs.~\eqref{eq:dot-tans-Ad} and~\eqref{eq:dot-pis-Ad} with the equations of motion for the true anomaly $\ell$ in Eq.~\eqref{eq:dot-psi-Ad}, 
we arrive at the following of ODEs 
\begin{equation}\label{eq:ddPsddPsi}
    \frac{\ud^2 \pi_s}{\ud \ell^2} + \left(1-\frac{3M}{r}\right) \, \pi_s= 0 
    \quand
    \tan s = \frac{L}{E\fK} \frac{\ud \pi_s}{\ud \ell} \,.
\end{equation}
%
A notable fact is that $\pi_s(\ell)$ satisfies a Hill differential equation
when $r(\ell)$ is a periodic function, which is always the case for bound orbits. 
Moreover, the spin-perpendicular configuration, defined in Eq.~\eqref{eq:spin-config},
is well-encapsulated in Eqs.~\eqref{eq:ddPsddPsi} as the $\fK\rightarrow 0$ limit; 
in this case we must have $s = {\pm}\pi/2$ so that $\pi_s$ is regular under $\fK = 0$ and evolves freely.

It is worth noting that the Hill equation in Eq.~\eqref{eq:ddPsddPsi} 
in the equatorial limit $\theta \to \pi/2$ bears resemblance to the other Hill equation 
for the TD spin $S^{a}_{\tTD}$ derived by Apostolatos 
in Ref.~\cite{TheocharisAApostolatos_1996}. 
This latter equation is given by 
[cf. Eqs.~(24) and (25) in Ref.~\cite{TheocharisAApostolatos_1996}]: 
\begin{equation}\label{eq:ddSddphi}
    \frac{\ud^2 (r S^{\phi}_{\tTD})}{\ud \ell^2} + \left(1-\frac{3M}{r}\right) \, 
    (r S^{\phi}_{\tTD}) = 0 
    \quand
    -S^{r}_{\tTD} = \frac{\ud (r S^{\phi}_{\tTD})}{\ud \ell} \,,
\end{equation}
where we used $\phi = \ell$ in the equatorial limit with $\nu = 0$.
Indeed, Eqs.~\eqref{eq:ddPsddPsi} and~\eqref{eq:ddSddphi} are mutually consistent. 
In the equatorial limit, $S^{r}_{\tTD}$ and $S^{\phi}_{\tTD}$ 
in terms of Andoyer variables $y_{\mcN}^{\textrm{A}}$ may be manipulated 
to take schematic form given by 
\begin{equation}\label{eq:ddSddphi}
    S^{A}_{\tTD} = (\ldots) \tan s + (\ldots) \pi_s
    \quad \text{with} \quad
    A = {(r, \phi)}\,,
\end{equation}
where the coefficients $(\ldots)$ are functions of $r$, $\pi_r$ and integrals of motion 
$\mu, E$ and $L$ [cf. Sec.~\ref{subsec:Andoyer-H}] 
Inserting this relation into Eq.~\eqref{eq:ddSddphi}, 
it then becomes evident that the result recovers Eq.~\eqref{eq:ddPsddPsi}.

Nevertheless, there is an essential distinction between the two Hill equations. 
Despite their formal identity, it can be demonstrated that $ r S^{\phi}_{\tTD} \neq \pi_s$, 
and similarly, $S^{r}_{\tTD}$ is not just proportional to $\tan s$.
In other words, the Andoyer variables $(\tan s, \pi_s)$ are not merely components 
of the TD spin vector; 
rather, they are a non trivial combination of the latter given by Eq.~\eqref{eq:ddSddphi} 
because the coefficients are not identically zero even in the equatorial limit. 
From this perspective, our Hill equation in Eq.~\eqref{eq:ddPsddPsi} may be considered 
an extension of Apostolatos's formulation to a non-equatorial orbit 
(and symplectic setups).

\subsubsection{The true anomaly as the Carter-Mino time}

The Hill differential equation in Eq.~\eqref{eq:ddPsddPsi} parametrizes the pair
$(s,\pi_s)$ with respect to the true anomaly $\ell$, instead of the (dimensionless) proper time ${\bar{\tau}}$. 
We point out that the true anomaly $\ell$ is very closely related to 
the (so-named) Carter-Mino time $\lambda$ introduced in Ref.~\cite{Carter.68,Mino:2003yg}, 
which is widely adopted in the literature on the celestial mechanics of a non-spinning compact object in Kerr spacetime. 

In the Schwarzschild limit, the Carter-Mino time $\lambda$ is defined by 
$\ud \lambda = \ud \tau / r^2$. 
With Eq.~\eqref{eq:dot-psi-Ad}, we have 
$\ud \ell = ({\mathcal {L}} / {\mu} )\,\ud \lambda$, 
in which ${\mathcal {L}} / {\mu}$ is a constant of motion with dimension $[\length]^{+1}$. 
Integration of this equation is trivial, and we obtain 
\begin{equation}\label{eq:psivsMino}
    \ell = \frac{\mathcal{L}}{\mu} \lambda + \text{cst} \,.
\end{equation}
Note, however, that $\lambda$ is not a time parameter in the strict sense
because it has the dimension $[\length]^{-1}$; 
recall the dimension of the angle $\ell$, which is $[\length]^{0}$.
In general relativity, it is widely acknowledged that the Carter-Mino time ``remarkably'' uncouples the radial from angular motion of Kerr or Schwarzschild geodesics. 
This is almost ``expected'' in the context of classical mechanics, where the true anomaly 
is related to a conjugate (conserved angular) momentum vector.
This decoupling is achieved through a process that is ``by construction" and is explained in detail in Ref.~\cite{Lask.17}. 
However, it remains to explore this analogy further in the context 
of the relativistic celestial mechanics in Kerr geometry.

\subsubsection{Final remarks for the effective framework}
From the preceding paragraphs, it appears that a simple three-steps recipe for solving the effective dynamics would be as follows: (i) Solve the integrable effective Hamiltonian given 
in Eq.~\eqref{eq:def-Heff-Ad}; (ii) Insert the solution $r(\ell)$ into the Hill equation given by Eq.~\eqref{eq:ddPsddPsi} and solve it; (iii) Express algebraically all remaining unknowns of interest in terms of the solutions $(r,\pi_r,s,\pi_s)$, 
using Eqs.~\eqref{eq:def-Heff-Ad} and \eqref{C0C1simple}. 
While steps (i) and (iii) are tractable analytically, step (ii) should, in principle, rely on numerical integration, as Hill equations like Eq.~\eqref{eq:ddPsddPsi} can seldom be solved analytically for a given $r(\ell)$. In the case of bound orbits, the circular or quasi-circular configurations can be solved analytically following this recipe.
The eccentric configuration, regrettably, necessitates the numerical integration of Hill  equation, and this is where our development would stop if we were to study the MPTD system in Eq.~\eqref{EElin} as the $12$D Hamiltonian formulation on $\mcN$. 

Nevertheless, these drawbacks of the formulation on $\mcN$ do not cause any fundamental obstacle to solve the MPTD dynamics. Ultimately, as we have already explained in Sec.~\ref{sub2sec:mcP}, this $12$D framework is incomplete, and we must move on to the $10$D formulation on $\mcP$ (recall Fig.~\ref{fig:PS}), accounting for the TD SSC to the dynamics. 
In fact, it was demonstrated that the $10$D system on $\mcP$ is \textit{integrable}, as evidenced in Paper I. As a result, it is plausible that for the specific $r(\ell)$ obtained here (which incorporates elliptic functions, as will be elucidated in a subsequent publication), there may still be a possibility to identify analytical solutions to the Hill effective framework. 
This will be investigated in future work.

\section{From $\mcN$ to $\mcP$: reduction from constraints} 
\label{sec:Reduc}

At this stage, we have a symplectic formulation of the dynamics 
on the $12$D manifold $\mcN$, 
with canonical Andoyer variables $y_{\mcN}$ in Eq.~\eqref{eq:def-rel-Andoyer} 
and the Hamiltonian $H_{\mcN}$ in Eq.~\eqref{Hred}. 
We will now impose the algebraic constraints $C^{0} = 0 = C^{1}$ in Eq.~\eqref{C0C1simple} 
originally obtained from the TD SSC in Eq.~\eqref{TDSSC}, 
and restrict to the 10D physical manifold $\mcP\subset\mcN$ where they hold identically.
Any solution to the equations of motion on $\mcP$ will thus correspond 
to a physically well-posed trajectory in real Schwarzschild spacetime, 
in the sense that the orbital dynamics of the spinning body satisfies 
the linear-in-spin, MPTD $+$ TD SSC system [Eqs.~\eqref{EElin}-\eqref{TDSSC}].

\subsection{Hamiltonian constraints} 
\label{subsec:DBalgo}

The $10$D sub-manifold $\mcP$ is defined as the subset of the $12$D manifold $\mcN$ 
where the two algebraic constraints $C^{0} = 0$ and $C^{1} = 0$ in Eq.~\eqref{C0C1simple} hold. 
From these constraints, 
we may wish to isolate (any) two variables among the $12$ variables 
($y_{\mcN}$ or $y_{\mcN}^{\text {A}}$) on $\mcN$, 
express them in terms of the remaining $10$ variables, 
and then recycle the $10$ variables as coordinates on $\mcP$. As explained in Sec.~V of Paper I, $\mcP$ admits a non-degenerate symplectic structure inherited from that on $\mcN$. 
The corresponding Poisson brackets on $\mcP$ are obtained 
by computing the restriction of the Poisson structure of $\mcN$ to $\mcP$.  
They admit the following explicit formula \cite{DiracQM,Bro.22,Derigl.22}
\begin{equation} \label{DB}
  \{ F, G \}^{\mcP} 
  := 
  \{ F, G \} - \frac{\{ F, C^0 \} \{ G, C^1 \}- \{ F, C^1 \} \{ G, C^0 \}}{\{ C^0, C^1 \}} \,,
\end{equation}
for any two functions $F$ and $G$ on $\mcP$. 
In order to distinguish the Poisson brackets in different phase spaces, 
the Poisson brackets on $\mcP$ will be called the ``$\mcP$-brackets'', denoted $\{,\}^\mcP$, 
while those on $\mcN$ will be called the ``$\mcN$-brackets'', simply denoted $\{,\}$. 

Two points must be made regarding the right-hand side of the aforementioned formula 
in Eq.~\eqref{DB}. 
First, the non-vanishing~\footnote{
The formula in Eq.~\eqref{DB} is singular when $\{C^0, C^1\} = 0$, 
which occurs when $\pi_{\alpha} = 0$ [cf. Eq.~\eqref{comucons}].
The singularity is only apparent, however, because 
both $\{C^0, C^1\}$ and the numerator of the formula is always proportional to $\pi_{\alpha}$, 
which we will see from Eq.~\eqref{constbrac}. 
The limit $\pi_{\alpha} \rightarrow 0$ of Eq.~\eqref{DB} is therefore regular.
} 
of the denominator $\{C_0,C_1\}$ is a consequence of the symplectic nature of the Poisson structure on $\mcP$, as explained in Sec.~V.A.2 of Paper I. Second, the right-hand side must be \textit{first} computed with the brackets on $\mcN$, 
and \textit{then} simplified the result, making use of the two constraints $C_0 = 0$ and $C_1=0$ 
if necessary. It is important to appreciate the order of these (non-commuting) operations; 
the redundant variables in Eq.~\eqref{DB} 
must be removed \textit{only after} the brackets on $\mcN$ are completely evaluated.

Technically, the evaluation of the $\mcP$-brackets is straightforward  
because its right-hand side is written in terms of only $\mcN$-brackets, 
for which we can make use of the canonical coordinates on $\mcN$ 
(e.g., the relativistic Andoyer variables $y_{\mcN}^{A}$ in Eq.~\eqref{eq:def-rel-Andoyer}) 
with the canonical brackets in Eq.~\eqref{PBCano}.
But the challenge does not lie in the implementation. 
They rather sit at the level of the actual construction of a sub-set of the $10$ variables, 
which should (i) cover $\mcP$ and (ii) be complete on $\mcP$, 
i.e., such that all their pairwise brackets can be expressed solely in terms of them.
Indeed, this task was previously identified as a significant challenge associated 
with the TD SSC, as outlined in Ref.\cite{WiStLu.19,Witz.MG.23}. 
The subsequent sections will address this issue and compute 
the $\mcP$-brackets in Eq.~\eqref{DB} accordingly.

\subsubsection{Andoyer spin variables} 
\label{sub2sec:AndSpin}

To begin, we introduce some notation for particular combinations of the Andoyer variables $y_{\mcN}^{\text {A}}$ on $\mcN$;  
referring to them as the Andoyer spin variables. They are as follows
\begin{subequations} \label{spinsympSSC-Ad}
    \begin{align}
        \Sigma^1 &:= {\pi_{s}}            \,, \\ 
        \Sigma^2 &:= \sqrt{\pi_{\zeta}^2 - \pi_{s}^2} \cos s        \,, \\
        \Sigma^3 &:= \sqrt{\pi_{\zeta}^2 - \pi_{s}^2} \sin s        \,, \\
        %
        %
        \Delta^1 &:= -\sqrt{1 - S_{\circ}^2 / \pi_{\zeta}^2} 
        \sqrt{\pi_{\zeta}^2 - \pi_{s}^2} \sin \zeta     \,, \\
        \Delta^2 &:= \sqrt{1 - S_{\circ}^2 / \pi_{\zeta}^2} \, 
        \bigl( {\pi_{s}} \sin \zeta \cos s + {\pi_{\zeta}} \cos \zeta \sin s \bigr) \,, \\
        \Delta^3 &:= \sqrt{1 - S_{\circ}^2 / \pi_{\zeta}^2} \, 
        \bigl( {\pi_{s}} \sin \zeta \sin s -  {\pi_\zeta} \cos \zeta \cos s \bigr) \,.
    \end{align}
\end{subequations}

We note that these Andoyer spin variables $(\vec{\Sigma},\vec{\Delta})$ are identical to $(\vec{S},\vec{D})$ in Eq.~\eqref{spinsympSSC}, 
except that the pair $(\sigma,\pi_\sigma)$ is replaced with $(s,\pi_s)$. 
Since $s$ is a combination [cf. Eq.~\eqref{Andoyerangles}] of the orbital angles $\theta,\phi$ and the spin angle $\sigma$ adapted to Schwarzschild's spherical symmetry, $(\vec{\Sigma},\vec{\Delta})$ will be much more practical to use than $(\vec{S},\vec{D})$, 
as we shall see.

The definition of $(\vec{\Sigma},\vec{\Delta})$ is \textit{algebraically} 
the same as $(\vec{S},\vec{D})$ (the only difference being  $(s,\pi_s)\mapsto(\sigma,\pi_\sigma)$). 
As a result, they too satisfy the Casimir relations in Eq.~\eqref{eq:def-Casimir}, namely 
\beq \label{CasimirSigDel}
S_\circ^2 =\vec{\Sigma}\cdot \vec{\Sigma} - \vec{\Delta}\cdot \vec{\Delta}  
\quad \text{and} \quad S_\star^2 = \vec{\Delta}\cdot \vec{\Sigma} \,;
\eeq
as well as the SO$(1,3)$ brackets \eqref{PBsSD}, namely
\beq \label{PBsΣΔ}
    \{\Sigma^I,\Sigma^J\} = \varepsilon^{IJ}_{\phantom{IJ}K} \Sigma^K \,, \quad
    \{\Delta^I,\Delta^J\} = - \varepsilon^{IJ}_{\phantom{IJ}K} \Sigma^K \,, \quad
    \{\Delta^I,\Sigma^J\} = \varepsilon^{IJ}_{\phantom{IJ}K} \Delta^K \,.
\eeq

Both results will be employed extensively in subsequent developments.

\subsubsection{Choice of 10 variables on $\mcP$}

Our next step is to chose $10$ variables to cover the phase space $\mcP$.
Any subset of the $12$ Andoyer variables $y_{\mcN}^{\text {A}}$ (or any combination thereof) 
on $\mcN$ is in principle possible, altough in practice there is no general rule 
as to which variables should be retained or discarded.
Our approach will be to retain the constants of motion  
$(E,\,L_z,\,L^2 = \fQ,\,\fK)$ (or, again, any combination thereof) 
introduced in Sec.~\ref{subsec:inv}. Indeed,
since we know that the system is integrable on $\mcP$ [cf. Sec.~VI in Paper I], these constants of motion can be readily promoted to momenta on $\mcP$. We just have to express them in terms of Andoyer variables. These expressions can readily be found below Eq.~\eqref{Hred} for $(E,L_z,L)$, 
while that for $\fK$ is simply an upgrade from Eq.~\eqref{Rud} to 
\beq\label{Rud-Ad}
    \fK = r \Delta^1\,,
\eeq
using the Andoyer spin variables in Eq.~\eqref{spinsympSSC-Ad}. 

All attempts to construct such $10$ variables are equally valid without loss of generality. 
While the set of constants of motion $(E,L_z,L)$ are highly practical on $\mcP$, 
it turns out that $\fK$ is not particularly well-adapted for our purposes. 
A number of alternative choices were considered as potential starting points, 
with the help of the discussion above 
and the Hill formulation in Sec.~\ref{subsec:Hill-EOMs}. 
The issue with $\fK$ is that, if it is to serve as the variable that encodes the spin-dependence, it must have the correct dimension. Through trial and error, we find that the quantity $\mu\fK/L$ was much more suitable choice than $\fK$: it induces much simpler $\mcP$-brackets and has the correct spin dimension $[\length]^{+2}$.

Ultimately, we decided to adopt the following $10$ variables 
$y_{\mcP}^{\text {C}}$ to cover $\mcP$: 
\begin{equation} \label{choiceP}
y_{\mcP}^{\text {C}}
:=
(t,\pi_t,r,\pi_r,\nu,\pi_\nu,\ell,\pi_\ell,\Gamma,\pi_{\varpi}) \,, 
\quad \text{with} \quad 
\Gamma := \frac{\pi_\ell}{\pi_t r\Delta^1}  \frac{\Sigma^1 \Sigma^2}{\Sigma^3} 
\quad \text{and} \quad 
\pi_\varpi := \mu \frac{r \Delta^1}{\pi_{\ell}} \,.
\end{equation}
It should be noted that $\mu$ in $\pi_{\varpi}$ is understood as a function of $12$ Andoyer variables, as throughout our work. It is not a fixed constant like the Casimir $S_\circ$, but a first integral.

Our choice in Eq.~\eqref{choiceP} was motivated by reasons described below. 
First, Eq.~\eqref{choiceP} allows us to obtain a ``closed'' set of $\mcP$-brackets 
between themselves, i.e., all brackets between any two pair of the set can be expressed 
solely in terms of this set. 
Second, it leads to a simple parametrization of the orbital quantities of interest, 
based on the constants of motions, i.e., Killing invariants 
described in Sec.~\ref{subsec:inv}. 
Third, it includes the invariant $\pi_{\varpi}$ with the dimension of spin $[\length]^{+2}$, 
which is directly related to the R\"{u}diger invaiant $\fK$ in Eq.~\eqref{Rud-Ad}  
through the combination $\pi_{\varpi} = \mu \fK / \pi_{\ell}$. 
Fourth, similarly, it introduces the ``angle'' variable $\Gamma$
with the physical dimension $[\length]^{0}$, which has the form of 
$\Gamma = [\pi_{\ell} / (\pi_{t} \fK)] [\pi_{s} / \tan s]$, 
yet involves the spin Andoyer variables.
The pair of (non-canonical) variables $(\Gamma, \pi_{\varpi})$ is very convenient 
to parameterize the spin dynamics in $\mcP$.
In practice, any one of these motives is crucial in the final reduction 
to the {canonical} coordinates, which we shall develop in Sec.~\ref{subsec:tocano}.

\subsubsection{Symplecticity of the $\mcP$-structure}
Next we proceed to compute the $\mcP$-brackets in Eq.~\eqref{DB} 
between any two variables in Eq.~\eqref{choiceP}.
To outline the general calculation strategy, we here go through a simple warm-up exercise: 
the $\mcN$-bracket between two constraints $C^0$ and $C^1$ in Eq.~\eqref{two}
that appears at the denominator of Eq.~\eqref{DB}.
We detail this particular calculation below 
because it is a prototypical example of every kind of bracket computation that will be performed in the forthcoming section (and in forthcoming works, notably the extension to Kerr).

The constraints $C^0$ and $C^1$ in Eq.~\eqref{C0C1simple} can be simply expressed 
in terms of the Andoyer spin variables in Eq.~\eqref{spinsympSSC-Ad} as 
\begin{equation} \label{two}
        C^0 = -\sqrt{f} \pi_r \Delta^1 - \frac{\pi_\ell}{r} \Delta^3\, 
        \quand
        C^1 =  \frac{\pi_t}{\sqrt{f}} \Delta^1 + \frac{\pi_\ell}{r} \Sigma^2\,.
\end{equation}
The computation of the $\mcN$-brackets results in a step-by-step process 
that is consistent to linear order in spin:
%
    \begin{align}\label{comucons}
        \{C^0,C^1\} 
        &=  -\sqrt{f} \pi_r \frac{\pi_{\ell}}{r} \{ \Delta^1, \Sigma^2 \} 
        -\frac{\pi_{\ell} \pi_t}{r \sqrt{f}} \{ \Delta^3, \Delta^1 \} 
        -\frac{\pi_{\ell}^2}{r^2} \{ \Delta^3, \Sigma^2 \} \cr
        &= -\sqrt{f} \pi_r \frac{\pi_{\ell}}{r} \Delta^3 + \frac{\pi_{\ell} \pi_t}{r \sqrt{f}} \Sigma^2 + \frac{\pi_{\ell}^2}{r^2}\Delta^1 \cr
        &= \left( - \frac{\pi_t^2}{f} + f \pi_r^2 + \frac{\pi_{\ell}^2}{r^2} \right) \Delta^1 \cr
        &= -\mu \frac{\pi_\ell\pi_{\varpi}}{r} \,,
    \end{align}
%
where, in the first line we applied the Leibniz rule and discarded quadratic-in-spin terms, in the second and the third lines, we used the SO$(1,3)$ brackets in Eq.\eqref{PBsΣΔ} 
to eliminate the brackets for $(\Sigma^I,\Delta^I)$, 
and simplified it with the SSCs in Eq.~\eqref{two} via
\begin{equation}\label{eq:sol-S2D3}
\Sigma^2 = -\frac{r \pi_t}{\sqrt{f} \pi_{\ell}} \Delta^{1}\,
\quand
\Delta^3 = -\frac{\sqrt{f} r \pi_r}{\pi_{\ell}} \Delta^{1}\,;
\end{equation}
recall that the simplification with constraints must be always performed 
\textit{after} all the brackets are evaluated. 
The final equality in Eq.~\eqref{comucons} is a mere rewriting, 
using the Andoyer Hamiltonian in Eq.~\eqref{Hred} in the form
\begin{equation}\label{eq:mu-G}
-\frac{\mu^2}{2} 
= H_{\mcN} 
= - \frac{\pi_t^2}{2 f} + \frac{f \pi_r^2}{2} + \frac{\pi_\ell^2}{2 r^2} + O(\epsilon)\,,
\end{equation}
as well as $\Delta^1 = {\pi_{\ell} \pi_{\varpi} } / {\mu r}$, 
which follows from the definition of $\pi_{\varpi}$ in Eq.~\eqref{choiceP}. 
Most importantly, the final expression in Eq.~\eqref{comucons} is solely expressed 
in terms of our choice of the $10$ variables $y_{\mcP}^{\text {C}}$ in Eq.~\eqref{choiceP}.

\subsubsection{$\mcP$-brackets}
\label{sub2sec:P-brackets}
We can now compute the $\mcP$-brackets of all pairs of 
the $10$ variables of $y_{\mcP}^{\text {C}}$ in Eq.~\eqref{choiceP} via Eq.~\eqref{DB}. 
To compute the $\mcN$-brackets in the right-hand side in the formula, 
we follow the general strategy adopted in the calculation of $\{C^0, C^1\}$ 
in the preceding section; 
first, apply the Leibniz rule to relate the expression to 
the brackets in terms of $(\Sigma^I,\Delta^I)$, 
discarding any terms that are quadratic or higher in spins. 
Next, employ the SO$(1,3), \mcN$-brackets in Eq.~\eqref{PBsΣΔ} 
to eliminate the brackets in the expression.
Finally, replace $(\Sigma^I,\Delta^I)$ with $\Gamma$ and $\pi_{\varpi}$ 
(using the relation in Eq.~\eqref{eq:sol-SigmaDelta} of Appendix~\ref{app:PB-TDSSCs}), 
to put the result in the form solely expressed 
in terms of $y_{\mcP}^{\text {C}}$ in Eq.~\eqref{choiceP}.
The key technical point here is to fully make use of the Andyer spin variables 
$(\Sigma^I,\Delta^I)$ in the intermediate stage, which we find to be the most optimal.\footnote{In principle, it is possible to directly calculate the $\mcP$-brackets 
from the Andoyer variables in Eq.~\eqref{eq:def-rel-Andoyer}, 
without relying on $(\Sigma^I,\Delta^I)$. 
However, it appears that such a strategy is computationally intractable in general 
(even with a computer algebra system).}

An explicit listing of the $\mcN$-brackets between the constraints $C^0$ and $C^1$ in Eq.~\eqref{two}
and the $10$ variables $y_{\mcP}^{\text {C}}$ in Eq.~\eqref{choiceP} 
is given in Appendix~\ref{app:PB-TDSSCs}.
Collecting the results from Eq.~\eqref{comucons} and Eq.~\eqref{constbrac}, 
and then inserting them into Eq.~\eqref{DB}, we arrive at 
\begin{subequations} \label{allPbrackets}
    \begin{align}
        \label{eq:tPt-P}
        \{t,\pi_t\}^\mcP &=1 \,, \\ 
        \label{eq:pPp-P}
        \{\ell,\pi_\ell\}^\mcP &=1 \,, \\
        \label{eq:nuPnu-P}
        \{\nu,\pi_\nu\}^\mcP &=1 \,, \\
        \label{eq:tr-P}
        \{ t, r \}^\mcP & = -\frac{\pi_{\ell} \pi_{\varpi}}{\mu^3r} \,, \\
        \label{eq:tPr-P}
        \{ t, \pi_r \}^\mcP & = 
        \left(1 - \frac{M}{r}\right) \frac{\pi_r \pi_{\ell} \pi_{\varpi}}{\mu^3 r^2 f} \,, \\
        \label{eq:tp-P}
        \{ t, \ell \}^\mcP & = \frac{\pi_r \pi_{\varpi}}{\mu^3 r} \,, \\
        \label{eq:rPr-P}
        \{ r, \pi_r \}^\mcP & = 1 - \left(1 - \frac{3M}{r}\right)  \frac{\pi_t \pi_{\ell} \pi_{\varpi}}{\mu^3 r^2 f} \,, \\
        \label{eq:rp-P}
        \{ r, \ell \}^\mcP & = -\frac{\pi_t \pi_{\varpi}}{\mu^3 r}\,,\\
        \label{eq:Prp-P}
        \{ \pi_r, \ell \}^\mcP & = \frac{2 M \pi_t \pi_r \pi_{\varpi}}{\mu^3 r^3 f} \,, \\
        \label{eq:tG-P}
        \{t,\Gamma\}^\mcP &=   \frac{1}{\pi_t \mu^2} 
        \left( - \frac{\pi_r \pi_{\ell}}{r} + \left( f \pi_r^2 - \frac{\pi_{\ell}^2}{r^2} \right) \Gamma + \frac{\pi_r \pi_{\ell} }{r} f \Gamma^2 \right) \,, \\
        \label{eq:rG-P}
        \{r,\Gamma\}^\mcP &= \frac{1}{\mu^2} \left( \frac{\pi_{\ell}}{r} - \pi_r f \Gamma \right) \,,\\
        \label{eq:PrG-P}
        \{\pi_r,\Gamma\}^\mcP &= - \frac{1}{r^2\mu^2} \left(\frac{2 M \pi_r \pi_{\ell} }{r f} 
        - \left( \frac{M\pi_t^2}{f^2} + \pi_r^2 M 
        + f_4 \frac{\pi_{\ell}^2}{f r} \right) \Gamma 
        + f_3 \pi_r \pi_{\ell} \Gamma^2 \right)\,,\\ 
        \label{eq:pG-P}
        \{\ell,\Gamma\}^\mcP &= \frac{1}{r \mu^2}
        \left( \frac{\pi_{\ell}}{r} \Gamma - \pi_{r} f \Gamma^2 \right) \,,\\ 
        \label{eq:PzG-P}
        \{\pi_{\varpi},\Gamma\}^\mcP &= 
        \frac{1}{\mu \pi_{t}} \left( - \frac{\pi_t^2}{f} + f \pi_r^2 
        + \frac{2 f \pi_{r} \pi_{\ell}}{r} \Gamma 
        - \left(\pi_t^2 - \frac{f\pi_{\ell}^2}{r^2} \right) \Gamma^2  \right) \,.
    \end{align}
\end{subequations}
where, in Eq.~\eqref{eq:PrG-P}, $f_3 := 1 - 3M/r$ and $f_4 := 1 - 4M/r$. 
All other $\mcP$-brackets either vanish or are obtained by anti-symmetry of the ones above. 
We note that the $\mcP$-brackets involving the ``angle'' variable $\Gamma$ are consistently truncated at $O(\epsilon^0)$ (while all the other brackets must be kept through $O(\epsilon)$);  we may write $\Gamma$ in the form of $\pi_s \propto \fK \Gamma + O(\epsilon^2)$, in which $\fK \Gamma$ is already at $O(\epsilon)$ with the the R\"{u}diger invariant $\fK = O(\epsilon)$.

The $\mcP$-brackets in Eq.~\eqref{allPbrackets} uniquely define the symplectic structure on $\mcP$ using the $10$ variables $y_{\mcP}^{\text {C}}$ in Eq.~\eqref{choiceP}. 
Even though they are, evidently, non-canonical, 
they are still ``complete'' in the sense that their $\mcP$-brackets involve 
these $10$ coordinates only. 
This is a significant and non-trivial aspect of our findings. 
Its tractability is primarily a consequence of the existence of a multitude of first integrals and the Schwarzschild symmetries.

\subsubsection{Generalized law of motion}
The $\mcP$-brackets presented in Eq.~\eqref{allPbrackets} show very clearly that 
the $10$ variables $y_{\mcP}^{\text{C}}$ in Eq.~\eqref{choiceP} are not canonical on $\mcP$. Still, we can compute the Hamilton equations using the general law of motion for a non-canonical formulation, as reviewed in App.~D of Paper I, which we also recall below.

In the phase space $\mcP$, we have  
the \textit{non-canonical} coordinates $y_{\mcP}^{\text {C}}$ given in Eq.~\eqref{choiceP},
in terms of which the Poisson structure is defined by the $\mcP$-bracket in Eq.~\eqref{allPbrackets}.
The Hamiltonian of the system on $\mcP$, denoted $H_\mcP$, is obtained from the Hamiltonian $H_\mcN$ on ${\mcN}$ in Eq.~\eqref{Hred} and re-expressing it in terms of the $10$ coordinates $y_{\mcP}^{\text {C}}$. The constraints can be used freely to remove any term that would not be expressed in terms of the $y_{\mcP}^{\text {C}}$, since they are Casimirs of the $\mcP$-brackets (this is really the point of using the $\mcP$-structure). After a straightforward calculation 
involving Eqs.~\eqref{spinsympSSC-Ad} and~\eqref{two}, 
we find that (also recall Eq.~\eqref{eq:mu-G})
\begin{equation}\label{eq:H-non-cano-P}
    H_{\mcP}
    =
    - \frac{\pi_t^2}{2 f} + \frac{f \pi_r^2}{2} + \frac{\pi_{\ell}^2}{2 r^2} 
    + \left(1-\frac{3M}{r}\right) \frac{\pi_t \pi_\ell \pi_{\varpi}}{\mu r^2 f} \,,
\end{equation}
where we used the relation between $(\Sigma^I,\Delta^I)$ and $(\Gamma, \pi_{\varpi})$ 
in Eq.~\eqref{eq:sol-SigmaDelta}. 
With these notions in place, the generalized law of motion for any coordinate
$y^{i}_{{\text {C}}} \in y_{\mcP}^{\text {C}}$ is given by 
\begin{equation} \label{loM}
\frac{\ud y_{{\text {C}}}^{i}}{\ud \bar{\tau}} 
= 
\{y^{i}, H_{\mcP} \}^{\mcP} 
= 
\sum_{j} \, \{ y_{{\text {C}}}^{i}, y_{{\text {C}}}^{j} \}^{\mcP}\,
\frac{\partial H_{\mcP}}{\partial y_{{\text {C}}}^{j}} \,.
\end{equation}
where we used the Leibniz rule at the second equality. It is then straightforward to explicitly evaluate this law of motion, 
inserting the $\mcP$ brackets in Eq.~\eqref {allPbrackets} within Eq.~\eqref{loM}. Performing the computation, we obtain\footnote{
As expected, we also get the trivial equations $\{\mathcal{I},H_\mcP\}^\mcP=0$ for all first integrals $\mathcal{I}\in(\nu, \pi_{t}, \pi_{\ell}, \pi_{\nu}, \pi_{\varpi})$.}
\begin{subequations}\label{eq:loM-y}
    \begin{align}
    \frac{\ud t}{\ud \bar{\tau}} &= -\frac{\pi_t}{f} - \frac{M \pi_\ell\pi_{\varpi} }{\mu r^3 f } \,, \label{oneone} \\
    \frac{\ud r}{\ud \bar{\tau}} &= f \pi_r \,, \\
    \frac{\ud \pi_r}{\ud \bar{\tau}} &= - \frac{\pi_t^2 M}{f^2 r^2} 
    - \frac{M\pi_r^2}{r^2} + \frac{\pi_\ell^2}{r^3} 
    + \left(1-\frac{4M}{r}\right) \left(2-\frac{3M}{r}\right) \frac{\pi_t \pi_\ell \pi_{\varpi} }{\mu r^3 f^2} \pi_{\varpi}\, \\
    \frac{\ud \ell}{\ud \bar{\tau}} &= 
    \frac{\pi_\ell}{r^2}  + \frac{\pi_t \pi_{\varpi} }{\mu r^2} \,, \\
    \frac{\ud \Gamma}{\ud \bar{\tau}} &= 
    - \frac{\pi_{\ell}}{r^2}\left( 1 +  \left(1-\frac{3M}{r}\right) \Gamma^2 \right)\,,
    \end{align}
\end{subequations}

The expressions in Eq.~\eqref{eq:loM-y} displayed here can be compared 
with the Hamilton's equations in terms of Andoyer variables in Sec.~\ref{subsec:Andoyer-H}.  
Inserting Eqs.~\eqref{choiceP} with Eq.~\eqref{spinsympSSC-Ad}, 
we find that Eq.~\eqref{eq:Heq-Ad} are then recovered from Eq.~\eqref{eq:loM-y}, 
except the equations for the canonical pair $(\zeta, \pi_\zeta)$ 
that are algebraically obtained from the constraints in Eq.~\eqref{C0C1simple}.

Nevertheless, it should be noted that there is an essential distinction 
between these two equations of motions at the level of the foundation. 
The Andoyer Hamilton's equations in Eq.~\eqref{eq:Heq-Ad} 
are derived in the 12D manifold $\mcN$ 
and the constraints in Eq.~\eqref{C0C1simple} must be applied 
\textit{after} we obtain the physical equations of motion.
On the other hand, the generalized law of motion in Eq.~\eqref{eq:loM-y} 
is directly derived in the 10D physical phase space $\mcP$ 
and the constraints are accounted for in the $\mcP$-brackets in Eq.~\eqref{DB} 
\textit{before} we compute the equations of motion. 

As a consequence, the agreement between Eqs.~\eqref{eq:Heq-Ad} and ~\eqref{eq:loM-y} is nontrivial, 
and gives us confidence that our reduction from $\mcN$ 
to the physical $12$D phase space $\mcP$ is valid.

\subsection{Towards canonical coordinates} 
\label{subsec:tocano}

In this section, we work towards our main goal, to obtain a canonical 
formulation on $\mcP$.
More precisely, we wish to transform from the $10$ non-canonical variables 
$y_{\mcP}^{\text {C}}$ in Eq.~\eqref{choiceP}
to another set of the ``final'' $10$ canonical variables $y_{\mcP}^{\text {F}}$. 
To achieve this, we shall follow the method introduced in App.~B. of Paper I, namely, 
find the map between $y_{\mcP}^{\text {C}}$ and $y_{\mcP}^{\text {F}}$ 
by directly solving the $\mcP$-brackets 
as a coupled system of partial differential equations (PDEs).

The strategy described above cannot be done in full generality, because there exist an infinite number of canonical systems of coordinates on $\mcP$. However, we can reduce this arbitrariness using (i) the existence of first integrals on $\mcP$ and (ii) the fact that the $\mcP$-bracket between any pair in $(\pi_t, r,\nu, \pi_\nu, \pi_\ell, \pi_{\varpi})$ vanishes (in words, they are all in pairwise involution under the symplectic structure on $\mcP$). 
Taking into account these considerations, and inspired by the theory of generating functions, 
we will (i) assume that the final canonical set contains the $6$ variables $(\pi_t,r,\nu,\pi_\nu,\pi_\ell,\pi_{\varpi})$ stay canonical, and (ii) complete the set $(\pi_t,r,\pi_\ell,\pi_{\varpi})$ with 4 conjugated coordinates, which we will construct explicitly. Accordingly, we shall denote the $10$ \textit{final} canonical variables $y_{\mcP}^{\text {F}}$ on $\mcP$ as follows: 
\begin{equation}\label{eq:def-yF}
    y_{\mcP}^{\text {F}} 
    :=
    (\tilde{t}, r, \nu, \tilde{\ell}, {\varpi}, \pi_t, \tilde{\pi}_r, \pi_\nu, \pi_\ell, \pi_{\varpi})\,,
\end{equation}
where the three new variables $(\tilde{t}, \tilde{\ell}, \tilde{\pi}_r)$  
will necessarily be shifted from the Schwarzschild-Droste time $t$, Andoyer-angle $\ell$ and 
the Schwarzschild-Droste radial momenta $\pi_r$, respectively, while the new variable ${\varpi}$ will be that conjugated to the first integral $\pi_{\varpi}$.

We assume that the first $5$ variables $(\tilde{t}, r, \nu, \tilde{\ell}, {\varpi})$ in Eq.~\eqref{eq:def-yF} 
represent the system's degrees of freedom while the following $5$ variables 
$(\pi_t, \tilde{\pi}_r, \pi_\nu, \pi_\ell, \pi_{\varpi})$ are their respective conjugate momenta; 
again their ordering is important. 
The $6$ variables $(r, \nu, \pi_t, \pi_\nu, \pi_\ell, \pi_{\varpi})$ stay unchanged, 
and the remaining $4$ unknowns $(\tilde{t},\tilde{\pi}_r,\tilde{\ell},{\varpi})$ are assumed to take the form
\begin{subequations}\label{eq:def-T-Pr-Psi-z}
    \begin{align}
        \tilde{t}     &= \tilde{t} (t, r, \ell, \Gamma, \pi_t, \pi_r, \pi_\ell, \pi_{\varpi})  \,, \\
        \tilde{\pi}_r &= \tilde{\pi}_r (t, r, \ell, \Gamma, \pi_t, \pi_r, \pi_\ell, \pi_{\varpi})  \,, \\
        \tilde{\ell}  &= \tilde{\ell} (t, r, \ell, \Gamma, \pi_t, \pi_r, \pi_\ell, \pi_{\varpi})  \,, \\
        {\varpi}     &= {\varpi} (t, r, \ell, \Gamma, \pi_t, \pi_r, \pi_\ell, \pi_{\varpi})  \,,
    \end{align}
\end{subequations}
as functions of $y_{\mcP}^{\text {C}}$; 
the dependence on the variables $\nu$ and $\pi_\nu$ is excluded here, as
they already form a canonical conjugate pair in Eq.~\eqref{allPbrackets}.
Our task in this section is, therefore, to find a transformation 
between $y_{\mcP}^{\text {C}}$ and $y_{\mcP}^{\text {F}}$, which takes the form 
\begin{equation} \label{looktransfo}
    (t,r, \ell, \Gamma, \pi_t, \pi_r, \pi_\ell, \pi_{\varpi})_{\text{non-cano}} 
    \mapsto 
    (\tilde{t},r, \tilde{\ell}, {\varpi}, \pi_t, \tilde{\pi}_r, \pi_\ell, \pi_{\varpi})_{\text{cano}}  \,,
\end{equation}
and determine the functional form of Eq.~\eqref{eq:def-T-Pr-Psi-z}
so that all the $\mcP$-brackets between $y_{\mcP}^{\text {F}}$ must be canonical. 
This requirement implies that $y_{\mcP}^{\text {F}}$ must satisfy 
the fundamental Poisson bracket relations given by 
\begin{equation} \label{PBCano-yF}
\{y_F^{i},y_F^{j}\}^{\mcP} = \,\Lambda^{ij} \,  
\quad \text{with} \quad 
\Lambda^{ij} := 
\mathbb{J}_{10} = \left( \begin{array}{cc}
  0 & \mathbb{I}_5 \\
-\mathbb{I}_5 & 0 
\end{array} \right)\,,
\end{equation}
where $y_F^{i} \in y_{\mcP}^{\text {F}}$ and $\mathbb{I}_5$ is the $5 \times 5$ identity matrix, 
such that $\mathbb{J}_{10}$ is the canonical $10 \times 10$ Poisson matrix 
(of the five pairs of canonical coordinates). 
Because the $\mcP$-bracket are anti-symmetric, there are $28$ independent brackets to be examined
\begin{equation}\label{eq:matrix-22}
\begin{matrix}
    \{ \tilde{t}, \pi_t \}^{\mcP} &  &  &  &  &  & \\
    \{ \tilde{t}, r \}^{\mcP} & {\{ \pi_{t}, r \}^{\mcP}} &  &  &  &  & \\
    \{ \tilde{t}, \tilde{\pi}_{r} \}^{\mcP} & \{ \pi_{t}, \tilde{\pi}_{r} \}^{\mcP} & \{ r, \tilde{\pi}_{r} \}^{\mcP} &  &  &  & \\
    \{ \tilde{t}, \tilde{\ell} \}^{\mcP} & \{ \pi_{t}, \tilde{\ell} \}^{\mcP} & 
    \{ r, \tilde{\ell} \}^{\mcP} & \{ \tilde{\pi}_{r}, \tilde{\ell} \}^{\mcP} &  &  & \\
    \{ \tilde{t}, \pi_{\ell} \}^{\mcP} & {\{ \pi_{t}, \pi_{\ell} \}}^{\mcP} & 
    {\{ r, \pi_{\ell} \}}^{\mcP} & \{ \tilde{\pi}_{r}, \pi_{\ell} \}^{\mcP} & 
    \{ \tilde{\ell}, \pi_{\ell} \}^{\mcP} &  & \\
    \{ \tilde{t}, {\varpi} \}^{\mcP} & \{ \pi_{t}, {\varpi} \}^{\mcP} & \{ r, {\varpi} \}^{\mcP} & 
    \{ \tilde{\pi}_{r}, {\varpi} \}^{\mcP} & \{ \tilde{\ell}, {\varpi} \}^{\mcP} & \{ \pi_{\ell}, {\varpi} \}^{\mcP} & \\
    \{ \tilde{t}, \pi_{\varpi} \}^{\mcP} & {\{ \pi_t, \pi_{\varpi} \}}^{\mcP} & 
    {\{ r, \pi_{\varpi} \}}^{\mcP} & \{\tilde{\pi}_r, \pi_{\varpi} \}^{\mcP} & 
    \{ \tilde{\ell}, \pi_{\varpi} \}^{\mcP} & {\{ \pi_{\ell}, \pi_{\varpi} \}}^{\mcP} & 
    \{ {\varpi}, \pi_{\varpi} \}^{\mcP}
\end{matrix}
\end{equation}
where, we note, the $6$ independent brackets between pairs in $(\pi_t,r,\pi_\ell,\pi_{\varpi})$ 
are already canonical 
(in the sense that they already satisfy the $\mcP$-bracket in Eq.~\eqref{PBCano-yF}).
The remaining $22$ brackets are non-trivial, and we demand that 
they must be all canonical. 
This requirement resulted in the brackets being incorporated into the coupled system of the partial differential equations (PDEs) for Eq.~\eqref{eq:def-T-Pr-Psi-z} 
in terms of non-canonical variables $y_{\mcP}^{\text {C}}$.

The objective of the remainder of this section is to solve these PDEs in a step-by-step manner. Readers who are disinterested in the technical details of our calculations may wish to skip directly to Eq.~\eqref{eq:sol-yF} without further delay, where the final solution of the aforementioned PDEs can be found, 
namely the explicit formula for the transformation map in Eq.~\eqref{looktransfo}. 

\subsubsection{The $t$ and $\ell$ dependence}

Our general approach to the problem is to expand the $\mcP$-brackets in Eq.~\eqref{eq:matrix-22} 
into those of the non-canonical variables $ y_{\mcP}^{\text {C}}$
displayed in Eq.~\eqref{allPbrackets}, making use of the Leibniz rule. 
The results of this computation produce the PDEs  
for $(\tilde{t},\tilde{\pi}_r,\tilde{\ell},{\varpi})$ in Eq.~\eqref{looktransfo}.  
Here, as an illustration, we detail the method to determine 
the $t$ and $\ell$ dependence of $(\tilde{t},\tilde{\pi}_r,\tilde{\ell},{\varpi})$.  

To begin, consider the canonical pair $(\tilde{t}, \pi_t)$. It should satisfy the canonical bracket $\{ \tilde{t}, \pi_t \}^{\mcP} = 1$. 
Applying the Leibniz rule, we obtain  
\begin{equation}\label{eq:DB-TPt}
    \{ \tilde{t}, \pi_t \}^{\mcP} = 1
    \quad \Leftrightarrow \quad 
    \frac{\partial \tilde{t}}{\partial t} \{ t,\pi_t \}^{\mcP}
    + 
    \sum_{y^i_{\text {C}} \neq t} \frac{\partial \tilde{t}}{\partial y^i_{\text {C}}} 
    \{ y^i_{\text {C}}, \pi_t \}^{\mcP} = 1\,.
\end{equation}
Substituting Eq.~\eqref{allPbrackets} into Eq.~\eqref{eq:DB-TPt} simplifies 
the expression to $\partial \tilde{t}/\partial t= 1$. 
This is integrated immediately to obtain some information about the new coordinate $\tilde{t}$, namely,
\begin{equation}\label{eq:sol-T0}
    \tilde{t} = t + \tilde{t}_0 (r, \ell, \Gamma, \pi_t, \pi_r, \pi_{\ell}, \pi_{\varpi}) \,,
\end{equation}
leaving the ``constant of integration'' $\tilde{t}_0$ unspecified. The same recipe works for the other brackets involving $\pi_t$. For example, the brackets $\{\tilde{\pi}_r, \pi_t \}^{\mcP} = 0$, $\{\tilde{\ell}, \pi_t \}^{\mcP} = 0$ and $\{{\varpi}, \pi_t \}^{\mcP} = 0$, imply, with the Leibniz rule and Eqs.~\eqref{allPbrackets}, the following PDEs 
\begin{equation}\label{eq:dyFdt}
    \frac{\partial \tilde{\pi}_r}{\partial t} = 0\,,
    \quad
    \frac{\partial \tilde{\ell}}{\partial t} = 0\,,
    \quand
    \frac{\partial {\varpi}}{\partial t} = 0\,.
\end{equation}
The integration of Eq.~\eqref{eq:dyFdt} is again straightforward. 
The general solutions are 
\begin{subequations}\label{eq:sol-y0}
    \begin{align}
    \tilde{\pi}_{r} &= \tilde{\pi}_{0} (r, \ell, \Gamma, \pi_t, \pi_r, \pi_{\ell}, \pi_{\varpi})\,, \\
    \tilde{\ell}    &= \tilde{\ell}_{0} (r, \ell, \Gamma, \pi_t, \pi_r, \pi_{\ell}, \pi_{\varpi})\,, \\
    {\varpi}       &= {\varpi}_{0} (r, \ell, \Gamma, \pi_t, \pi_r, \pi_{\ell}, \pi_{\varpi})\,, 
    \end{align}
\end{subequations}
which are independent of $t$ and leave the functions 
$\tilde{\pi}_{0},\,\tilde{\ell}_{0}$ and ${\varpi}_{0}$ unspecified. At this point, the $t$-dependence of $\tilde{t},\tilde{\pi}_r,\tilde{\ell},{\varpi}$ in Eq.~\eqref{looktransfo} is completely identified.

We next turn to the determination of the $\ell$ dependence of 
$\tilde{t}_0,\,\tilde{\pi}_0,\,\tilde{\ell}_0$ and ${\varpi}_0$ in Eqs.~\eqref{eq:sol-T0} and~\eqref{eq:sol-y0}, 
respectively. PDEs for these functions are obtained 
by following the same strategy as in the case of the $t$ dependence, the relevant $\mcP$-brackets being 
$\{\tilde{t}, \pi_\ell\}^{\mcP} = 0 ,\, \{\tilde{\pi}_r, \pi_\ell \}^{\mcP} = 0,\,
\{{\varpi}, \pi_\ell \}^{\mcP} = 0$ 
and
$\{\tilde{\ell}, \pi_\ell\}^{\mcP} = 1$. 
These brackets lead to the PDEs
\begin{equation}\label{eq:dyFdpsi}
    \frac{\partial \tilde{t}_0}{\partial \ell} = 0\,,
    \quad
    \frac{\partial \tilde{\pi}_0}{\partial \ell} = 0\,,
    \quad
    \frac{\partial {\varpi}_0}{\partial \ell} = 0\,,
    \quand
    \frac{\partial \tilde{\ell}_0}{\partial \ell} = 1\,,
\end{equation}
where we substituted Eq.~\eqref{allPbrackets} as well as the solutions 
in Eqs.~\eqref{eq:sol-T0} and~\eqref{eq:sol-y0} within the target $\mcP$-brackets.
After the integration, we find that $\tilde{t}_0,\tilde{\pi}_0,{\varpi}_0$ are independent of $\ell$, but $\tilde{\ell}_0$ is linear in $\ell$. The temporary solutions in Eqs.~\eqref{eq:def-T-Pr-Psi-z} are 
thus upgraded to 
\begin{subequations}\label{eq:sol-y1}
    \begin{align}
        \tilde{t}     &= t + \tilde{t}_{1}(r, \Gamma, \pi_t, \pi_r,\pi_\ell, \pi_{\varpi}) \,, \\
        \tilde{\pi}_r &= \tilde{\pi}_{1}(r, \Gamma, \pi_t, \pi_r,\pi_\ell, \pi_{\varpi}) \,, \\
        \tilde{\ell}  &= \ell + \tilde{\ell}_{1}(r, \Gamma, \pi_t,r,\pi_r,\pi_\ell, \pi_{\varpi}) \,, \\
        {\varpi}     &= {\varpi}_{1}(r, \Gamma, \pi_t, \pi_r,\pi_\ell, \pi_{\varpi}) \,,
        \label{eq:sol-z1}
    \end{align}
\end{subequations}
again leaving the functions $\tilde{t}_1,\tilde{\pi}_1,\tilde{\ell}_1$ and ${\varpi}_1$ unspecified. 
At this stage, the $t$ and $\ell$ dependence in Eq.~\eqref{looktransfo} 
is completely identified.

\subsubsection{The $\Gamma$ dependence}

We proceed with the determination of the $\Gamma$ dependence in Eq.~\eqref{eq:sol-y1}.  
To achieve this, we look at the remaining $\mcP$-brackets involving its $\pi_{\varpi}$, namely
$\{\tilde{t}, \pi_{\varpi}\}^{\mcP} = 0 ,\, \{\tilde{\pi}_r, \pi_{\varpi} \}^{\mcP} = 0,\,
\{\tilde{\ell}, \pi_{\varpi} \}^{\mcP} = 0$ 
and
$\{{\varpi}, \pi_{\varpi}\}^{\mcP} = 1$. Using Eq.~\eqref{eq:sol-y1} turns the first three brackets into the trivial PDEs
\begin{equation}\label{eq:dyFdG}
    \frac{\partial \tilde{t}_1}{\partial \Gamma} = 0\,,
    \quad
    \frac{\partial \tilde{\pi}_1}{\partial \Gamma} = 0\,,
    \quand
    \frac{\partial \tilde{\ell}_1}{\partial \Gamma} = 0\,,
\end{equation}
implying that $\tilde{t}_1,\tilde{\pi}_1$ and $\tilde{\ell}_1$ are all $\Gamma$-independent. 
The fourth bracket, $\{{\varpi}, \pi_{\varpi}\}^{\mcP} = 1$, is slightly different 
from that in previous calculations,  
because it now involves the non-trivial bracket $\{\Gamma,\pi_{\varpi}\}^\mcP$ 
derived in Eq.~\eqref{eq:PzG-P}. 
Using Eq.~\eqref{eq:sol-y1} the Leibniz rule on $\{{\varpi}, \pi_{\varpi}\}^{\mcP} = 1$ gives the PDE
\begin{equation} \label{eq:dzdG}
    \frac{\partial {\varpi}_1}{\partial \Gamma} 
    = 
    \frac{1}{\{\Gamma,\pi_{\varpi}\}^\mcP}
    =
    \frac{1}{a\,\Gamma^2 + 2b\,\Gamma + c}\,,
\end{equation}
with the $\Gamma$-independent coefficients $a,b,c$ given by
\begin{equation}\label{eq:dzdG-coeff-abc}
    a  =  \frac{f}{\mu\pi_t} 
    \left( -\frac{\pi_t^2}{f} + \frac{ \pi_\ell^2}{r^2} \right)\,,
    \quad
    b = \frac{f \pi_r \pi_\ell}{\mu r \pi_t}\,,
    \quand
    c = \frac{1}{\mu \pi_t} \left( - \frac{\pi_t^2}{f} + f \pi_r^2 \right)\,.
\end{equation}
The integration of Eq.~\eqref{eq:dzdG} is facilitated by noting that $b^2-ac=-1+O(\epsilon)$ and completing the square. 
The simultaneous resolution of Eqs.~\eqref{eq:dyFdG} and \eqref{eq:dzdG} yields 
the following solutions
\begin{subequations}\label{eq:sol-y2}
    \begin{align}
        \tilde{t}     &= t + \tilde{t}_{2}(r, \pi_t, \pi_r,\pi_\ell, \pi_{\varpi}) \,, \\
        \tilde{\pi}_r &= \tilde{\pi}_{2}(r, \pi_t, \pi_r,\pi_\ell, \pi_{\varpi}) \,, \\
        \tilde{\ell}  &= \ell + \tilde{\ell}_{2}(r, \pi_t, \pi_r,\pi_\ell, \pi_{\varpi}) \,, \\
        {\varpi}     &=  \arctan\left(a \,\Gamma+b\right)
                + {\varpi}_2(r, \pi_t, \pi_r, \pi_\ell, \pi_{\varpi}) \,,
        \label{eq:sol-z2}
    \end{align}
\end{subequations}
which leave the new functions $\tilde{t}_2,\tilde{\pi}_2,\,\tilde{\ell}_2$ and ${\varpi}_2$ unspecified; 
at this stage the $t,\ell,\Gamma$ dependence of Eq.~\eqref{looktransfo} 
is completely identified.

\subsubsection{The $\pi_r$ dependence}

We turn to the $\pi_{r}$-dependence in Eq.~\eqref{eq:sol-y2}.
This time, we use the remaining $\mcP$-brackets involving the coordinate $r$, namely $\{r,\tilde{t}\}^{\mcP} = 0,\, \{r,\tilde{\pi}_r\}^{\mcP} = 1,\, \{r,\tilde{\ell}\}^{\mcP} = 0$
and $\{r,{\varpi}\}^{\mcP} = 0$. 
As in the preceding subsections, the Leibniz rule, in conjunction with 
the solutions in Eq.~\eqref{eq:sol-y2} and Eqs.~\eqref{allPbrackets}, 
yields three PDEs for $\tilde{t}_2,\,\tilde{\pi}_2$ and $\tilde{\ell}_2$. These are given by 
\begin{subequations}\label{eq:dyFdPr}
    \begin{align}
    \frac{\partial \tilde{t}_2}{\partial \pi_r} 
    &= \frac{\{t,r\}^\mcP}{\{r,\pi_r\}^{\mcP}}
    =  -\frac{\pi_{\ell} \pi_{\varpi}}{r \mu^3}\,, \\
    \frac{\partial \tilde{\pi}_2}{\partial \pi_r} 
    &=  \frac{1}{\{r,\pi_r\}^{\mcP}}
    =  \frac{\pi_{t} \pi_{\varpi}}{r \mu^3}\,, \\
    \frac{\partial \tilde{\ell}_2}{\partial \pi_r} 
    &= \frac{\{\ell,r\}^\mcP}{\{r,\pi_r\}^{\mcP}} 
    = 1 + \left(1 - \frac{3M}{r}\right)  \frac{\pi_t \pi_{\ell} \pi_{\varpi}}{\mu^3 r^2 f}\,. 
\end{align}
\end{subequations}

The situation for $\{r, {\varpi}\}^{\mcP} = 1$ is less straightforward,
because the solution ${\varpi}$ in Eq.~\eqref{eq:sol-z2} depends on $\pi_{r}$ both through the function ${\varpi}_2$ itself and 
the coefficients $a$ and $b$ in Eq.~\eqref{eq:dzdG-coeff-abc}. 
However, under Eqs.~\eqref{allPbrackets} and~\eqref{eq:sol-z2}, this bracket gives rise to a trivial PDE
for ${\varpi}_2$, namely
\begin{equation}\label{eq:dzdPr}
    \frac{\partial {\varpi}}{\partial \pi_r} 
    = 
    \frac{\{\Gamma,r\}^\mcP}{\{r,\pi_r\}^\mcP} \frac{\partial {\varpi}}{\partial \Gamma} 
    \quad   \Rightarrow \quad
    \frac{\partial {\varpi}_2}{\partial \pi_r} = 0\,.
\end{equation}
As a consequence, the function ${\varpi}_2$ is $\pi_r$-independent.

Moving on to the integration Eqs~\eqref{eq:dyFdPr} and~\eqref{eq:dzdPr} 
for $\tilde{t}_2,\tilde{\pi}_2,\,\tilde{\ell}_2$ and ${\varpi}_2$, we must recall the fact that $\mu$ is actually 
a function of $\pi_{r}$ given in Eq.~\eqref{eq:mu-G} [cf. Sec.~\ref{subsec:Ham-N}].
With this, we still find that the general solutions in Eq.~\eqref{eq:sol-y2} 
can be expressed as
\begin{subequations}\label{eq:sol-y3}
    \begin{align}
        \tilde{t}     &= t - \frac{\pi_r \pi_\ell \pi_{\varpi}}{r \mu \nu_0^2} 
        + \tilde{t}_{3}(r, \pi_t, \pi_\ell, \pi_{\varpi}) \,, \\
        \tilde{\pi}_r &= \pi_r + \left(1 - \frac{3M}{r} \right) 
            \frac{\pi_r \pi_t \pi_\ell \pi_{\varpi}}{r^2 f \mu \nu_0^2}
            + \tilde{\pi}_{3}(r, \pi_t, \pi_\ell, \pi_{\varpi}) \,, \\
        \tilde{\ell}  &= \ell +\frac{\pi_r \pi_t \pi_{\varpi}}{r \mu \nu_0^2} 
            + \tilde{\ell}_{3}(r, \pi_t, \pi_\ell, \pi_{\varpi}) \,, \\
        {\varpi}     &=  \arctan\left(a \,\Gamma+b\right)
                + {\varpi}_3(r, \pi_t, \pi_\ell, \pi_{\varpi}) \,,
    \end{align}
\end{subequations}
where we introduced the shorthand notation 
\begin{equation}\label{eq:def-nu0}
    \nu_0^2 := \pi_t^2/f - \pi_\ell^2 / r^2 \,. 
\end{equation}
These solutions, although they do not specify the integration constants $\tilde{t}_3$, $\tilde{\pi}_3$, $\tilde{\ell}_3$ and ${\varpi}_3$, nevertheless contain all the dependence on $t$, $\ell$, $\Gamma$ and $\pi_r$ of Eq.~\eqref{looktransfo}.

\subsubsection{Fixing the residual dependence}

Finally, the functional form of $\tilde{t}_3,\tilde{\pi}_3,\tilde{\ell}_3$ and ${\varpi}_3$ in Eqs.~\eqref{eq:sol-y3} 
may be determined by the remaining six independent brackets for $(\tilde{t},\tilde{\pi}_r,\tilde{\ell},{\varpi})$, 
which produce a set of coupled PDEs for these functions. 
In this particular problem, however, this approach is not viable because the resulting PDEs become trivial, producing the identities $0 = 0$. 

The determination of $\tilde{t}_3,\tilde{\pi}_3,\tilde{\ell}_3$ and ${\varpi}_3$ 
as a function of variables $(r, \pi_t, \pi_\ell, \pi_{\varpi})$ is instead based on 
an alternative approach, which we detail in Appendix.~\ref{app:homogeneous_PDEs}. 
The main conclusion of this analysis is that there is no unique solution, 
due to the freedom in choosing canonical coordinates in symplectic mechanics. 
Nevertheless, there is a trivial homogeneous solution expressed as  
[cf. Eqs.~\eqref{eq:hom-sol-Pi3z3} and~\eqref{eq:hom-sol-T3Psi3}] 
\beq
\label{eq:hom-sol}
    (\tilde{t}_3,\tilde{\pi}_3,\tilde{\ell}_3,{\varpi}_3) = (0, 0, 0, 0)\,,
\eeq
and we adopt these solutions for Eq.~\eqref{eq:sol-y3}.  
This selection was motivated by the fact that 
it led to the simplest expression for the transformation in Eq.~\eqref{looktransfo} 
to the canonical coordinates $y_{\mcP}^{\text {F}}$;    
at this stage, the $y_{\mcP}^{\text {C}}$ dependence of Eq.~\eqref{eq:sol-y3}, 
namely Eq.~\eqref{looktransfo}, is completely identified.

To sum up, we have thus found $5$ pairs of canonical coordinates $y_{\mcP}^{\text {F}}$ 
on $\mcP$, given by 
\beq
    y_{\mcP}^{\text {F}} = (\tilde{t},r,\nu,\tilde{\ell},{\varpi},\pi_t,\tilde{\pi}_r,\pi_\nu,\pi_\ell,\pi_{\varpi}) \,,
\eeq
where $\nu,\pi_t,\pi_\nu,\pi_\ell$ and $\pi_{\varpi}$ are all the first integrals mentioned previously, $r$ is the Schwarzschild-Droste radius, and 
\begin{subequations}\label{eq:sol-yF}
    \begin{align}
        \tilde{t}     &= t - \frac{\pi_r \pi_\ell \pi_{\varpi}}{r \mu \nu_0^2} \,, \\
        \tilde{\pi}_r &= \pi_r + \left(1 - \frac{3M}{r} \right) 
            \frac{\pi_r \pi_t \pi_\ell\pi_{\varpi}}{r^2 f \mu \nu_0^2} \,, \\
        \tilde{\ell}  &= \ell +\frac{\pi_r \pi_t\pi_{\varpi}}{r \mu \nu_0^2}  \,, \\
        \tan {\varpi}     &=  \frac{f}{\mu \pi_t} \left( \frac{\pi_r \pi_{\ell}}{r} + \left( -
\frac{\pi_t^2}{f} + \frac{\pi_{\ell}^2}{r^2} \right) \Gamma \right) \,. 
    \end{align}
\end{subequations}
are combinations of the Andoyer coordinates in Eqs.~\eqref{eq:def-rel-Andoyer} that were canonical on $\mcN$ (but not on $\mcP$); 
we recall that $\nu_0$ is defined in Eq.~\eqref{eq:def-nu0}.

It is also straightforward to obtain the inverse relations associated 
with Eq.~\eqref{eq:sol-yF}, i.e., the expressions $(t,\pi_r,\ell,\Gamma)$ 
in terms of $(\tilde{t},\tilde{\pi}_r,\tilde{\ell},{\varpi})$. 
Limiting ourselves to linear-in-$\pi_{\varpi}$ order, we have that 
\begin{subequations} \label{inverse}
    \begin{align}
        t &= \tilde{t} + \frac{\tilde{\pi}_r\pi_\ell\pi_{\varpi}}{\nu_0^2 r \mu_0} \,, \\
        \pi_r &=  \tilde{\pi}_r - \left(1-\frac{3M}{r} \right)\frac{\tilde{\pi}_r\pi_t\pi_\ell\pi_{\varpi}}{r^2 f \nu_0^2\mu_0} \,, \\
        \ell &= \tilde{\ell} - \frac{\tilde{\pi}_r\pi_t\pi_{\varpi}}{\nu_0^2 r \mu_0} \,, \\
        \Gamma &= \frac{1}{\nu_0^2} 
        \left( \frac{\pi_\ell\tilde{\pi}_r}{r} + \frac{\mu_0 \pi_t}{f} \tan {\varpi} \right)\,,
    \end{align}
\end{subequations}
where $\pi_r$ in $\mu_0$ has to be replaced by $\tilde{\pi}_r$. 
The explicit formulae in Eq.~\eqref{inverse} allow a direct verification 
of the map in Eq.~\eqref{looktransfo}. This map can now be evaluated using the expression 
of Poisson brackets for \textit{canonical} coordinates, 
which is given by 
$\{F,G\}^\mcP = \sum_i \frac{\partial F}{\partial q^i}  \frac{\partial G}{\partial p_i} - \frac{\partial G}{\partial q^i}  \frac{\partial F}{\partial p_i}$ 
[$F$ and $G$ are arbitrary functions of canonical pairs $(q^i,\,p_i)$ ], 
and we find exact matching of the expressions in Eq.\eqref{allPbrackets}. 

Our main task in Sec.~\ref{sec:Reduc}, namely, the reduction of the Hamiltonian system 
from the $12$D phase space $\mcN$ to the $10$D one $\mcP$, is now completed.

\section{Canonical Hamiltonian system on $\mcP$} 
\label{sec:canofin}

In this section, we present a canonical Hamiltonian formulation of the linear-in-spin, MPTD $+$ TD SSC system in Eqs.~\eqref{TDSSC}-\eqref{EElin} based on the canonical coordinate $y_{\mcP}^{\text {F}}$ in Eq.~\eqref{eq:def-yF}, which is derived in Sec.~\ref{sec:Reduc}.
This Hamiltonian system will be demonstrated to be \textit{integrable} (as previously established in Paper I), as it exhibits the five functionally independent integrals of motion, namely, $\mu$, $E$, $L_z$, $L$, and $\pi_{\varpi}$. 
These are directly related to the Killing(-Yano) invariants introduced in Sec.~\ref{subsec:inv}. 
The Liouville-Arnold theorem in Refs.~\cite{Liou1855,Arn}, which states that integrability implies the existence of action-angle variables and the gauge-invariant notion of Hamiltonian frequencies, allows us to derive their explicit forms. Implementing them in subsequent works in the most practical manner should therefore be straightforward.  

We begin in Sec.~\ref{subsec:can-HP} with a summary of the canonical Hamiltonian system: the phase space, the Poisson structure and the Hamiltonian. Hamilton's equations of the motion of this system are also derived there. In Sec.~\ref{subsec:spin-HP}, we explain how to get back to the spin dynamics from this picture. The action-angle variables and the Hamilton frequencies are then defined in Sec.~\ref{subsec:AA}.
As a first application, we use in Sec.~\ref{subsec:demo-HP} the canonical formulation to the particular case of circular orbits (but with arbitrary orbital inclination and spin), and compare them with known results in literature. The application to eccentric orbits will be studied in a forthcoming paper.

As previously stated in the introductory section~\ref{sub2sec:intro-reduction}, 
the canonical Hamiltonian system presented in Sec.~\ref{subsec:can-HP} is completely 
equivalent to the original linear-in-spin, MPTD $+$ TD SSC system 
in Eqs.~\eqref{TDSSC}-\eqref{EElin}.
While the canonical coordinate system $y_{\mcP}^{\text {F}}$
used in the canonical Hamiltonian system differs apparently 
from $(x^\alpha,p_\beta,S^{\gamma\delta})$ used in the original system, 
they are, in fact, related through a one-to-one correspondence 
[cf. Eq.~\eqref{eq:intro-reduction-proc}], 
a fact that was established through discussions in all preceding sections.

From this point onward, we take the $10$D canonical Hamiltonian system as our fundamental starting point, and we shall omit the ``$\mcP$'' and ``F'' used in Sec.~\ref{sec:Reduc} on brackets and variables labels, to unclutter the notation (unless otherwise specified).

\subsection{The canonical Hamiltonian system} 
\label{subsec:can-HP}

The Hamiltonian system consists of three key ingredients: 
the phase space, the Poisson structure and the Hamiltonian 
[see the review in App.~D of Paper I]. 
In our case, they are defined as follows: 
\begin{itemize}
    \item the $10$D physical phase space $\mcP$, equipped 
        with the $10$ canonical coordinates [the ordering is important below]
        \begin{equation}\label{eq:def-yc}
            (\tilde{x}^i ; \tilde{\pi}_i)_{i\in\{1,\ldots,5\}} 
            := 
            \left( \tilde{t}, r, \nu, \tilde{\ell}, {\varpi} ; \pi_t, \tilde{\pi}_r, \pi_{\nu}, \pi_\ell, \pi_{\varpi} \right)
            \in \RR^5 \times \RR^5 \,;
        \end{equation}
    \item the Poisson structure, given by the canonical Poisson brackets between two smooth functions 
    $F: \mathcal{\mcP} \rightarrow \RR$ and $G: \mathcal{\mcP} \rightarrow \RR$, namely, 
        \begin{equation}\label{eq:PB-c}
            \{F,G\} := \sum_{i=1}^{5} \, 
            \left( \, \frac{\partial F}{\partial \tilde{x}^i}  \frac{\partial G}{\partial \tilde{\pi}_i}
            - \frac{\partial G}{\partial \tilde{x}^i}  \frac{\partial F}{\partial \tilde{\pi}_i} \,  \right) \,;
        \end{equation}
    \item the Hamiltonian $H_c: {\mcP} \rightarrow \RR$ defined by 
    \begin{equation} \label{eq:def-Hc}
            H_c (\tilde{x}^i,\tilde{\pi}_i)
            := 
            -\frac{\pi_t^2}{2 f} + \frac{f }{2}\tilde{\pi}_r^2 + \frac{\pi_\ell^2}{2 r^2} 
            + \left( 1 - \frac{3 M}{r} \right) 
            \frac{\mu_0 \pi_t \pi_{\ell} \pi_{\varpi}}{f r^2\nu_0^2} \,,  \\
    \end{equation}
    where [cf. Eq.~\eqref{eq:def-nu0}]
    \begin{equation}\label{eq:def-mu0-nu0}
    \mu_0^2 = \pi_t^2/f - f\tilde{\pi}_r^2 - \pi_\ell^2 / r^2 = \nu_0^2 - f\tilde{\pi}_r^2\,. 
    \end{equation}
\end{itemize}
Notice that the canonical coordinates $(\tilde{t}, \tilde{\ell}, \tilde{\pi}_r)$ are different 
from the Schwarzschild-Droste time $t$, Andoyer angle $\ell$ and 
the  Schwarzschild-Droste radial momenta $\pi_r$, respectively. 
As we saw back in Eq.~\eqref{inverse}, 
they acquire linear-in-spin corrections. This is inevitable if one is to perform the Hamiltonian reduction consistently, as the spin degrees of freedom couple to every aspect of the orbital motion (time, radial, polar and azimuthal).

The canonical coordinates $(\tilde{t},\nu,\pi_\nu,\tilde{\ell},{\varpi})$ are ``cyclic'', in the sense that they do not appear explicitly in the canonical Hamiltonian $H_{c}$ 
in Eq.~\eqref{eq:def-Hc}, which is also autonomous. As a result, the system possesses six (linearly independent) first integrals:  
\begin{itemize}\itemsep0em
    \item $\mu^2 = -2H_c$, the mass of the particle,
    \item $E = -\pi_t$, the total energy of the spinning body\,,
    \item $L = \pi_\ell$, the norm of the angular momentum vector $\vec{L}$ \,,
    \item $\nu$, a constant angle defining the line of nodes of the fixed plane orthogonal to $\vec{L}$ \,,
    \item $L_{z} = \pi_\nu$, the $z$-component of the angular momentum vector $\vec{L}$, or, equivalently, $\cos\iota = \pi_{\nu} / \pi_{\ell}$, which is the (cosine of) the inclination angle of that plane\,, 
    \item $S_{\parallel} = \pi_{\varpi}$, the spin component of the small body ``parallel''
   to the Killing-Yano angular momentum vector ${\cal L}^{a}$ in Eq.~\eqref{eq:def-KYL}.   
\end{itemize}
The first five of them carry the same physical interpretations 
as those in the geodesic case, and only $S_{\parallel}$ appears 
as a new integral of motion specific to the spinning case.
In this sense, $H_c$ is a natural generalization of the well-established 
(covariant) geodesic Hamiltonian introduced in, e.g., Refs.~\cite{Schm.02,HiFl.08} 
[cf. Eq.~\eqref{H0} in Appendix.~\ref{app:geo}]. 
It is important to note that only five out of the six integrals of motion are 
in pairwise involution,~\footnote{Two first integrals $\mathcal{I}_1$ and $\mathcal{I}_2$ 
are in involution whenever $\{\mathcal{I}_1,\mathcal{I}_2\}=0$.}
because the canonical pair $(\nu,\,\pi_\nu)$ satisfies the bracket $\{\nu , \pi_\nu\} = 1$ 
by definition.

\subsubsection{Spin configurations} 
\label{sub2sec:cano-spin-config.}

The interpretation of $\pi_{\varpi}$ as the ``parallel'' component of the spin  
is motivated by the fact that it is related to the spin-orbit coupling 
in Eq.~\eqref{eq:KQ-L} 
between the TD spin vector $S_{a}^{\tTD}$ in Eq.~\eqref{eq:def-TD-spins} 
and the Killing-Yano angular momentum vector ${\mathcal {L}}^{a}$ in Eq.~\eqref{def:KY-AM}. 
By definition, we simply write $\pi_{\varpi}$ in Eq.~\eqref{choiceP} as  
\begin{equation}\label{eq:Pz-SO-coupling}
    \pi_{\varpi} 
    = 
    \frac{{\mathcal {L}}^{a} S_{a}^{\tTD}}{\pi_{\ell}} \,,
\end{equation}
which is also related to the R\"{u}diger invariant $\fK$ introduced in Sec.~\ref{subsec:inv}. 
This is a geometrical manifestation of $\pi_{\varpi}$ being the ``parallel'' component 
of the spin to ${\mathcal {L}}^{a}$ when viewed in Schwarzschild spacetime.

In particular, Eq.~\eqref{eq:Pz-SO-coupling} implies that 
the special configurations of the spin: 
(aligned, perpendicular or anti-aligned) with respect to ${\mathcal {L}}^{a}$ 
are also given in terms of $\pi_{\varpi}$ via Eq.~\eqref{eq:spin-config}, namely:
\begin{equation}\label{eq:spin-config-Pz}
    \pi_{\varpi} 
    = 
    \begin{cases}
		+S_{\circ} \quad &\text{[aligned]} \,, \\
        0 \quad &\text{[perpendicular]} \,,\\ 
        -S_{\circ} \quad &\text{[anti-aligned]}  \,, 
    \end{cases}
\end{equation}
We shall attempt more precise identification of the (anti)aligned spin 
in Sec.~\ref{subsec:spin-HP} below.

\subsubsection{Hamilton's canonical equations}

The evolution of any function $F:\mcP\rightarrow\RR$ along a phase space trajectory 
is the flow of the Hamiltonian $H_c$ in Eq.~\eqref{eq:def-Hc} on $\mcP$, 
which is given by 
\beq\label{eq:Hc-flow}
    \frac{\ud F}{\ud \bar {\tau}} = \{F, H_c\}\,
\eeq
from the definition of the Poisson bracket in Eq.~\eqref{eq:PB-c}. 
Setting $F = \tilde{x}^{i}$ or $\tilde{\pi}_{i}$, we simply obtain the evolution 
of the canonical coordinates, namely, Hamilton's canonical equations; 
we shall not display the trivial Hamilton's equations for the constants of motion below.

For the orbital sector $( \tilde{t}, r, \nu, \tilde{\ell},\, \pi_t, \tilde{\pi}_r, \pi_{\nu}, \pi_\ell)$, 
we have 
\begin{subequations} \label{Hameqcano}
    \begin{align}
        \frac{\ud \tilde{t}}{\ud \bar{\tau}} &= 
        -\frac{\pi_t}{f} 
        + \left(1 - \frac{3M}{r}\right) \,\frac{ \left( 
        \left(f \nu_0^{2} -  2 \pi_t^{2} \right) \mu_0^{2} + \pi_t^{2} \,\nu_0^{2} 
        \right) \pi_\ell}
        {\mu_0  \,r^{2} \nu_0^{4} f^{2}} \pi_{\varpi} 
        \,, \\
        \frac{\ud r}{\ud \bar{\tau}} &= 
        f\tilde{\pi}_r - \left(1 - \frac{3M}{r}\right) 
        \frac{\tilde{\pi}_r\pi_t \pi_{\ell}}{r^2\nu_0^2\mu_0} \pi_{\varpi}
        \,, \label{HamEqr}\\
        \frac{\ud \tilde{\pi}_r}{\ud \bar{\tau}} &= 
        -\frac{M \,\pi_t^{2}}{r^2 f^{2}} 
        -\frac{M \,\tilde{\pi}_r^{2}}{r^{2}} 
        + \frac{\pi_\ell^{2}}{r^{3}} + \left(1-\frac{4M}{r}\right) \,
        \frac{\mu_0 \left(2 r - 3 M \right) \pi_t \pi_{\ell}}{r^{4} \nu_0^{2} f^{2}} \pi_{\varpi}
        \,, \label{HamEqPir}\\
        \frac{\ud \tilde{\ell}}{\ud \bar{\tau}} &= 
        \frac{\pi_\ell}{r^{2}}
        - \left(1-\frac{3M}{r}\right)  \frac{ \left(
        \left( {f \nu_0^{2}} - 2 \pi_t^{2} \right)\, \mu_0^{2}
        + \left( \pi_t^{2} - {f \nu_0^{2}} \right) \,\nu_0^{2}
        \right) \pi_t}
        {\mu_0  \,r^{2} \nu_0^{4} f^{2}} \pi_{\varpi}\,. 
    \end{align}
\end{subequations}
For the spin sector $({\varpi},\,\pi_{\varpi})$, we have
\begin{equation}\label{HamEqz}
        \frac{\ud {\varpi}}{\ud \bar{\tau}} = 
        \left(1-\frac{3M}{r}\right)  \frac{\mu_0 \pi_t \pi_{\ell} }{f r^2\nu_0^2} 
        + O(\epsilon)\,,
\end{equation}
which does not involve the term linear in spin. 
We shall see in Sec.~\ref{subsec:spin-HP} that Eq.~\eqref{HamEqz} is closely related 
to the description of the spin precession.

The complicated expressions on the right-hand sides in Eqs.~\eqref{Hameqcano} and~\eqref{HamEqz}
are only apparent. 
The Hamilton's equations just depend on the canonical pair $(r, \tilde{\pi}_r)$, 
and all the other canonical variables are constants of motion, 
which are completely decoupled from the differential equations. 
This is a manifestation of the fact that the Hamiltonian system is \textit{integrable}.
In principle, Eqs.~\eqref{Hameqcano} and~\eqref{HamEqz} are now solvable by quadrature 
(i.e., algebraic manipulations and one-dimensional integrals); 
the closed-form solutions of Eqs.~\eqref{Hameqcano} and~\eqref{HamEqz} 
will be presented in a forthcoming paper.\footnote{See also Refs.~\cite{WiPi.23,Mathew:2024prep} 
for other non-Hamiltonian attempts to obtain the analytical solutions 
of the linear-in-spin, MPTD $+$ TD SSC system in Eqs.~\eqref{TDSSC}-\eqref{EElin}.} 

It should be noted that the explicit spin correction to Eq.~\eqref{Hameqcano}
comes with $\pi_{\varpi}$ only, which is the parallel component of the spin $S_{\parallel}$, 
and another spin related quantity such as the spin norm $S_{\circ}$ 
or ``perpendicular'' component of the spin is not featured here.  
This agrees with the results from the independent analysis 
in Refs.~\cite{WitzHJ.19,DruHug.I.22,DruHug.II.22,WiPi.23}. 

\subsubsection{Inverse transformation: from $\mcP$ to $\mcN$}
\label{sub2sec:P-to-N}

We finally return to our pending task in Sec.~\ref{subsec:Hill-EOMs}: 
the determination of the solution $(s, \pi_s)$ of the Hill differential equation 
in an explicit functional form.
Now that we have the integrable system,
of which the Hamilton's equations are solvable in quadrature, 
this can be simply achieved with the inverse transformation 
from 10D phase space $\mcP$ to the 12D phase space $\mcN$: 
$(\tilde{x}^i,\tilde{\pi}_i) \mapsto y^{\mcN}_A$, augmented with the two TD SSCs in Eq.~\eqref{C0C1simple}; 
we recall the relativistic Andoyer variables $y^{\mcN}_A$ in Eq.~\eqref{eq:def-rel-Andoyer}. 

With the inverse transformation obtained in Eq.~\eqref{inverse}, 
converting $(s, \pi_s)$ to $(\tilde{x}^i,\tilde{\pi}_i)$ (and vice versa) is a simple exercise. 
We find $\tan s = \Sigma^3 / \Sigma^2$ and $\pi_s = \Sigma^1$ 
from the definition in Eq.~\eqref{choiceP}, 
which are written in terms of $y^{F}_{\mcP}$ as 
\begin{equation}\label{eq:sol-Hill-noncano}
    \tan s = \frac{\mu \sqf}{\pi_t \pi_{\varpi}}\, \Sigma^3
    \quand
    \pi_s = - \sqf \Gamma \Sigma^3\,,
\end{equation}
with [cf. Appendix~\ref{app:PB-TDSSCs}]
\begin{equation}\label{eq:sol-Sigma3}
    {\Sigma^3} 
    =
    \frac{r^2\pi_t}{f}
    \frac{S_{\circ}^2 - \pi_{\varpi}^2}
    {\left( r^2 \pi_{t}^2 - f \pi_{\ell}^2 \right)\Gamma^2 
    - 2 r f \pi_{r} \pi_{\ell} \Gamma 
    + r^2 \mu^2 + \pi_{\ell}^2}\,,
\end{equation}
making use of the transformation 
between $({\vec \Sigma},\,{\vec \Delta})$ and $y^{F}_{\mcP}$ in Eq.~\eqref{eq:sol-SigmaDelta}; 
here $S_{\circ}$ is the spin norm, or the Casimir invariant [cf. Eq.~\eqref{CasimirSigDel}].
We then apply the map in Eq.~\eqref{inverse}
to Eq.~\eqref{eq:sol-Hill-noncano} and~\eqref{eq:sol-Sigma3}. 
The transformation for $\Sigma^3$ is obtained by inserting Eq.~\eqref{inverse} 
into Eq.~\eqref{eq:sol-Sigma3}. This yields 
$\Sigma^3 = ({\nu_0} / {\mu}) \sqrt{S_{\circ}^2 - \pi_{\varpi}^2}\,\cos {\varpi}$.  
After some algebra, we find
\begin{equation}\label{eq:sol-Hill-cano}
    \tan s = \frac{\nu_0 \sqf}{\pi_t \pi_{\varpi}} \sqrt{S_{\circ}^2 - \pi_{\varpi}^2}\,\cos {\varpi}
    \quand
    \pi_s 
     = 
     - \frac{ \sqrt{S_{\circ}^2 - \pi_{\varpi}^2} }{\nu_0 \sqf} 
     \left(
        \frac{\tilde{\pi}_r \pi_{\ell} f}{ r \mu} \cos {\varpi}
        +
        {\pi_t} \sin {\varpi}
     \right)\,.
\end{equation}
Therefore, we have the analytical solutions $(s, \pi_s)$ of the Hill equation 
in Eq.~\eqref{eq:ddPsddPsi} free-of-charge, 
once the (analytical) solutions $(\tilde{x}^i,\tilde{\pi}_i)$ are given as input. 
This is another advantage of working with the integrable system.

\subsection{Parameterizing the body's spin: phase space VS spacetime}
\label{subsec:spin-HP}

When viewed in the phase space $\mcP$ (i.e., the symplectic manifold), 
the configuration of the body's spin is completely parameterized 
by the canonical pair $({\varpi}, \pi_{\varpi})$ in Eq.~\eqref{eq:def-yc}. 
At the same time, the body's spin in the Schwarzschild spacetime 
(i.e., the spacetime manifold) is described by the spin tensor $S^{a b}$ 
or equivalently the TD spin 1-form $S_{a}^{\tTD}$ in Eq.~\eqref{eq:def-TD-spins}. 
We now establish the link between these two pictures.


We first examine $S_{t}^{\tTD}$ and $S_{r}^{\tTD}$. 
In this case, we begin with the expressions of the canonical variables $y_{\mcN}\in\mcN$ and of the spin and mass dipole $3$-vectors $(\vec S, \vec D)$ in Eq.~\eqref{eq:def-vecSD}, 
which were introduced in Sec.~\ref{sub2sec:mcN}.  
From Eqs.~\eqref{eq:def-TD-spins} and~\eqref{StoS}, we have 
\beq\label{eq:TD-spins-tr-N}
    S_{t}^\tTD = 
        - f \bar{p}_{r} S^1 - \frac{\bar{p}_{\theta} \sqf }{r} S^2
        - \frac{\bar{p}_{\phi} \sqf}{{r}\, \sin \theta} S^3\quand
        S_{r}^\tTD = \frac{\bar{p}_{\phi} D^2}{r \sqf \sin \theta}
        - \frac{\bar{p}_{\theta} D^3}{r \sqf} - \frac{\bar{p}_{t}}{f} S^1\,, 
\eeq
where $\bar{p}_{\alpha} = p_{\alpha} / \mu$ are the components 
of the $4$-velocity along the worldline $\scL$. 
We next transform the right-hand side of Eq.~\eqref{eq:TD-spins-tr-N} 
from $y_{\mcN}$ to the Andoyer variable $y_{\mcN}^A$ in Eq.~\eqref{eq:def-rel-Andoyer} 
[cf. Sec.~\ref{sec:Ando}]. 
Inserting the identity in Eq.~\eqref{magic} into Eq.~\eqref{eq:TD-spins-tr-N}, 
we find 
\begin{equation}\label{eq:TD-spins-tr-Ad}
        \mu\,S_{t}^\tTD 
        = 
        -{f \pi_r} \Sigma^1 -\frac{\sqrt{f} \pi_{\ell}}{r} \Sigma^3
        \quand
        \mu\, S_{r}^\tTD 
         = 
         \frac{\pi_{\ell}}{r \sqrt{f}} \Delta^2 -  \frac{\pi_t}{f} \Sigma^1\,, 
\end{equation}
where $\pi_{i} = \mu \bar{p}_{i}  + O(\epsilon)$ 
and we used the Andoyer spin variables $({\vec \Sigma},\,{\vec \Delta})$ 
in Eq.~\eqref{spinsympSSC-Ad} to simplify the expressions.
Finally, we write the right-hand side of Eq.~\eqref{eq:TD-spins-tr-Ad} 
in terms of the canonical variables $(\tilde{x}^i,\tilde{\pi}_i)$ 
via the (intermediate) non-canonical variables $y_{\mcP}^C$ in Eq.~\eqref{choiceP}.
The expressions for ${\vec \Sigma}(y_{\mcP}^C)$ and ${\vec \Delta}(y_{\mcP}^C)$ 
are displayed in Eqs.~\eqref{eq:sol-SigmaDelta} and ~\eqref{eq:sol-Sigma3}, respectively, 
and $y_{\mcP}^C(\tilde{x}^i,\tilde{\pi}_i)$ can be found in Eq.~\eqref{inverse}. 
Collecting the results and inserting those within Eq.~\eqref{eq:TD-spins-tr-Ad}, 
simple manipulations reveal a strikingly simple correspondence given by 
\begin{subequations}\label{eq:TD-spin-cano}
    \begin{align}
        S_t^{\tTD} 
        &= \frac{\sqf}{\nu_0} \sqrt{ S_{\circ}^2 - \pi_{\varpi}^2 } 
        \left(
        - \frac{\pi_{\ell}}{r} \cos {\varpi} + \frac{\pi_{t} \tilde{\pi}_{r}}{\mu} \sin {\varpi}
        \right)\,,\\ 
        S^{\tTD}_r
        &= \frac{\nu_{0}}{\mu \sqf} \sqrt{S_{\circ}^2 - \pi_{\varpi}^2} \sin {\varpi}\,
    \end{align}
\end{subequations}
with the spin norm, or the Casimir invariant $S_{\circ}$ [cf. Eq.~\eqref{CasimirSigDel}].
These equations clearly show that $S_{t}^{\tTD}$ and $S_{r}^{\tTD}$  
are proportional to the (square of the) ``perpendicular'' component of the spin 
$S_{\circ}^2 - \pi_{\varpi}^2 = S_{\circ}^2 - S_{\parallel}^2$ 
[cf. Sec.~\ref{sub2sec:cano-spin-config.}],
while the variable ${\varpi}$ is closely related to the spin ``precession phase'' along the orbit ${\scL}$ in Schwarzschild spacetime.

The mapping procedure described in the preceding paragraph
could be recycled to write down $S_{\theta}^{\tTD}$ 
and $S_{\phi}^{\tTD}$ for $(\tilde{x}^i,\tilde{\pi}_i)$
\footnote{There are three first integrals $(L_z,L, {\pi_{\varpi}})$, a Casimir invariant $S_{\circ}$
and an algebraic relation: the TD SSC in Eq.~\eqref{eq:norm-TD-spins} 
for five unknowns ${\bar p}_{\theta},\,{\bar p}_{\phi}$, 
$S_{\theta}^{\tTD},\,S_{\phi}^{\tTD}$ and $\sin \theta$. 
Given that we already have $S_{t}^{\tTD}(\tilde{x}^i,\tilde{\pi}_i)$ 
and $S_{r}^{\tTD}(\tilde{x}^i,\tilde{\pi}_i)$ in Eq.~\eqref{eq:TD-spin-cano}, 
these five equations are solvable to obtain 
$S_{\theta}^{\tTD}(\tilde{x}^i,\tilde{\pi}_i)$ and $S_{\phi}^{\tTD}(\tilde{x}^i,\tilde{\pi}_i)$. 
}.
In this case, however, it is more instructive to express instead 
the constants of motion $\pi_{\varpi}$ in terms of $S_{\alpha}^{\tTD}$. 
From Eqs.~\eqref{eq:def-KYL} and~\eqref{eq:Pz-SO-coupling}, 
we trivially find that 
\begin{equation}\label{eq:TD-spin-Pz}
       \pi_{\varpi} = 
        \frac{\csc \theta}{r L}
        \left( -\pi_{\phi} S_{\theta}^{\tTD} + \pi_{\theta} S_{\phi}^{\tTD} \right)\,.
\end{equation}
In words, the specific combination of $S_{\theta}^{\tTD}$ and $S_{\phi}^{\tTD}$
is directly related with the parallel component of the spin $\pi_{\varpi} = S_{\parallel}$.
This agrees with the finding in Ref.~\cite{WiPi.23}.

\subsection{Action-angle variables}
\label{subsec:AA}

A significant implication of our integrable Hamiltonian system is 
the existence of a set of canonical coordinates, designated action-angle variables, $(\vartheta^i,\mathcal{J}_i)_{i=1\ldots5}$ on $\mcP$. 
This is a consequence of The Liouville–Arnold theorem, as presented 
in Refs.~\cite{Liou1855,Arn} and its subsequent generalisation 
in the case of the non-compact level set, 
as outlined in Refs.~\cite{2003JPhA...36L.101F,HiFl.08}. 
The angles $\vartheta^i$ are cyclic 
(in the sense that the Hamiltonian $H_c$ does not depend on them by definition)  
and the actions $\mathcal{J}_i$ are constants of motion, 
which are in a $1$-to-$1$ correspondence with the set $(H_{c},E,L_z,L,\fK)$.
In particular, action-angle variables offer a practical means 
of obtaining the frequencies of oscillatory (or rotational) motion 
in the system. 

The Liouville-Arnold theorem guarantees the existence 
of action-angle variables $(\vartheta^i, \mathcal{J}_i)$, 
yet the difficulty remains of determining their explicit functional forms. 
Once again, our previously established canonical Hamiltonian system 
in Sec. ~\ref{subsec:can-HP} provides a practical method for achieving this objective.
The key point is that four (out of five) canonical momenta $(\pi_t, \pi_\nu, \pi_\ell, \pi_{\varpi})$ 
in Eq.~\eqref{eq:def-yc} are all constants of motion, 
and the Hamiltonian $H_c$ in Eq.~\eqref{eq:def-Hc} is already reduced 
to the ``radial'' Hamiltonian for $(r,\,\tilde{\pi}_r)$ essentially. 
The overall structure of the Hamiltonian system is therefore identical 
to the well-known one for the geodesic dynamics in Schwarzschild 
[cf. Appendix~\ref{app:geo}], 
except for the addition of the spin canonical pair $({\varpi},\,\pi_{\varpi})$. 
We can immediately obtain the action-angle variables for the spinning-body 
in much the same way as those defined in the case of the non-spinning body 
[cf. the Kerr geodesic results in Refs~\cite{Schm.02,HiFl.08}].

In this section, we obtain the action-angle variables $(\vartheta^i,\mathcal{J}_i)_{i=1\ldots5}$ 
of the physical phase space $\mcP$. 
They can be defined in generality in Schwarzschild, 
but, as a first application, we restrict ourselves to a bound periodic orbit. 
The methods exploited here originates from Schmitt~\cite{Schm.02} and Hinderer and Flanagan~\cite{HiFl.08}, 
implemented for geodesic dynamics in Kerr spacetime. Our approach is a generalization of their construction to a spinning body. 

\subsubsection{Derivation}

To begin we denote the set of five constants of motion as [cf. Sec.~\ref{subsec:can-HP}]
\begin{equation}\label{eq:def-P}
  (P_i)_{i=1\ldots5}
  :=  \left( H_c, E, L_z, L, S_{\parallel} \right) 
  = 
  \left( - \frac{\mu^2}{2}, E, L_z, L,  S_{\parallel} \right)\,,
\end{equation}
which can be inverted to express the canonical momenta $\tilde{\pi}_{i}$ 
as a function of $r$ and $P_{\alpha}$. We find that 
\begin{equation}\label{eq:def-pi-x-P}
    \pi_t = -E \,, \quad 
    \tilde{\pi}_r = V(r;P_\alpha) \,, \quad
    \pi_{\nu} = L_z \,, \quad 
    \pi_{\ell} = L \,, \quad 
    \pi_{\varpi} = S_{\parallel} \,,
\end{equation}
where 
\begin{equation}\label{eq:def-Pr-x-P}
        V(r;P_\alpha)
         := \pm \sqrt{\frac{E^2}{f^2} - \frac{L^2}{f r^2} - \frac{\mu^2}{f} 
        + 
        \frac{r - 3 M}{r - 2 M} \frac{2 \mu E L }{E^2 r^2 - L^2 f} S_{\parallel}}\,.
\end{equation}

Following \cite{Schm.02,HiFl.08}, 
we define the generalized action $\mathcal{J}_i$ by 
\begin{subequations}\label{eq:def-spin-J}
    \begin{align}
    \label{eq:def-JT}
    \mathcal{J}_1 & := \frac{1}{2 \mathpi} \int_0^{2 \mathpi} \pi_t \, \mathd \tilde{t} = - E\,,\\
    \label{eq:def-Jr}
    \mathcal{J}_2 & :=  \frac{1}{2 \mathpi} \oint V(r;P_\alpha)\,  \mathd r 
    = 
    \frac{1}{2 \mathpi} \oint 
    \sqrt{\frac{E^2}{f^2} - \frac{L^2}{f r^2} - \frac{\mu^2}{f} 
        + 
        \frac{r - 3 M}{r - 2 M} \frac{2 \mu E L }{E^2 r^2 - L^2 f} S_{\parallel}}\, \ud r\,,\\
    \mathcal{J}_3 & :=  \frac{1}{2 \mathpi} \oint \pi_{\nu} \, \ud \nu = L_z\,,\\
    \mathcal{J}_4 & :=  \frac{1}{2 \mathpi} \oint \pi_{\ell} \, \ud \tilde{\ell} = L\,, \\
    \mathcal{J}_5 & := \frac{1}{2 \mathpi} \oint \pi_{\varpi} \, \ud {\varpi} = S_{\parallel}\,,
\end{align}
\end{subequations}
where the integral path is given by one periodic orbit in the $(r,\,\tilde{\pi}_r)$ space,
except for $\mathcal{J}_1$ which involves the non-compact ``time'' level set 
and is integrated over the curve of the length $2 \pi$ [cf. the usual prescription described in Refs.~\cite{HiFl.08,Schm.02}].
In practice, it would be convenient to only keep $\mathcal{J}_2$ in Eq.~\eqref{eq:def-Jr}
to first order in the spin expansion; any $O(\epsilon^2)$ terms can be discarded 
in our linear-in-spin treatment. After some simple algebra, we have 
\begin{equation}\label{eq:lin-Jr}
    \mathcal{J}_2 
    = 
    \frac{1}{2 \mathpi} \oint 
        \left(
        \frac{\sqrt{V_{0}}}{f r^2}
        + 
       \frac{r - 3 M}{r - 2 M} \frac{\mu E L}{\nu_0^2 \sqrt{V_0}} S_{\parallel}
        \right) \ud r\,
\end{equation}
with [cf. Eq.~\eqref{eq:def-nu0}]
\begin{equation}\label{eq:def-V0}
    V_{0}(r;P_\alpha) := r^4 E^2 - f r^2 \left( \mu^2 r^2 + L^2 \right)\,.
\end{equation}
It is, however, crucial to distinguish 
between the function $V_0$ and the ``radial geodesic potential,'' 
which is defined, e.g., in Eq. (2.26a) in Ref.~\cite{HiFl.08} 
(after taking the Schwarzschild limit). 
This distinction is important as the former accounts for spin effects implicitly 
through corrections to the constants of motion, $E$, $L$, and $\mu$.

The spin actions $\mathcal{J}_i$ in Eqs.~\eqref{eq:def-spin-J} are a generalization 
of geodesic actions defined in Refs.~\cite{Schm.02,HiFl.08} to the spinning setting, 
through linear order in the body's spin;  
see also Refs.~\cite{WitzHJ.19,WiPi.23,Stein:2024prep,Gonzo:2024zxo,Witzany:2024ttz} for other attempts 
to define action variables for a spinning body.
These spin actions are the functions of constants of motion $P_{\alpha}$ 
in Eq.~\eqref{eq:def-P}. 
Because the two sets $P_{\alpha}$ and $\mathcal{J}_i$ are bijective by definition, 
we immediately find that 
\begin{equation}\label{eq:PvsJ}
    P_{\alpha}  = P_{\alpha} \left( \mathcal{J}_i \right)\,.
\end{equation}

Next we define the conjugate angle variables $\vartheta^i$. 
Among the various definitions, a (type-$2$) canonical transformation 
$(\tilde{x}^i, \tilde{\pi}_i) \mapsto (\vartheta^i,\mathcal{J}_i)$  
is widely used in the classical literature, 
whose generating function is the Hamilton's characteristic function 
${\mathcal {W}} ( \tilde{x}^i ,\, \mathcal{J}_i ) $. 
In our spinning case, we define 
\begin{equation}\label{eq:def-W}
    {\mathcal {W}} ( \tilde{x}^i ,\, \mathcal{J}_i ) 
    :=
    {\mathcal {J}}_{1}  \tilde{t} +  {\mathcal {J}}_{3} \nu + {\mathcal {J}}_{4} \tilde{\ell} + {\mathcal {J}}_{5} {\varpi} + {\mathcal W}_r(r)
    \quad \text{with} \quad
    {\mathcal W}_r(r) := \int^r V(r;\mathcal{J}_i)\, \ud r\,,
\end{equation}
from which we obtain the desired canonical transformation as  
\begin{equation}\label{eq:cano-to-AA}
    \tilde{\pi}_i = \frac{\partial \mathcal {W}}{\partial \tilde{x}^{i}} ( \tilde{x}^i ,\, \mathcal{J}_i )
    \quand
    \vartheta^i 
    = \frac{\partial \mathcal {W}}{\partial {\mathcal {J}_i}} ( \tilde{x}^i ,\, \mathcal{J}_i )\,.
\end{equation}
The first equation is consistent with the definition of ${\mathcal {J}}_i$ 
in Eq.~\eqref{eq:def-spin-J}, while the second equation defines the angle variables.

It is easy to verify that the action-angle variables in the form 
of $\vartheta^j(\tilde{x}^i,\tilde{\pi}_i)$ and ${\mathcal {J}_j}(\tilde{x}^i,\tilde{\pi}_i)$
all satisfy the Poisson brackets in Eq.~\eqref{eq:PB-c}: 
\begin{equation}
\{\vartheta^i,\vartheta^j \} = 0\,,
\quad
\{\mathcal{J}_i,\mathcal{J}_j \} = 0
\quand
\{\vartheta^i,\mathcal{J}_j \} = \delta^{i}_{j}\,,
\end{equation}
where $\delta^{i}_{j}$ is the Kronecker delta.
Therefore, action-angle variables $(\vartheta^i,\mathcal{J}_i)$ defined 
in Eqs.~\eqref{eq:def-spin-J} and~\eqref{eq:cano-to-AA} are indeed 
canonical variables. Their Hamilton's equations are then simply given by  
\begin{equation}\label{eq:Heq-AA}
    \frac{\ud \mathcal{J}_i} {\ud \bar \tau} 
    = 
    -\frac{\partial H_c}{\partial \vartheta^i} 
    = 0
    \quand
   \frac{\ud \vartheta^i}{\ud \bar \tau}
    = \frac{\partial H_c}{\partial \mathcal{J}_i}\,,
\end{equation}
with the canonical Hamiltonian $H_c({\mathcal{J}}_i)$; 
recall Eqs.~\eqref{eq:def-P} and~\eqref{eq:PvsJ}.

\subsubsection{Hamiltonian frequencies}
\label{sub2sec:H-freq}

As a physical application of our spin action-angle variables, 
we here compute the invariant Hamilton's frequencies $(\omega^{i})_{i=1\ldots5}$  
for the spin angle variables $\vartheta^i$ defined by 
\begin{equation}\label{eq:def-H-freq}
    \omega^{i} (\mathcal{J}) 
    := \frac{\partial H_c}{\partial \mathcal{J}_i}\,.
\end{equation}
Physically, they correspond to the oscillation (or rotation) frequencies 
of bound orbits. 
The formalism described here is adapted from Appendix. A 
of Schmidt~\cite{Schm.02}, originally developed for geodesic frequencies in Kerr; 
see also Appendix A of Ref.~\cite{HiFl.08} as well as 
Sec. IV of Ref.~\cite{Tanay:2021bff}. 

The Key observation of this approach is the bijective map in Eq.~\eqref{eq:PvsJ}, 
which returns the simple chain rule given by
\begin{equation}\label{eq:dPdJ-dJdP}
    \frac{\partial P_{i}}{\partial {\mathcal J}_{k}}  
    \frac{\partial {\mathcal J}_{k}}{\partial P_{j}} 
    = 
    \delta^{i}_{j}\,.
\end{equation}
We then take advantage of the fact that the component $P_1$ 
in Eq.~\eqref{eq:def-P} is identical to the canonical Hamiltonian 
$H_c$, and that the $i = 0$ component of Eq.~\eqref{eq:dPdJ-dJdP} 
just becomes 
$\omega^{k}  ( {\partial {\mathcal {J}}_{k}} / {\partial P_{j}} ) 
= \delta^{1}_{j}$.
Inverting this equation, we have  
\begin{equation}\label{eq:H-freq}
    \omega^{1} =   \omega^2  \frac{\partial {\mathcal J}_{2}}{\partial E}\,, \quad
    \omega^{2}  = \left( \frac{\partial {\mathcal J}_{2}}{\partial H_c} \right)^{- 1}\,, \quad
    \omega^{3}  = 0\,, \quad
    \omega^{4}  = - \omega^2  \frac{\partial {\mathcal J}_{2}}{\partial J}\,, \quad
    \omega^{5}  = - \omega^2  \frac{\partial {\mathcal J}_{2}}{\partial S_{\parallel}}\,.
\end{equation}
This can be solved one step at a time, simply from the partial derivatives of
the radial action ${\mathcal {J}}_2$ in Eq.~\eqref{eq:def-Jr}.
After linearization, they are given explicitly by [cf. Eq.~\eqref{eq:def-V0}]
\begin{subequations}\label{eq:dJdP}
    \begin{align}
        \frac{\partial {\mathcal J}_{2}}{\partial E}
        &=  \frac{1}{2 \mathpi} \oint 
        \frac{E r^2}{ f\sqrt{V_0}}
        \left(
        1 
        + 
        \frac{\mu L \left(r - 3 M \right)}{r^3 \nu_0^2 E}
        \left(
        1 - \frac{r^4 E^2}{V_0} -\frac{2 E^2}{f \nu_0^2}
        \right)
        S_{\parallel}
        \right) \ud r\,,\\
        \frac{\partial {\mathcal J}_{2}}{\partial H_c}
        &=  \frac{1}{2 \mathpi} \oint 
        \frac{r^2}{\sqrt{V_{0}}}
         \left(
        1
        -
       \frac{E L r \left( r - 3 M \right)}{\mu V_0} S_{\parallel}
        \right) \ud r\,,\\
        \frac{\partial {\mathcal J}_{2}}{\partial L}
        &= -\frac{1}{2 \mathpi} \oint 
        \frac{L}{\sqrt{V_0}}
        \left(
        1
        +
        \frac{r - 3M}{r - 2 M}\frac{\mu E}{\nu_0^2 L}
        \left(
        1 - \frac{r^4 E^2}{V_0} - \frac{2 E^2}{f \nu_0^2}
        + \frac{r^4 f \nu_0^2}{V_0} 
        \right) S_{\parallel}
        \right) \ud r\,,\\
        \frac{\partial {\mathcal J}_{2}}{\partial S_{\parallel}}
        &= - \frac{1}{2 \mathpi} \oint 
        \left(
        \frac{r - 3M}{r - 2M}\frac{\mu E L}{\nu_0^2 \sqrt{V_0}}
        \right) \ud r\,. \label{eq:dJrdS}
    \end{align}
\end{subequations}
Note that there is no linear-in-$S_{\parallel}$ term in Eq.~\eqref{eq:dJrdS} 
because we know that $S_{\parallel} = O(\epsilon)$.
The integration of Eqs.~\eqref{eq:dJdP} is straightforward in terms of elliptic functions, and will be presented in a forthcoming paper. 
%

\subsection{A demonstration of the formalism: The circular case}
\label{subsec:demo-HP}

As a preliminary illustration of our canonical formulation, 
we examine the model problem of circular orbits, 
which are otherwise general orbital and spin configurations. 
In this case, the spin precesses, as does the orbital plane. 
This will highlight the core ideas of our canonical formulation,
as well as illustrate a practical advantage over the traditional way.

Firstly, we shall set some values of the invariants of motion 
$(\pi_t,\pi_\nu,\pi_\ell,\pi_{\varpi})=(-E,L_z,L,S_{\parallel})$, 
and consider the canonical Hamiltonian in Eq.~\eqref{eq:def-Hc} 
as a one-dimensional one for the radial pair $(r,\tilde{\pi}_r)$ only, written as
\beq \label{Hrad}
        H_{\text{rad}} (r,\tilde{\pi}_r):= 
        -\frac{E^2}{2 f} + \frac{f \tilde{\pi}_r^2}{2} + \frac{L^2}{2 r^2} 
        - \left( 1 - \frac{3 M}{r} \right) 
        \frac{\mu_0 E L S_\parallel}{f r^2\nu_0^2}\,,
\eeq
where the subscript ``rad'' refers to the fact that we only look 
at the ``radial'' $2$D phase space $(r,\tilde{\pi}_r)$.
One should picture the physical phase space $\mcP$ 
through which one selects a $2$D slice 
when fixing $(\pi_t,\pi_\nu,\pi_\ell,\pi_{\varpi})=(-E,L_z,L,S_{\parallel})$. 
The principles of symplectic geometry guarantee the consistency of fixing these values 
before computing Hamilton's equation for the pair $(r, \tilde{\pi}_r)$, 
provided that the fixed values correspond to integrals of motion. 
This is indeed the case in the context at hand.

\subsubsection{Orbital parameters}
\label{sub2sec:orb-param-circ}

The $2$D slice $(r,\tilde{\pi}_r)$ in $\mcP$ is dependent of the choice 
of values $(E,L_z,L,S_{\parallel})$.
The primary difference comes from the existence (or lack thereof) of fixed points, corresponding to critical points of the function $(r,\tilde{\pi}_r)\mapsto H_{\text{rad}}(r,\tilde{\pi}_r)$. 
Upon examination of Hamilton's canonical equations in Eq.~\eqref{Hameqcano}, 
it becomes evident that the fixed points will be located at $(r, \tilde{\pi}_r)=(r_\text{c},0)$, 
where $r_c$ satisfies the algebraic equation resulting 
from equating Eq.~\eqref{HamEqPir} to zero. 
When combined with the on-shell value of the Hamiltonian in Eq.~\eqref{Hrad} at circular orbits, which is $-\mu^2/2 = H_{\text {rad}} (r_\text{c},0)$, 
this yields two algebraic equations in the form 
\begin{subequations} \label{eq1eq2}
    \begin{align}
        & \frac{1}{2}- \frac{\varepsilon^2}{2 f} + \frac{l^2}{2 \rho^2} - \left( 1 -\frac{3}{\rho} \right)  \frac{\varepsilon l  s_{\parallel}}{f \rho^2} =0 \,, \\
        &\frac{\varepsilon^2}{f^2} - \frac{l^2}{\rho} - \left( \frac{3}{\rho} -\left( \frac{2}{\rho f} + \frac{l^2}{\rho^2} - \frac{\varepsilon^2}{\rho f^2} + 2 \right) \left( 1 - \frac{3}{\rho} \right) \right) \frac{\varepsilon l s_{\parallel}}{\rho f} = 0 \,, 
    \end{align}
\end{subequations}
where we introduced the normalization according to  
\beq \label{normalization}
    r_c = M \rho  \,, \quad E = \mu \varepsilon \,, \quad 
    L = \mu M l \,, \quad S_{\parallel} = \mu M s_{\parallel} \,
\eeq
so that $\rho,\,\varepsilon,\,l$ and $s_{\parallel}$ become new dimensionless variables.
It can be seen that Eqs.\eqref{eq1eq2} represent two simultaneous equations 
for the five unknown variables, $(r_c, E, L, S_\parallel,\mu)$, 
despite the fact that the dimensionless variables in Eq.~\eqref{normalization} 
has absorbed any dependence on the particle's mass $\mu$, 
which is in line with the universality of the dipolar approximation. 
In the traditional approach, the system of equations~\eqref{eq1eq2} is solved 
for $E$ and $L$ as a function of the circular radius $r_c$ at linear order 
in the spin parameter $S_\parallel$. 
This solution can be expressed as follows: 
\begin{subequations} \label{sol circ}
    \begin{align}
        \varepsilon_\text{circ}
        &= 
        \frac{\rho - 2}{\sqrt{\rho^2 - 3 \rho}} 
        - \frac{s_{\parallel}}{2}\frac{1}{ \rho\, (\rho - 3)^{3/2}} \,, \\
        l_\text{circ} 
        &= 
        \frac{\rho}{\sqrt{\rho - 3}} 
        + \frac{s_{\parallel}}{2} \frac{(\rho - 2)(2 \rho - 9)}{\sqrt{\rho}\,(\rho - 3)^{3 / 2} }  \,.
    \end{align}
\end{subequations}
These solutions remain valid regardless 
of the spin evolution. 
This is due to the fact that the spin evolution is completely encoded 
in the parallel component of the spin $s_{\parallel}$, 
which is a constant of motion. 

All the circular values of the parameters $\varepsilon,l$ derived above were 
first analytically computed for equatorial orbits and spin-aligned configurations 
in a Schwarzschild spacetime;see, e.g., in Appendix B.3 of \cite{Favata:2010ic}. 
These findings have since been corroborated by other authors 
[cf. Refs.~\cite{Jefremov:2015gza,DruHug.I.22,DruHug.II.22,Mathew:2024prep}.]

\subsubsection{Spin dynamics}
\label{sub2sec:circ-TDspin}

The motion of the spin is easily determined in the case of a circular orbit of radius $r_c$. 
In the circular case, the integration of Eq.~\eqref{HamEqz} for the spin angle ${\varpi}$ 
is straightforward, as the right-hand side becomes a constant. 
Inserting the circular-orbit solutions of $E$ and $L$ in Eq.~\eqref{sol circ} 
into Eq.~\eqref{HamEqz}, we find the simple result (at leading order $O(\epsilon^0)$)
\beq \label{spinanglecirc}
  {\varpi}_\text{circ} 
  = 
  -\frac{1}{\rho^{3/2}}\, \frac{\tau}{M}\,.
\eeq
In particular, applying the relationship between the spin angle ${\varpi}$ 
and the Marck (precession) angle $\psi_{p}$ established in Eq.~\eqref{eq:z-vs-psi_p} 
to the circular orbit demonstrates that the solution in Eq.~\eqref{spinanglecirc} is 
identical to the Marck precession angles $\psi_{p}$ (modulo the offset phase $\phi_{s}$) 
derived in Refs.~\cite{DruHug.I.22,DruHug.II.22,WiPi.23}; 
see, e.g., Eqs.~(4.25) and (4.26) in Ref.~\cite{DruHug.I.22} (in Schwarzschild limit) 
\footnote{
The minus sign on the right-hand side of Eq.~\eqref{spinanglecirc} is a consequence 
of the \textit{right-handed} tetrad in our formalism, 
in contrast to the \textit{left-handed} Marck tetrad that is typically employed 
in the literature [cf. Refs.~\cite{VdM.20,DruHug.I.22,DruHug.II.22,Mathew:2024prep}]. 
A more detailed discussion of this distinction can be found in Appendix~\ref{app:Marck}.}.
This result is a direct consequence of the fact that the angle $\Theta$ 
in Eq.~\eqref{eq:z-vs-psi_p} can be set to zero for circular orbits, 
given that we have $\cos \Theta = 0$ (and $\sin \Theta = 1$) 
when $\tilde{\pi}_r = 0$.

By inserting Eqs.~\eqref{sol circ} and \eqref{spinanglecirc} 
into Eq.~\eqref{eq:TD-spin-cano}, we can also derive expressions 
for the TD spin $4$-vector $S_a^{\tmop{TD}}$ for circular orbits:  
[For simplicity, we only display their $t$ and $r$ components here.]  
\begin{subequations}\label{TDspincirc}
    \begin{align}
         S_t^{\tmop{TD}} &= - \sqrt{\frac{\rho - 2}{\rho (\rho - 3)}} 
        \sqrt{S_{\circ}^2 - S_{\parallel}^2}\, \cos {\varpi} ,\\
        S^{\tmop{TD}}_r &= \sqrt{\frac{\rho - 2}{\rho}}  
        \sqrt{S_{\circ}^2 - S_{\parallel}^2}\,\sin {\varpi} \,.
    \end{align}
\end{subequations}
Once more, these results agree with those presented 
in Refs.~\cite{DruHug.I.22,DruHug.II.22,WiPi.23}  
(after taking into account the left-handedness of the tetrad in these works); 
see, e.g., Eq. (25) in Ref.~\cite{WiPi.23}.

\subsubsection{ISCO parameters}
\label{sub2sec:ISCO-param}

The physical stability of circular orbits in response to an arbitrary spacetime perturbation 
is directly related to the spectral stability of the corresponding fixed point in phase space,
which is determined by the two eigenvalues of the linearized system around it. 
These eigenvalues depend on the parameters $(-E,L,L_z,S_{\parallel})$ of the slice, 
but they are always opposite and purely imaginary for circular orbits,  
which correspond to elliptic fixed point. 
For specific values of the parameters, however, they can be shown to vanish identically. 
This corresponds to the two circular orbits (stable and unstable) colliding 
as we traverse the two-dimensional slices of constant parameters 
$\pi_t$, $\pi_\ell$, $\pi_\nu$, and $\pi_{\varpi}$. 
The resulting orbit is neither stable nor unstable; 
it is semi-stable and corresponds to a parabolic fixed point.

For $2$D canonical systems, the eigenvalues in question are easily seen 
to be those of the matrix $\mathbb{J}_2 {\mathcal{H}}$, 
where $\mathbb{J}_2$ is the canonical $2\times 2$ symplectic matrix 
and $\mathcal{H}$ is the Hessian of the Hamiltonian. 
This is the matrix that encodes the linearized dynamics around a given point in phase space. 
In order to demonstrate this concept within the context of our problem, 
we return to our one-dimensional Hamiltonian $H_{\text {rad}}$ 
in Eq.~\eqref{Hrad}, for which we compute its Hessian and evaluate it at a circular orbit, 
namely $(r,\tilde{\pi}_r)=(M\rho, 0)$, utilizing also Eq.~\eqref{sol circ}. 
The result is
\begin{subequations}
    \begin{align}
    \frac{\partial^2 H_{\text {rad}}}{\partial r^2} 
    &= 
    \frac{\mu^2}{M^2} \frac{\left(18 - 9\rho + \rho ^2\right) \rho ^{3/2}-\left(18-3 \rho -2 \rho^2\right) s_\parallel}{(\rho - 3)^2 (\rho - 2) \rho ^{7/2}} \,, \\
    \frac{\partial^2 H_{\text {rad}}}{\partial \tilde{\pi}_r^2} 
    &= 
    \frac{(\rho - 2 ) ( \rho^{3/2} + s_\parallel)}{\rho^{5/2}} \,,
\end{align}
\end{subequations}
and the mixed second partial derivative vanishes identically. 
The Hessian $\mathcal{H}$ is thus diagonal at circular orbits, 
and the eigenvalues of the matrix $\mathbb{J}_2 \mathcal{H}$ vanish 
if and only if either of the two partial derivatives in Eq.~\eqref{Hrad} vanish. 
Clearly, this result can only be derived from the first equation 
(because, in any case, $\rho > 2$ outside the Schwarzschild event horizon), 
which, upon solving for $\rho$, yields the radius of the semi-stable circular orbit 
given by 
\begin{equation}
    \rho_{\text{isco}} = 6 - 2 \sqrt{\frac{2}{3}} s_\parallel \,, 
\end{equation}
such that 
\begin{equation}
    \frac{\partial^2 H_{\text {rad}}}{\partial r^2} = 0 
    \quand 
    \frac{\partial^2 H_{\text {rad}}}{\partial \tilde{\pi}_r^2} 
    = 
    \frac{2}{3} - \frac{s_\parallel}{9 \sqrt{6}} \,.
\end{equation}
These formulae are the ones commonly found in the literature 
to define the so-called innermost stable circular orbit (ISCO), whence the notation. 
If we insert the value $\rho_\text{isco}$ 
into the circular-orbit parameters in Eqs.~\eqref{sol circ}, 
which are valid for any circular orbit, we obtain 
\beq
    \varepsilon_\text{isco}= \frac{2 \sqrt{2}}{3}- \frac{s_\parallel}{36 \sqrt{3}}  
    \quand 
    l_\text{isco}= 2 \sqrt{3} +\frac{\sqrt{2}}{3} s_\parallel\,.
\eeq

 We have verified that our ISCO formulae obtained in this section are all in perfect agreement 
 with the results presented in Eqs.~(B17) through (B20b) in Ref.~\cite{Favata:2010ic}
 \footnote{There is, however, a typo in Eq.~(B.18) of Ref.~\cite{Favata:2010ic}, where the denominator of the second term in the right-hand side should read $r-3m_2$ instead of $r-2m_2$.}; given that their small parameter (denoted $s$ there) is our $-K/\pi_\phi$. See also Refs.~\cite{Tanaka:1996ht,Suzuki:1996gm,Suzuki:1997by,Saijo:1998mn} 
 for earlier examinations of the ISCO of a spinning body in Kerr spacetime.

\acknowledgments
We thank David Brown, Lisa Drummond, Scott Hughes, Alexandre Le Tiec, 
Jacques F\'ejoz, Josh Mathews, Leo Stein, Sashwat Tanay and Vojt\v{e}ch Witzany 
for useful discussions. 
S.I. is also grateful to Leor Barack, Alvin J. K. Chua, Ryuichi Fujita, 
Hiroyuki Nakano and Norichika Sago for their continuous encouragement.
S.I. is supported by the Ministry of Education, Singapore,
under the Academic Research Fund Tier 1 A-8001492-00-00 (FY2023).
%

\clearpage


\appendix

\section{The non-spinning (geodesic) Hamiltonian} 
\label{app:geo}
This Appendix presents an analysis of the geodesic Hamiltonian 
for the Schwarzschild background. 
The results presented here are fairly standard, 
but our derivation of these results may offer new insights. 

We consider the Schwarzschild spacetime covered 
with the Schwarzschild-Droste coordinates $x^\alpha=(t,r,\theta,\phi)$.\footnote{Technically, we are only interested 
in the Schwarzschild-Droste region (i.e. outside the event horizon) of the Schwarzschild spacetime, where $r>2M$, denoted $\mathcal{M}_\text{I}$ in \cite{GourgoulhonBH}.}
The equations of motion, whose physical solutions describe timelike geodesics of the Schwarzschild spacetime, are equivalent to Hamilton's equation for the Hamiltonian 
$H_{\mcG}=\tfrac{1}{2}g^{\alpha\beta}(x) p_\alpha p_\beta$, 
where $x^{\alpha}$ denote the Schwarzschild-Droste coordinates 
and $p_{\alpha}$ are the component of the four-momentum $1$-form. 
With the contravariant metric coefficients $g^{\alpha\beta}(x)$ in Eq.~\eqref{metrics}, 
we can write the geodesic Hamiltonian $H_{\mcG}$ 
in terms of the symplectic coordinates $(x^\alpha, p_\alpha)$ [recall $f := 1 - 2M / r$],
\beq \label{H0}
H_{\mcG} = -\frac{p_t^2}{2f} + \frac{f p_r^2}{2} + \frac{1}{2r^2} \left( p_\theta^2 + \frac{p_\phi^2}{\sin^2\theta} \right) \,.
\eeq
Along any solution, the Hamiltonian $H_{\mcG}$ evaluates to the constant $-\mu^2/2$, and the ``time'' parameter conjugated to $H_{\mcG}$ is the proper time per unit mass $\bar{\tau}=\tau/\mu$, a dimensionless affine parameter. 

In the following sections, we first perform a reduction of the problem 
through the application of the invariants of motion. 
We then proceed to analytically solve the dynamics of the system, 
with particular focus on the properties 
that will persist in our description of a spinning body.

\subsection{Schwarzschild spherical symmetry} 
\label{app:redgeo}

Let us think for a moment in Euclidean terms and consider a Cartesian orthonormal triad $(\vec{e}_x, \vec{e}_y,\vec{e}_z)$ associated with the usual spherical coordinates $(r,\theta,\phi)$. A given orbital plane intersects the equatorial plane $\text{Span} (\vec{e}_x, \vec{e}_y)$ (or $\theta = \pi / 2$ in spherical coordinates), thereby defining a particular line, called the \textit{line of nodes}. 
Let that line be defined by the relation $\phi = \nu$ for some fixed angle $\nu$ (this actually defines half that line, the other half being $\phi = \nu + \pi$). 
The line of nodes defines a new axis $\vec{f}_x$, which can be rotated counterclockwise within the equatorial plane to construct another one, say $\vec{f}_y$. 
Setting $\vec{f}_z = \vec{e}_z$, we get a new orthonormal triad $(\vec{f}_x, \vec{f}_y, \vec{f}_z)$, explicitly given by,  
\beq
  \vec{f}_x =  \cos \nu \, \vec{e}_x + \sin \nu \,\vec{e}_y \,, \quad 
  \vec{f}_y =  - \sin \nu \, \vec{e}_x + \cos \nu \,\vec{e}_y \,, \quad 
  \vec{f}_z =  \vec{e}_z \,.
\eeq
Next we perform another rotation, this time leaving the direction $\vec{f}_x$
unchanged and inclining the (originally) equatorial plane 
by an angle $\iota \in [0, \pi]$. 
This sends $\vec{f}_y$ and $\vec{f}_z$ to
$\vec{g}_y = \cos \iota \vec{f}_y + \sin \iota \vec{f}_z$ and $\vec{g}_z = -
\sin \iota \vec{f}_y + \cos \iota \vec{f}_z$, respectively. Finally, setting $\vec{g}_x =
\vec{f}_x$, and expressing everything 
in terms of the original triad $(\vec{e}_x, \vec{e}_y,\vec{e}_z)$, 
we arrive at 
\begin{subequations}
\begin{align}
  \vec{g}_x & =  \cos \nu \,\vec{e}_x + \sin \nu \,\vec{e}_y\,,\\
  \vec{g}_y & = - \cos \iota \sin \nu \,\vec{e}_x + \cos \iota \cos \nu\,
  \vec{e}_y + \sin \iota \,\vec{e}_z \,, \\
  \vec{g}_z & =  \sin \iota \sin \nu \,\vec{e}_x - \sin \iota \cos \nu\,
  \vec{e}_y + \cos \iota \,\vec{e}_z \,.
\end{align}
\end{subequations}
Now let a unit position vector be defined as usual as 
$\vec{n} = \cos \phi \sin \theta \vec{e}_x +
\sin \phi \sin \theta \,\vec{e}_y + \cos \theta \,\vec{e}_z$. 
Using some trigonometric identities, in the new triad, the vector $\vec{n}$ reads
%
\begin{multline}\label{n}
  \vec{n} =  \cos (\phi - \nu) \sin \theta \,\vec{g}_x 
   +  (\cos \theta \sin \iota + \sin (\phi - \nu) \sin \theta \cos
  \iota)  \,\vec{g}_y\\
   +  (\cos \theta \cos \iota - \sin (\phi - \nu) \sin \theta \sin
  \iota)  \,\vec{g}_z \,.
\end{multline}
%
By construction, the invariant orbital plane is now $\text{Span} (\vec{g}_x, \vec{g}_y)$
and the Schwarzschild geodesic motion is confined within it. Therefore, one must have $\vec{n} \cdot \vec{g}_z =
0$, leading to the condition
\begin{equation} \label{noz}
\sin (\phi - \nu) \sin \theta \sin \iota = \cos \theta \cos \iota\,.
\end{equation}
Since the angles $\iota$ and $\nu$ are both constant, 
this gives an important algebraic relation between $\theta$ and $\phi$, 
which is the source of the $(\theta-\phi)$-resonance mentioned earlier. 
In the orbital plane $\text{Span}(\vec{g}_x,\vec{g}_y)$, 
let us define an angle $\ell$, called the \textit{true anomaly}, 
such that the position vector is $\vec{n} = \cos \ell\vec{g}_x + \sin \ell \vec{g}_y$. 
Then, combining this definition as well as Eqs.~\eqref{n} and \eqref{noz}, we obtain the following three relations
\begin{subequations}\label{newanglesapp}
    \begin{align}
    \cos \theta & = \sin \iota \sin \ell \,, \\
     \sin \theta \cos (\nu - \phi) & = \cos \ell \,, \\
     \sin \theta \sin (\nu - \phi) & = -\sin \ell \cos \iota \,.
    \end{align}    
\end{subequations} 

These equations define the new coordinate $\ell$ (true anomaly) 
in terms of the old ones $\theta$ and $\phi$ (polar and azimuthal angles), 
and two fixed parameters $\iota$ and $\Omega$ 
(orbital inclination and an angle defining the line of nodes).
Along with these definitions for the new angles $\ell$ and $\nu$, 
we define the usual quantities
\beq \label{newmomapp}
    p_\ell^2 := p_\theta^2+p_\phi^2\csc^2\theta \quand p_\nu := p_\phi \,,
\eeq
in which $p_\ell$ is simply the norm of the angular-momentum.
Together, formulae Eqs.~\eqref{newanglesapp} and \eqref{newmomapp} 
define a canonical change of coordinates
\begin{equation}
    (t,p_t,r,p_r,\theta,p_\theta,\phi,p_\phi) 
    \mapsto 
    (t,p_t,r,p_r,\ell,p_\ell,\nu,p_\nu) \,.
\end{equation}

\subsection{The orbital plane as a $1:1$ resonance} 
\label{app:reso}

The squared norm of the angular momentum
\beq \label{J2}
    L^2 = p_{\theta}^2 + L_z^2 \csc^2 \theta\,,
\eeq
which is a geodesic integral of motion, places a stringent constraint 
on the $(\theta, p_\theta)$-sector of dynamics. 
In a plane where one plots the variables $p_{\theta}$ VS $L_z\csc \theta$, 
this curve depicts a circle of radius $L$. 
Therefore, $\theta$ varies between $\arcsin (L_z / L)$ and $\pi -\arcsin (L_z / L)$. 
This curve is modified slightly in the plane $p_\theta$ VS $\theta$, 
with the circle mapped into a squircle curve $\mathcal{C}=\mathcal{C} (L_z,L)$. 

The Poincar{\'e} invariant (i.e., an action) $J_{\theta}$ associated 
with the $(\theta,p_\theta)$ motion is defined by the area enclosed 
by this curve $\mathcal{C}$ :
\beq
    J_{\theta} 
    := \frac{1}{2 {\pi}} \oint_{\mathcal{C}}\, p_{\theta} \ud \theta 
    = \frac{| L |}{{\pi}} 
    \int_{\arcsin (L_z/L)}^{{\pi} - \arcsin(L_z/L)} 
    \sqrt{1 - \frac{(L_z/L)^2}{\sin^2 
   \theta}}\, \ud \theta \,,
\eeq
where $L_z/L\in[-1;1]$. Using Eq.~\eqref{J2} and the horizontal symmetry of the curve $\mathcal{C}$, we can write 
(we assume that $L_z/L \geq 0$; the other case $L_z/L\leq 0$ leads to $L + L_z$ instead.)
\beq
J_{\theta} = \frac{2 L}{{\pi}} \int_{\arcsin (L_z/L)}^{{\pi} / 2}
   \sqrt{1 - \frac{(L_z/L)^2}{\sin^2 \theta}} \ud \theta = L-L_z
\eeq
Therefore, in all cases, we have 
\begin{equation}
  J_{\theta} =  
  \sqrt{p_{\theta}^2 + \frac{p_{\phi}^2}{\sin^2 \theta}} - | p_{\phi} | 
  \quand 
  J_{\phi} =  p_{\phi} \,.
\end{equation}

Now, let us consider the radial action, defined as the integral 
of the radial momentum $p_r$ over the loop $\mathcal{C}_r$ 
in the plane $(r,p_r)$, $J_r:=\tfrac{1}{2\pi}\oint p_r\ud r$. 
In practice, this quantity can be integrated by inverting the Hamiltonian 
in favour of the radial momentum, $p_r$, namely, [where we use $E = -p_t$].
\beq
    H_{\mcG} = 
    - \frac{E^2}{2 f} + \frac{f p_r^2}{2} + \frac{L^2}{2r^2} 
    \quad \Rightarrow \quad 
    p_r = \pm \sqrt{\frac{E^2}{f^2} + \frac{2 H_{\mcG}}{f} - \frac{L^2}{f r^2}}\,, 
\eeq
and we obtain $J_r$ as a function of the constants $H_{\mcG},E^2$ and $L^2$. 
In particular, the energy $E$ and the norm of the angular momentum in Eq.~\eqref{J2} can be 
expressed as $E = - J_t$ and $L^2 = (J_\theta + |J_\phi|)^2$ in terms of actions, respectively. 
Consequently, the radial action can be expressed as a function of only other actions. 
Schematically, we have that
\beq\label{Jr-G}
    J_r = \mathcal{F} (H_{\mcG}, J_t, J_{\theta} + | J_{\phi} |) \,,
\eeq
where the function $\mathcal{F}$ would involve some kind of elliptic integrals 
and inverse trigonometric functions. 
By solving Eq.~\eqref{Jr-G} for the geodesic Hamiltonian $H_{\mcG}$, 
we can formally express it in terms of actions of the form
\beq
H_{\mcG} (J_t, J_r, J_{\theta}, J_{\phi}) = \mathcal{G} (J_t, J_r, J_{\theta} + |
   J_{\phi} |) \,,
\eeq
for a function $\mathcal{G}$.

What is crucial here is that $H_{\mcG}$ depends on the actions $J_{\theta}, J_{\phi}$
only through the combination $J_{\theta} + | J_{\phi} |$. 
Therefore, from the Hamiltonian frequencies defined by $\Omega_{\ui} =
\partial H_{\mcG} / \partial J_{\ui}$, we obtain 
\begin{equation}\label{eq:res-G-theph}
    \Omega_{\theta} 
    = 
    \left\{
    \begin{array}{lll}
     \Omega_{\phi} & \text{if} & J_{\phi} \leq 0\,, \\
     - \Omega_{\phi} & \text{if} & J_{\phi} \geq 0\,.
    \end{array} \right. 
\end{equation}
In either case, the motion is \textit{resonant};  
there exists a non-trivial linear combination of the frequencies 
$\Omega_{\theta}$ and $\Omega_{\phi}$ with integer coefficients that vanishes. 
The plus (minus) sign in the right-hand side of Eq.~\eqref{eq:res-G-theph} 
corresponds to prograde (retrograde) orbits. 
This resonance is reminiscent of the existence of an orbital plane.

\section{TD spin vectors and parallel transport}
\label{app:TD-spin-vec}

The dynamical degrees of freedom associated with the small body's spin can be described 
at the level of the TD spin $S_{\alpha}^{\tTD}$ in Eq.~\eqref{eq:def-TD-spins}. 
It follow from Eqs.~\eqref{eq:SD-decomposition},~\eqref{eq:norm-TD-spins} and~\eqref{EElin} 
that $S_{\alpha}^{\tTD}$ satisfies 
\begin{equation}\label{eq:TDspin-p-tansport}
    (\nabla_u S_{\tTD})^a = 0\, 
    \quad \text{with} \quad
    p^{\alpha} S_{\alpha}^{\tTD} = 0\,.
\end{equation}
This equation implies that the TD spin is purely spatial in a freely-falling frame 
that moves along the orbit $\scL$ with four momenta $p_{\alpha}$, 
and that $S_{\alpha}^{\tTD}$ is parallel transported 
along $\scL$ as well (in the liner-in-spin system).  
As such, we can write the TD spin in terms of an orthonormal tetrad 
with the constant spin amplitude, 
by demanding that the tetrad is also parallel transported along $\scL$.

The methods summarized above originate from Marck~\cite{Marck.83}, 
are widely implemented in literature. 
Details in the case of Kerr spacetime are provided 
in Refs.~\cite{Marck.83,1986JMP....27.1589K,VdM.20,MaPoWa.22}, 
and we review the main points of Marck's method here; 
see also Sec. III.D of Ref.~\cite{DruHug.I.22}.
In this appendix, we work only with the geodesic orbit $\scL_0$ in Schwarzschild spacetime,  
where spin effects to the orbital dynamics are negligible in Eq.~\eqref{eq:TDspin-p-tansport},  
and omit the ``g'' label to unclutter the notation
%
\footnote{The difference between the true orbit $\scL$ and $\scL_0$ is negligible here 
because $S_{a}^{\tTD}$ is already a quantity of $O(\epsilon)$.}.

\subsection{The Marck tetrad}
\label{app:Marck}

A way to obtain the orthonormal tetrad is to make use of the solution 
to the geodesic orbit $\scL_{0}$ 
%
Their explicit construction and resulting expressions in Kerr spacetime are displayed 
in, e.g., Sec. 3 of Ref.~\cite{Marck.83} and Eqs.(47) - (51) in Ref.~\cite{VdM.20}.
We specialize these results to Schwarzschild spacetime 
with the Schwarzschild-Droste coordinate $x^\alpha = (t,\,r,\,\theta,\,\phi)$ 
and four constants of motion $(\mu, E, L_z, L)$ of $\scL_{0}$ [cf. Appendix~\ref{app:geo}].
It is helpful to note that the Marck tetrad is distinct 
from the natural-basis tetrad introduced in Eq.~\eqref{tetrads} 
and utilized in our Hamiltonian formulation. 

Following the tradition in the literature, we use the assignment 
[it is understood that $(r,\,\theta)$ = $(r({\bar \tau}),\,\theta({\bar \tau}))$ below]
\begin{equation}\label{eq:def-Marck-tetrad-03}
({e}_{0})_{\alpha}
:= \frac{p_{\alpha}}{\mu} \,
\quand
({e}_{3})_{\alpha}
:=
\frac{{\mathcal {L}}_{\alpha}}{\sqrt{\fQ}}
=
\left( 
0,\, 0,\, -\frac{r}{L \sin \theta} p_{\phi},\, 
\frac{r \sin \theta }{L} p_{\theta}\right)\,,
\end{equation}
where $p_{a} = p_{a}(\bar \tau)$ are the four momenta along $\scL_{0}$, 
while ${\fQ} = L^2 $ and ${\mathcal L}_{\alpha}$ are the R\"{u}diger invariant in Eq.~\eqref{RudsPS} 
and the Killing-Yano angular momentum vector in Eq.~\eqref{eq:def-KYL}, respectively, 
in the geodesic limit $S_{\circ} \to 0$. 
The remaining legs of the tetrad, which are orthogonal to $({e}_{0})_{\alpha}$ 
and $({e}_{3})_{\alpha}$, are 
\begin{subequations}\label{eq:def-Marck-tetrad-12}
\begin{align}
(\tilde{e}_{1})_{\alpha}
&=
\left( -\frac{r f}{\sqrt{L^2 + r^2 \mu^2}} p_{r},\, 
 \frac{rE}{f \sqrt{L^2 + r^2 \mu^2}},\, 
0,\, 0 \right)\,,\\
({\tilde e}_{2})_{\alpha}
&=
\left( \frac{L E}{\mu \sqrt{L^2 + r^2 \mu^2}},\, 
-\frac{L}{\mu \sqrt{L^2 + r^2 \mu^2}} p_{r},\, 
-\frac{\sqrt{L^2 + r^2 \mu^2}}{\mu L} p_{\theta},\, 
-\frac{\sqrt{L^2 + r^2 \mu^2}}{\mu L} p_{\phi} \right)\,,
\end{align}
\end{subequations}
where we understood $f = 1 -  2 M  / r({\bar \tau})$ and $E = -p_t$.

While the tetrad $[({e}_{0})_{\alpha}, 
({\tilde e}_{1})_{\alpha},  
({\tilde e}_{2})_{\alpha},
({e}_{3})_{\alpha}]$ in Eqs.~\eqref{eq:def-Marck-tetrad-03} and~\eqref{eq:def-Marck-tetrad-12} 
is orthonormal, it is incomplete to describe the TD spin 
because only two legs $({e}_{0})_{\alpha}$ and $({e}_{3})_{\alpha}$ 
are parallel transported along $\scL_{0}$.
The other two legs $({\tilde e}_{1})_{\alpha}$ 
and $({\tilde e}_{2})_{\alpha}$ must be rotated 
by the precession phase $\psi_{p} = \psi_{p}({\bar \tau})$, 
so that the resulting legs are parallel transported along $\scL_{0}$. 

This idea to obtain the other parallel transported tetrad from 
$({\tilde e}_{1})_{\alpha}$ and $({\tilde e}_{2})_{\alpha}$
are initially developed by Marck~\cite{Marck.83}.
For historical reason, however, much of the recent literature, 
e.g., Refs.~\cite{WitzHJ.19,VdM.20,MaPoWa.22,DruHug.I.22,DruHug.II.22,WiPi.23} 
implemented Marck's method for the \textit{left-handed} tetrad 
(in the standard sense of the text book, e.g., Ref.~\cite{Wald}), 
``oppsite'' to the right-handed tetrad that we used in the bulk of this paper. 
In fact, Eqs.~\eqref{eq:def-Marck-tetrad-03} and~\eqref{eq:def-Marck-tetrad-12} 
are evidently left handed, satisfying $\varepsilon_{a b c d}
({e}_{0})^{a} 
({\tilde e}_{1})^{b}  
({\tilde e}_{2})^{c}
({e}_{3})^{d}
= -1$, which may bring confusion.\footnote{We thank Scott A Hughes for very helpful discussion about
the right- or left-handed representation of the Marck tetrad.}

For that reasons, we implement Marck's method for the \textit{right-handed} tetrad  
$({\epsilon}_{A})_{\alpha}$ defined in terms of 
Eqs.~\eqref{eq:def-Marck-tetrad-03} and~\eqref{eq:def-Marck-tetrad-12} by 
\begin{equation}\label{eq:def-Marck-tetradR-03}
    ({\epsilon}_{0})^a := ({e}_{0})^{a}\,,
    \quad
    ({\epsilon}_{3})^a := ({e}_{3})^{a}\,,
\end{equation}
and 
\begin{equation}\label{eq:def-Marck-tetradR-12}
\begin{bmatrix}
   ({\epsilon}_{1})_{\alpha} \\ ({\epsilon}_{2})_{\alpha}
\end{bmatrix}
    :=
\begin{bmatrix}
     \cos \psi_{p} & \sin \psi_{p} \\
    -\sin \psi_{p} & \cos \psi_{p} \\
\end{bmatrix}
\begin{bmatrix}
   ({\tilde e}_{2})_{\alpha}  \\  ({\tilde e}_{1})_{\alpha}
\end{bmatrix}\,,
\end{equation}
with the precession phase $\psi_{p}$. 
This tetrad is orthonormal, it is right-handed in the sense that $\varepsilon_{abcd}({\epsilon}_{0})^{a} 
({\epsilon}_{1})^{b}  
({\epsilon}_{2})^{c}
({\epsilon}_{3})^{d}
= +1$, and it is parallel transported along $\scL_0$ 
only if $\psi_{p}$ satisfies the differential equation 
\begin{equation}\label{eq:ODE-Mark-angleR}
    \frac{\ud \psi_{p}}{\ud \bar{\tau}}
    =
    - \frac{\mu E L} {L^2 + r^2 \mu^2}\,. 
\end{equation}
We will refer to the tetrad $({\epsilon}_{A})_{\alpha}$ 
in Eqs.~\eqref{eq:def-Marck-tetradR-03} and~\eqref{eq:def-Marck-tetradR-12} 
as the \textit{right-handed} Marck tetrad. 
In particular, it should be noted the overall minus sign in Eq.~\eqref{eq:ODE-Mark-angleR}, 
compared with the same differential equations for $\psi_{p}$ 
in the left-handed presentation 
[cf. Eq.(24) of Ref.~\cite{WiPi.23}]; 
the left-handed Mark tetrad $({\epsilon}_{A})_{\alpha}^{\rm{L}}$ 
used in literature is simply obtained from our right handed representation by 
$({\epsilon}_{0})_{\alpha}^{\rm{L}} = ({\epsilon}_{1})_{\alpha}$
and 
$({\epsilon}_{3})_{\alpha}^{\rm{L}} = ({\epsilon}_{3})_{\alpha}$ 
as well as 
$({\epsilon}_{1})_{\alpha}^{\rm{L}} = ({\epsilon}_{2})_{\alpha}$
and 
$({\epsilon}_{2})_{\alpha}^{\rm{L}} = ({\epsilon}_{1})_{\alpha}$,
but the precession phase in Eq.~\eqref{eq:def-Marck-tetradR-12} 
must be replaced with the ``reversed'' one $-\psi_p$.  
As a result, the differential equation for $\psi_p$ in the left-handed representation 
also differs by a minus sign from Eq.~\eqref{eq:ODE-Mark-angleR}.

\subsection{Parallel and perpendicular components}
\label{app:TD-spin-para}

We now use the (right-handed) Marck tetrad $({\epsilon}_{A})_{\alpha}$ 
to decompose the TD spin evaluated along the orbit $\scL_{0}$. We have
\begin{equation}\label{eq:TD-spin-decomposition-Marck}
S_{\alpha}^{\tTD} = {\mathcal S}^{A} ({\epsilon}_{A})_{\alpha}\,,
\end{equation}
where ${\mathcal S}^{A}$ is the constant spin amplitude 
associated with each leg of Marck tetrad, 
and it has the dimension of spins $[\length]^{+2}$. 
Because the $({\epsilon}_{A})_{\alpha}$ are mutually orthogonal, 
Eq.~\eqref{eq:TDspin-p-tansport} implies that we can set ${\mathcal S}^{0} = 0$ 
without loss of generality. 
In addition, we observe that $({\epsilon}_{3})_{\alpha}$ is proportional 
to the Killing-Yano angular momentum vector ${\mathcal L}_{\alpha}$ in Eq.~\eqref{eq:def-KYL}. 
This motivates to define the ``parallel'' and ``perpendicular'' components 
of $S_{\alpha}^{\tTD}$ to ${\mathcal L}_{\alpha}$ by (see, e.g., Ref.~\cite{DruHug.I.22}) 
\begin{equation}\label{eq:def-S-paralell-perp}
    \begin{pmatrix}
        {\mathcal S}^{1} \\ {\mathcal S}^{2} \\ {\mathcal S}^{3}
    \end{pmatrix}
    :=
    \begin{pmatrix}
        {\mathcal S}_{\perp} \cos \phi_{s} \\ 
        {\mathcal S}_{\perp} \sin \phi_{s}  \\ 
        {\mathcal S}_{\parallel}
    \end{pmatrix}
\end{equation}
where the constant-angle $\phi_{s}$ defines the orientation of $S_{\alpha}^{\tTD}$. 
Here, the spin magnitude $S_{\circ}^2$ is decomposed into the components 
parallel and normal to ${{\mathcal L}_{\alpha}}$ 
with ${\mathcal S}_{\parallel}$ and ${\mathcal S}_{\perp}$, respectively, 
so that ${\mathcal S}_{\parallel}^2 + {\mathcal S}_{\perp}^2 = {\mathcal S}_{\circ}^2$.
Inserting Eqs.~\eqref{eq:def-Marck-tetradR-03},~\eqref{eq:def-Marck-tetradR-12} 
and~\eqref{eq:def-S-paralell-perp} into Eq.~\eqref{eq:TD-spin-decomposition-Marck}, 
we find that the TD spin in terms of the Marck tetrad $({\epsilon}_{A})_{\alpha}$ 
can be expressed as 
\begin{equation}\label{eq:TD-spin-MarckR}
    S_{\alpha}^{\tTD} 
    =
    {\mathcal S}_{\perp} \left(
    \cos \phi_{s}  ({\epsilon}_{1})_{\alpha} + \sin \phi_{s}  ({\epsilon}_{2})_{\alpha}
    \right )
    + 
    {\mathcal S}_{\parallel} ({\epsilon}_{3})_{\alpha}\,.
\end{equation}
We note that the spin aligned and perpendicular cases are simply defined 
by setting ${\mathcal S}_{\perp} = 0$ and ${\mathcal S}_{\parallel} = 0$, respectively, 
in Eq.~\eqref{eq:TD-spin-MarckR}.

\subsection{Comparison with the canonical representation}
\label{app:Marck-TD-spin}



The TD spins $S_{a}^{\tTD}$, as expressed in Eq.~\eqref{eq:TD-spin-MarckR} in terms of the Marck tetrad, can be directly compared with an alternative expression 
for $S_{a}^{\tTD}(\tilde{x}^i,\tilde{\pi}_i)$
in terms of the canonical coordinates 
$(\tilde{x}^i,\tilde{\pi}_i)$ in Eq.~\eqref{eq:def-yc} in the phase space $\mcP$. 
This allows us to gain more insight into the physical interpretation 
of the spin canonical variables $({\varpi},\,\pi_{\varpi})$ [cf. Eqs.~\eqref{eq:def-yc}]
in the context of Schwarzschild spacetime. 


The relevant Marck's expressions to the comparison 
with Eq.~\eqref{eq:TD-spin-cano} and~\eqref{eq:TD-spin-Pz} are
\footnote{Notice that we here adopt the \textit{right-handed} representation 
of the Marck tetrad, instead of the left-handed one widely used 
in the literature [cf. Refs.~\cite{WitzHJ.19,VdM.20,MaPoWa.22,DruHug.I.22}]; 
the details are explained in Appendix.~\ref{app:Marck}.}
%
\begin{subequations}\label{eq:TD-spin-MarckR-tr}
    \begin{align}
        S_t^{\tTD} & = 
        -\frac{{\mathcal S}_{\perp}}{\sqrt{r^2 \mu^2 + L^2}}
        \left(
        \frac{L p_{t}}{\mu} \cos \left( \phi_{s} + \psi_{p} \right) 
        +
        r f p_{r} \sin \left( \phi_{s} + \psi_{p} \right) 
        \right)\,,\\
        S_r^{\tTD} & = 
        -\frac{{\mathcal S}_{\perp}}{\sqrt{r^2 \mu^2 + L^2}}
        \left(
         \frac{L p_{r}}{\mu} \cos \left( \phi_{s} + \psi_{p} \right) 
        +
        \frac{r p_{t}}{f} \sin \left( \phi_{s} + \psi_{p} \right) 
        \right)\,,
    \end{align}
\end{subequations}
and 
\begin{equation}\label{eq:TD-spin-MarckR-thph}
       p_{\phi} S_{\theta}^{\tTD} - p_{\theta} S_{\phi}^{\tTD}
       =
       -\frac{r L}{\csc \theta}  {\mathcal S}_{\parallel}\,.
\end{equation}
where we used $\pi_{\alpha} = p_{\alpha} + O(\epsilon)$. 
Here, the constant-angle $\phi_s$ defines the orientation of the TD spin (in the Marck frame), 
and ${\psi_p} = {\psi_p}({\bar \tau})$ describes the Marck's precession phase 
that grows with the proper time $\bar \tau$ 
according to the differential equation in Eq.~\eqref{eq:ODE-Mark-angleR}. 

%
%
%

Equation~\eqref{eq:TD-spin-MarckR-thph} immediately allows us to explain 
why the constant of motion $\pi_{\varpi}$ in Eq.~\eqref{eq:Pz-SO-coupling} 
is interpreted as the ``parallel component'' of the spin.
A direct comparison of these equations demonstrates that
\begin{equation}\label{eq:TD-spin-para-perp}
    \pi_{\varpi} = {\mathcal S}_{\parallel}\,,
    \quad \text{with} \quad 
    S_{\circ}^2 - \pi_{\varpi}^2 = {\mathcal S}_{\perp}\,,
\end{equation}
and we find that $\pi_{\varpi}$ is equivalent to the amplitude of the third leg of Marck tetrad 
${\mathcal S}_{\parallel}$ 
aligned with the direction of the Killing-Yano angular momentum vector ${\cal L}^{a}$.
Similarly, the overall amplitude $\sqrt{S_{\circ}^2 - \pi_{\varpi}^2}$ 
in Eq.~\eqref{eq:TD-spin-cano} can be interpreted as the ``perpendicular'' component 
of the TD spin ${\mathcal S}_{\perp} = \sqrt{S_{\circ}^2 - {\mathcal S}_{\parallel}^2}$, 
in the sense that it is perpendicular to ${\cal L}^{a}$.

Upon inserting Eq.~\eqref{eq:TD-spin-para-perp} within Eq.~\eqref{eq:TD-spin-cano}, 
$S_{t}^{\tTD}(\tilde{x}^i,\tilde{\pi}_i)$ and $S_{r}^{\tTD}(\tilde{x}^i,\tilde{\pi}_i)$
can be directly compared with each corresponding component 
in Eq.~\eqref{eq:TD-spin-MarckR-tr} through the covariance of the TD spin $S^a_\tTD$ 
(in contrast to the non-covariant $3$-spin vectors $\vec{S}$). 
This enables a comparison of the phases $\varpi$ and $\phi_{s} + \psi_{p}$ 
observed in the TD spin. 
Using $p_{a} = \pi_{a} + O(\epsilon)$ and $L = \pi_{\ell}$, 
this comparison gives rise to two linear relations 
between the phases of the TD spin, which are expressed in the matrix form 
\begin{equation}\label{eq:Marck-vs-cano}
    \left(\begin{matrix}
        \cos \Theta & -\sin \Theta \\
        \sin \Theta & \cos \Theta \\
    \end{matrix}\right)
    \left(\begin{matrix}
        \cos {\varpi}  \\  \sin {\varpi}
    \end{matrix}\right)
    =
    \left(\begin{matrix}
        \cos \left( \phi_{s} + \psi_{p} \right)  \\  
        \sin \left( \phi_{s} + \psi_{p} \right) 
    \end{matrix}\right)\,,
\end{equation}
where the auxiliary $\bar {\tau}$-dependent angle $\Theta(\bar \tau)$ is defined by 
\begin{equation}\label{eq:def-Theta}
    \cos \Theta 
    =  \frac{\mu r \pi_t}{\nu_{0} \sqrt{ f ( r^2 \mu^2 + \pi_\ell^2 )}} + O(\epsilon)
    \quad \text{and} \quad
    \sin \Theta 
    = -\frac{\pi_{\ell} \tilde{\pi}_{r} \sqf}{\nu_{0} \sqrt{ r^2 \mu^2 + \pi_\ell^2 }} + O(\epsilon)\,,
\end{equation}
that satisfies $\cos^2 \Theta + \sin^2 \Theta = 1 + O(\epsilon^2)$. 
Because the matrix in the left-hand side of Eq.~\eqref{eq:Marck-vs-cano} represents a counterclockwise rotation by an angle $\Theta$, 
Eq.~\eqref{eq:Marck-vs-cano} implies that 
\begin{equation}\label{eq:z-vs-psi_p}
    \Theta + {\varpi} = \phi_{s} + \psi_{p}  
    \quad \text{(mod $2 \pi$)}\,.
\end{equation}
It is in this sense that the variable ${\varpi}$ is related to the ``spin precession''.
When the canonical variable ${\varpi}$ is rotated 
by the angle $\Theta$ (in the phase space $\mcP$), 
it coincides with the Marck precession phase $\psi_p$ with the phase offset $\phi_s$ 
(in the Schwarzschild spacetime).

The link in Eq.~\eqref{eq:z-vs-psi_p} is further supported 
by consistency among the three differential equations for ${\varpi},\,\Theta$ and $\psi_p$. 
Taking into account the Hamilton's equation in Eqs.~\eqref{Hameqcano}, 
the ${\bar {\tau}}$-derivative of Eq.~\eqref{eq:def-Theta} returns  
\begin{equation}\label{eq:dot-Theta}
    \frac{\ud \Theta}{\ud \bar{\tau}}
    =
    \frac{\mu \pi_t \pi_{\ell}\left( 
    (2 \nu_0^2 + 3 f \tilde{\pi}_r^2) f^2 - (f \tilde{\pi}_r^2 + \pi_t^2) f + \pi_t^2
    \right)}
    {2 \nu_0^2 r^2 f \left( \pi_t^2  - f^2 \tilde{\pi}_r^2  \right)}\,.
\end{equation}
Combining this equation and the Hamilton's equation for ${\varpi}$ in Eq.~\eqref{HamEqz}, 
we correctly recover the differential equation for the classical Marck angle $\psi_{p}$
in Eq.~\eqref{eq:ODE-Mark-angleR}. 
This result should not be taken for granted because ${\varpi}$ and $\Theta$ in the phase space $\mcP$ 
are irrelevant to any notion of parallel transport along a Schwarzschild geodesic $\scL_0$, 
from which the differential equation for $\psi_{p}$ is defined. 

As a conclusion of this subsection, we need to comment on the angle $\Theta$ 
that appears in Eq.~\eqref{eq:z-vs-psi_p}. 
The physical meaning of this angle is currently unknown. 
We have attempted to derive Eqs.~\eqref{eq:def-Theta} and ~\eqref{eq:dot-Theta} 
within our Hamiltonian formalism 
(i.e. without relying on the comparison with Marck's representation of $S_{a}^{\tTD}$), 
but, for reasons that remain unclear to us, this approach does not seem to
not seem to be effective in this case.\footnote{
Technically, it was challenging to infer the specific combination $r^2 \mu^2 + \pi_{\ell}$ that enters the denominator of Eq.~\eqref{eq:def-Theta} (as well as Eq.~\eqref{eq:ODE-Mark-angleR}) from the canonical Hamiltonian system.}
The question of why the Marck's precession phase $\psi_p$ must be a superposition of the canonical variable ${\varpi}$ and "the" angle $\Theta$ (when viewed from the phase space $\mcP$) is a profound one; we leave it open for the time being.

\section{Details relative to the reduction from $\mcN$ to $\mcP$} 
\label{app:detailsNtoP}
This appendix provides a detailed account of some technical developments 
associated with the reduction from $\mcN$ to $\mcP$, 
which have been omitted from the main texts. 

\subsection{Computing Poisson brackets involving the TD SSCs}
\label{app:PB-TDSSCs}

We give here explicit expressions for the $\mcN$-brackets (i.e., Poisson brackets on $\mcN$) 
between the constraints (i.e. TD SSC) $C_0$ and $C_1$ in Eq.~\eqref{two} 
and the $10$ non-canonical variables $y_{\mcP}^{\text {C}}$ in Eq.~\eqref{choiceP}, 
using the definition in Eq.~\eqref{DB}.
We shall follow the general strategy adopted in Sec.~\ref{sub2sec:P-brackets}. 

A key technical input here is the relation between the Andoyer spin variables 
$(\Sigma^I,\Delta^I)$ in Eq.~\eqref{spinsympSSC-Ad} and $(\Gamma, \pi_{\alpha})$, 
which are obtained from the combination of five equations: 
the two constraints in Eq.~\eqref{two}, 
the definition of $(\Gamma, \pi_{\varpi})$ in Eq.\eqref{choiceP} 
and the Casimir invariant $S_\star$ in Eq.~\eqref{CasimirSigDel}; 
crucially it satisfies $S_\star^2 = 0$.
We find that 
\begin{subequations}\label{eq:sol-SigmaDelta}
    \begin{alignat}{6}
        \Delta^1 &= -\frac{ \pi_{\ell} \pi_{\varpi} }{\mu r}\,,
        ~ & ~
        &{\Delta^2} = 
        -\frac{f} {\pi_{t}}\left(\pi_{r} + \frac{\pi_{\ell}}{r} \Gamma \right) {\Sigma^3}\,,
        ~ & ~
        & \Delta^3 = \frac{\pi_{r} \pi_{\varpi} \sqf}{\mu}\,,\\
        {\Sigma^1} &= - \sqf \Gamma \, {\Sigma^3}\,,
        ~&~
        & \Sigma^2 = \frac{\pi_{t} \pi_{\varpi} }{\mu \sqf}\,.
        ~ & ~
        ~ & ~
    \end{alignat}
\end{subequations}
The remaining $\Sigma^3$ is also related to $(\Gamma, \pi_{\varpi})$
by the other Casimir invariant (or the spin norm) $S_{\circ}$
with the help of Eq.~\eqref{eq:sol-SigmaDelta}, 
and this is calculated as 
\begin{equation}
    {\Sigma^3} 
    =
    \frac{1}{f}
    \left(
    \frac{r^2 \pi_{t}^2 \left( S_{\circ}^2 - \pi_{\varpi}^2\right)}
    {\left( r^2 \pi_{t}^2 - f \pi_{\ell}^2 \right)\Gamma^2 
    - 2 r f \pi_{r} \pi_{\ell} \Gamma 
    + \left( r^2 \mu^2 + \pi_{\ell}^2\right)}
    \right)\,,
\end{equation}
However, the explicit form of $\Sigma^3$ is not always required here. 
As a consequence of our choice of $(\Gamma, \pi_{\varpi})$, 
$\Sigma^3$ encountered in the calculation mostly takes the form of $\Delta^{2} / \Sigma^{3}$ 
and $\Sigma^{1} / \Sigma^{3}$, 
which is directly obtained by Eq.~\eqref{eq:sol-SigmaDelta} only; 
we again find it advantageous our choice of $y_{\mcP}^{\text {C}}$. 

With Eq.~\eqref{eq:sol-SigmaDelta}, a straightforward calculation reveals that 
the non-vanishing brackets are given by 
\begin{subequations} \label{constbrac}
    \begin{align}
        \{t,C^1\} &= -\frac{\pi_{\ell} \pi_{\varpi}}{r \mu \sqrt{f}} \,, \\
        \{r,C^0\} &=\frac{\sqrt{f} \pi_{\ell} \pi_{\varpi}}{r \mu}  \,, \\
        \{\pi_r,C^0\} &= -\left(1 - \frac{M}{r}\right) \frac{\pi_{r} \pi_{\ell} \pi_{\varpi}}{r^2 \mu \sqrt{f}} \,, \\
        \{\pi_r,C^1\} &= \left(1 - \frac{3M}{r}\right) \frac{\pi_{t} \pi_{\ell} \pi_{\varpi}}{r^2 \mu f^{3/2}}\,, \\
        \{\ell,C^0\} &= -\frac{\sqrt{f} \pi_{r} \pi_{\varpi}}{r \mu}  \,, \\
        \{\ell,C^1\} &=  \frac{\pi_{t} \pi_{\varpi}}{ r \mu \sqrt{f}} \,, \\
        \{\Gamma,C^0\} &= \frac{\pi_r \pi_{\ell} \sqrt{f}}{\pi_t r} 
        + \left( \frac{2 \pi_{\ell}^2}{r^2}- \frac{\pi_t^2}{f} \right) \frac{\sqrt{f}}{\pi_t} \Gamma - \frac{\pi_r \pi_{\ell} f^{3 / 2}}{\pi_t r} \Gamma^2 \,, \\
        \{\Gamma,C^1\} &= - \frac{\pi_{\ell}}{r \sqrt{f}} + \pi_r \sqrt{f}\, \Gamma \,, \\
        \{\pi_{\varpi},C^0\} &= \frac{\pi_t \pi_{\varpi}}{\sqrt{f}} \,, \\
        \{\pi_{\varpi},C^1\} &=  - \sqf \pi_r  \pi_{\varpi}\,.
    \end{align}
\end{subequations}
All the unlisted brackets vanishes or are obtained from the anti-symmetry relation 
for the brackets above. 
We consistently carried out the bracket calculation in Eq.~\eqref{constbrac} 
through linear-in-spin order, 
except for that involving $\Gamma$, for which is suffices to work at zero-th order in spin; 
recall the relation $\pi_s \propto \fK \Gamma + O(\epsilon^2)$ from Eq.~\eqref{choiceP}.

\subsection{Fixing the homogeneous freedom in the transformation 
in Eq.~\eqref{eq:sol-y3}} 
\label{app:homogeneous_PDEs}

This Appendix provides the derivation of the remaining unknown functions 
$\tilde{t}_3,\tilde{\pi}_3,\,\tilde{\ell}_3$ and ${\varpi}_3$ in Eq.~\eqref{eq:sol-y3}.

In principle, these four functions can be determined 
by the remaining six $\mcP$-brackets:  
$
\{ \tilde{t}, \tilde{\ell} \}^{\mcP} = 0,\,
\{ \tilde{t}, {\varpi} \}^{\mcP} = 0,\,
\{ \tilde{t}, \tilde{\pi}_{r} \}^{\mcP} = 0,\,
\{ \tilde{\pi}_r, \tilde{\ell} \}^{\mcP} = 0,\,
\{ \tilde{\pi}_r, {\varpi} \}^{\mcP} = 0,\,
$
and
$\{ \tilde{\ell}, {\varpi} \}^{\mcP} = 0$, which we did not use yet. 
However, the calculation strategy adopted in previous subsections 
cannot be applied here because these brackets produce trivial PDEs 
for $\tilde{t}_3,\tilde{\pi}_3,\,\tilde{\ell}_3$ and ${\varpi}_3$ (after applying the Leibniz rule).  
The $\mcP$-brackets in Eq.~\eqref{allPbrackets} 
between $(r, \pi_t, \pi_\ell, \pi_{\varpi})$ are all zero, 
and we only find $0 = 0$ identically. 
To overcome this limitation, a suitable alternate must be employed. 

Our alternative approach to determine the functions $\tilde{t}_3,\tilde{\pi}_3,\,\tilde{\ell}_3$ and ${\varpi}_3$ 
is to solve the ``inverse'' problem;  
we rather express the $\mcP$-brackets for the non-canonical variables $y_{\mcP}^{\text {C}}$ 
in Eq.~\eqref{allPbrackets} 
as a function of our target canonical variables $y_{\mcP}^{\text {F}}$ 
in Eq.~\eqref{eq:def-yF}, and obtain the PDEs 
for $\tilde{t}_3,\tilde{\pi}_3,\,\tilde{\ell}_3$ and ${\varpi}_3$. 
The key observation is that $\tilde{t}_3,\tilde{\pi}_3,\,\tilde{\ell}_3$ and ${\varpi}_3$ have only four arguments 
$r, \pi_t, \pi_\ell$ and $\pi_{\varpi}$, 
which are common variables used for both the non-canonical variables $y_{\mcP}^{\text {C}}$ 
in Eq.~\eqref{choiceP} and our target canonical variables $y_{\mcP}^{\text {F}}$ 
in Eq.~\eqref{eq:def-yF}. 
As a result, we can now express $y_{\mcP}^{\text {C}}$ 
in terms of $y_{\mcP}^{\text {F}}$ by inverting Eq.~\eqref{eq:sol-y3} 
(through linear order in spin). We have that 
\begin{subequations}\label{eq:yF-to-yC}
    \begin{align}
        t &= (\tilde{t}  - \tilde{t}_3 )
        - \frac{\left(\tilde{\pi}_r - \tilde{\pi}_3 \right) \pi_\ell \pi_{\varpi}}{\nu_0^2 r \mu} \,, \\
        \pi_r &= \left( \tilde{\pi}_r - \tilde{\pi}_3 \right)
        \left( 1 + \left( 1 - \frac{3M}{r} \right) \frac{\pi_t \pi_\ell \pi_{\varpi}}{f \nu_0^2 r^2 \mu} \right) \,, \\
        \ell &= (\tilde{\ell} - \tilde{\ell}_3)
        + \frac{\left(\tilde{\pi}_r - \tilde{\pi}_3 \right) \pi_t \pi_{\varpi}}{\nu_0^2 r \mu} \,, \\
        \Gamma &= \frac{1}{\nu_0^2} 
        \left( \frac{\left(\tilde{\pi}_r - \tilde{\pi}_3 \right) \pi_\ell}{r} 
        + \frac{\mu \pi_t}{f} \tan ({\varpi} - {\varpi}_3 ) \right)\,,
        \label{eq:yF-to-Gamma}
    \end{align}
\end{subequations}
where we understood that $\pi_r$ in $\mu$ in Eq.~\eqref{eq:mu-G} is replaced with 
$\tilde{\pi}_r = \pi_r + O(\epsilon)$.
With this relation at hand, it is a simple matter to obtain an alternative expression 
for the $\mcP$-brackets displayed in Eq.~\eqref{allPbrackets} 
but in terms of $y_{\mcP}^{\text {F}}$, 
using the standard definition of the Poisson brackets 
for any two functions $F$ and $G$ of canonical variables: 
\begin{multline}\label{eq:PBCano-yF}
    \{F ,\, G\}^{\mcP} 
    = 
  \frac{\partial F}{\partial \tilde{t}} \frac{\partial G}{\partial\pi_t} - 
  \frac{\partial F}{\partial \pi_t} \frac{\partial G}{\partial \tilde{t}} +
  \frac{\partial F}{\partial r} \frac{\partial G}{\partial \tilde{\pi}_r} -
  \frac{\partial F}{\partial \tilde{\pi}_r} \frac{\partial G}{\partial r}
   +
  \frac{\partial F}{\partial \tilde{\ell}} \frac{\partial G}{\partial \pi_{\ell}} -
  \frac{\partial F}{\partial \pi_{\ell}} \frac{\partial G}{\partial \tilde{\ell}} +
  \frac{\partial F}{\partial {\varpi}} \frac{\partial G}{\partial \pi_{\varpi}} - 
  \frac{\partial F}{\partial \pi_{\varpi}} \frac{\partial G}{\partial {\varpi}}\,,
\end{multline}
by setting $F = y_{\text {C}}^i$ and $G = y_{\text {C}}^j$ 
(where  $y_{\text {C}}^i \in y_{\mcP}^{\text {C}}$);  
we excluded the canonical pair $(\nu, \pi_\nu)$ from Eq.~\eqref{eq:PBCano-yF} 
that is irrelevant to the transformation in Eq.~\eqref{looktransfo}. 
The results are too unwieldy to be presented here, 
but a bracket by bracket comparison between the resulting expressions 
and Eq.~\eqref{allPbrackets} reveals that $\tilde{t}_3,\tilde{\pi}_3,\,\tilde{\ell}_3$ and ${\varpi}_3$ 
must satisfy 
\begin{equation}\label{eq:hom-sol-Pi3z3}
    \tilde{\pi}_3 = 0\,, \quad {\varpi}_3 = 0\,,
    \quand
    \frac{\partial}{\partial \pi_\ell} \tilde{t}_3(\pi_t, \pi_\ell) 
    - \frac{\partial}{\partial \pi_t} \tilde{\ell}_3 (\pi_t, \pi_\ell) = 0\,.
\end{equation}
to ensure the consistency of the same $\mcP$-brackets 
from two different calculations.

Still, there is freedom of choice with $ \tilde{t}_3(\pi_t, \pi_\ell) $ and $ \tilde{\ell}_3 (\pi_t, \pi_\ell)$ 
in Eq.~\eqref{eq:hom-sol-Pi3z3}. 
We are unable to identify a physical argument for controlling this freedom, 
which rather seems a built-in property of our canonical coordinate $y_{\mcP}^{\text {F}}$.
We therefore leave these two functions arbitrary for the time being, 
but fix them in this paper by make a specific choice 
\begin{equation}\label{eq:hom-sol-T3Psi3}
    \tilde{t}_3 = 0\,, \quand \tilde{\ell}_3 = 0\,,
\end{equation}
which best simplify the expressions in Eq.~\eqref{eq:sol-y3} 
(though different choices of such $\tilde{t}_3$ and $ \tilde{\ell}_3$ are possible). 
This is the solution that we adopted in Eq.~\eqref{eq:hom-sol} 
in the main text.

\bibliography{ListeRef.bib,ListeRef-Sis}

\end{document}